\documentclass[twocolumn, twocolappendix, trackchanges]{aastex63}

\usepackage{amsmath}
\usepackage{booktabs}
\usepackage{longtable}
\usepackage{rotating}
\usepackage{graphicx}
\usepackage{hyperref}

\usepackage{xstring} 

\newcommand{\val}[1]{%
  \IfEqCase{#1}{%
{nstars-legacy}{719}
{nstars-new}{135}
}[\PackageError{tree}{Undefined option to tree: #1}{}]%
}%

\newcommand{\nstars}{719}

\newcommand{\knownplanets}{164}
\newcommand{\newplanets}{14}

\newcommand{\dBIC}{$\Delta\mathrm{BIC}$}

\newcommand{\rsun}{R$_{\odot}$}
\newcommand{\msun}{M$_{\odot}$}

\newcommand{\mearth}{$M_{\oplus}$}

\newcommand{\mjup}{$M_{\textrm{J}}$}

\newcommand{\msini}{$M \sin{i}$}

\newcommand{\caii}{\ion{Ca}{2}~H \& K}

\newcommand{\ms}{\text{m\,s}$^{-1}$}

\newcommand{\msd}{\text{m\,s}$^{-1}$\,\text{d}$^{-1}$}
\newcommand{\msdd}{\text{m\,s}$^{-1}$\,\text{d}$^{-2}$}

\graphicspath{{./}{figures/}}

\received{July 3, 2020}
\revised{November 17, 2020}
\accepted{February 2, 2021}
\submitjournal{AAS Journals}

\shorttitle{California Legacy Survey I: The Catalog}
\shortauthors{Rosenthal et al.}

\begin{document}
\pagenumbering{arabic}

\title{The California Legacy Survey I. A Catalog of 178 Planets from Precision Radial Velocity Monitoring of 719 Nearby Stars over Three Decades}

\correspondingauthor{Lee J.\ Rosenthal}
\email{lrosenth@caltech.edu}

\author[0000-0001-8391-5182]{Lee J.\ Rosenthal}
\affiliation{Cahill Center for Astronomy $\&$ Astrophysics, California Institute of Technology, Pasadena, CA 91125, USA}

\author[0000-0003-3504-5316]{Benjamin J.\ Fulton}
\affiliation{Cahill Center for Astronomy $\&$ Astrophysics, California Institute of Technology, Pasadena, CA 91125, USA}
\affiliation{IPAC-NASA Exoplanet Science Institute, Pasadena, CA 91125, USA}

\author[0000-0001-8058-7443]{Lea A.\ Hirsch}
\affiliation{Kavli Institute for Particle Astrophysics and Cosmology, Stanford University, Stanford, CA 94305, USA}

\author[0000-0002-0531-1073]{Howard T.\ Isaacson}
\affiliation{Department of Astronomy, University of California Berkeley, Berkeley, CA 94720, USA}

\author[0000-0001-8638-0320]{Andrew W.\ Howard}
\affiliation{Cahill Center for Astronomy $\&$ Astrophysics, California Institute of Technology, Pasadena, CA 91125, USA}

\author[0000-0001-9408-8848]{Cayla M.\ Dedrick}
\affiliation{Cahill Center for Astronomy \& Astrophysics, California Institute of Technology, Pasadena, CA 91125, USA}
\affiliation{Department of Astronomy \& Astrophysics, The Pennsylvania State University, 525 Davey Lab, University Park, PA 16802, USA}

\author[0000-0001-7730-0202]{Ilya A.\ Sherstyuk}
\affiliation{Cahill Center for Astronomy $\&$ Astrophysics, California Institute of Technology, Pasadena, CA 91125, USA}

\author[0000-0002-3199-2888]{Sarah C.\ Blunt}
\affiliation{Cahill Center for Astronomy $\&$ Astrophysics, California Institute of Technology, Pasadena, CA 91125, USA}
\affiliation{NSF Graduate Research Fellow}

\author[0000-0003-0967-2893]{Erik A.\ Petigura}
\affiliation{Department of Physics $\&$ Astronomy, University of California Los Angeles, Los Angeles, CA 90095, USA}

\author[0000-0002-5375-4725]{Heather A.\ Knutson}
\affiliation{Division of Geological and Planetary Sciences, California Institute of Technology, Pasadena, CA 91125, USA}


\author[0000-0003-0012-9093]{Aida Behmard}
\affiliation{Division of Geological and Planetary Sciences, California Institute of Technology, Pasadena, CA 91125, USA}
\affiliation{NSF Graduate Research Fellow}

\author[0000-0003-1125-2564]{Ashley Chontos}
\affiliation{Institute for Astronomy, University of Hawai‘i, Honolulu, HI 96822, USA}
\affiliation{NSF Graduate Research Fellow}

\author[0000-0003-0800-0593]{Justin R.\ Crepp}
\affiliation{Department of Physics, University of Notre Dame, Notre Dame, IN, 46556, USA}

\author[0000-0002-1835-1891]{Ian J.\ M.\ Crossfield}
\affiliation{Department of Physics and Astronomy, University of Kansas, Lawrence, KS, USA}

\author[0000-0002-4297-5506]{Paul A.\ Dalba}
\affiliation{Department of Earth and Planetary Sciences, University of California, Riverside, CA 92521, USA}
\affiliation{NSF Astronomy $\&$ Astrophysics Postdoctoral Fellow}

\author[0000-0003-2221-0861]{Debra A.\ Fischer}
\affiliation{Department of Astronomy, Yale University, New Haven, CT 06511, USA}

\author[0000-0003-4155-8513]{Gregory W.\ Henry}
\affiliation{Center of Excellence in Information Systems, Tennessee State University, Nashville, TN 37209 USA}

\author[0000-0002-7084-0529]{Stephen R.\ Kane}
\affiliation{Department of Earth and Planetary Sciences, University of California, Riverside, CA 92521, USA}

\author[0000-0002-6115-4359]{Molly Kosiarek}
\affiliation{Department of Astronomy and Astrophysics, University of California, Santa Cruz, CA 95064, USA}
\affiliation{NSF Graduate Research Fellow}

\author[0000-0002-2909-0113]{Geoffrey W. Marcy}

\author[0000-0003-3856-3143]{Ryan A.\ Rubenzahl}
\affiliation{Cahill Center for Astronomy $\&$ Astrophysics, California Institute of Technology, Pasadena, CA 91125, USA}
\affiliation{NSF Graduate Research Fellow}

\author[0000-0002-3725-3058]{Lauren M.\ Weiss}
\affiliation{Institute for Astronomy, University of Hawai‘i, Honolulu, HI 96822, USA}

\author[0000-0001-6160-5888]{Jason T.\ Wright}
\affiliation{Department of Astronomy and Astrophysics, The Pennsylvania State University, University Park, PA 16802, USA}
\affiliation{Center for Exoplanets and Habitable Worlds, The Pennsylvania State University, University Park, PA 16802, USA}
\affiliation{Penn State Extraterrestrial Intelligence Center, The Pennsylvania State University, University Park, PA 16802, USA}

\begin{abstract}

We present a high-precision radial velocity (RV) survey of \nstars\ FGKM stars, which host \knownplanets\ known exoplanets and \newplanets\ newly discovered or revised exoplanets and substellar companions. This catalog updated the orbital parameters of known exoplanets and long-period candidates, some of which have decades-longer observational baselines than they did upon initial detection. The newly discovered exoplanets range from warm sub-Neptunes and super-Earths to cold gas giants. We present the catalog sample selection criteria, as well as over 100,000 radial velocity measurements, which come from the Keck-HIRES, APF-Levy, and Lick-Hamilton spectrographs. We introduce the new RV search pipeline \href{https://california-planet-search.github.io/rvsearch/}{\texttt{RVSearch}} that we used to generate our planet catalog, and we make it available to the public as an open-source Python package. This paper is the first study in a planned series that will measure exoplanet occurrence rates and compare exoplanet populations, including studies of giant planet occurrence beyond the water ice line, and eccentricity distributions to explore giant planet formation pathways. We have made public all radial velocities and associated data that we use in this catalog.

\end{abstract}


\section{Introduction} \label{sec:intro}

 Expanding and characterizing the population of known exoplanets with measured masses, orbital periods, and eccentricities is crucial to painting a more complete picture of planet formation and evolution. A census of diverse exoplanets sheds light on worlds radically different than Earth, and can provide insight into how these planets, as well our own Solar System, formed. For instance, the mass, semi-major axis, and eccentricity distributions of giant planets can be used to constrain formation scenarios for these objects. \cite{Nielsen19} and \cite{Bowler20} used mass and eccentricity constraints from direct imaging surveys to show that planetary-mass gas giants likely form via core accretion \citep{Pollack96}, while more massive brown dwarfs and other substellar companions likely form via gravitational instability in protoplanetary disks \citep{Boss97}. The present-day architectures and orbital properties of planetary systems can also be used to constrain their migration histories. \cite{Dawson13} used a sample of giant planets with minimum masses and orbits constrained by radial velocity (RV) observations to provide evidence that giant planets orbiting metal-rich stars are more likely to be excited to high eccentricities or migrate inward due to planet-planet interactions. Many related questions remain unanswered. What is the mass--period distribution of planets out to 10 AU? How abundant are cold gas giants beyond the water ice line, and what can this abundance tell us about planet formation across protoplanetary disks? How do small, close-in planets arrive at their final masses and system architectures? What is the relationship between these warm small planets and cold gas giants; are their formation processes related? These questions can only be answered with an expansive and rigorously constructed census of exoplanets with measured masses and well-constrained orbits.

The community has made substantial progress on these fronts over the past few decades via targeted RV surveys. For instance, \cite{Bryan16} surveyed 123 known giant hosts to study outer giant companions; they found that half of all giants have an outer companion, with tentatively declining frequency beyond 3 AU. Similarly, \cite{Knutson14} found a 50$\%$ companion rate for transiting hot Jupiters using a sample of 51 stars. These two results suggest a planet formation process that favors giant multiplicity. On the small-planet front, \cite{Bryan19} constructed an RV survey of 65 super-Earth hosts and found a giant companion rate of $39\pm7\%$. This suggests that these two populations are related in some way. Some questions have seen conflicting answers, requiring further work with a more expansive RV survey. For instance, \cite{Fernandes19} studied planet occurrence as a function of orbital period by extracting the planetary minimum masses and periods, as well as completeness contours, from a catalog plot shown in \cite{Mayor11}, which presented a HARPS and CORALIE blind radial velocity survey of 822 stars and 155 planets over 10 yr (corresponding to a 4.6 AU circular orbit around a solar-mass star). The HARPS and CORALIE radial velocities were not published in \cite{Mayor11}, which measured giant planet occurrence as a function of orbital period out to 4000 days, in the range of the water ice line. \cite{Fernandes19} pushed out to low-completeness regimes and estimated a sharp falloff in occurrence beyond the water ice line. In sharp contrast, \cite{Wittenmyer20} used their radial velocities from the Anglo-Australian Planet Search to construct a blind survey of 203 stars and 38 giant planets over 18 yr. They found that giant planet occurrence is roughly constant beyond the water ice line, out to almost 10 AU. The discrepancy between these two results needs to be resolved.

The California Planet Search team \citep[CPS;][]{Howard10} has conducted many RV surveys over the past three decades, in order to find exoplanets, measure their minimum masses, and characterize their orbits. Many of these surveys were designed explicitly for the purpose of studying planet occurrence. Therefore, they used stellar samples that were constructed without bias toward stars with known planets, or an increased likelihood of hosting planets, such as metal-rich stars \citep{Gonzalez97}. For instance, the Keck Planet Search \citep{Cumming08} used 8 yr of Keck-HIRES data collected from 585 FGKM stars to study the occurrence of gas giants with periods as long as the survey baseline, measured the mass--period distribution of giant planets out to 5 AU, and found an increase in gas giant occurrence near the water ice line. The Eta-Earth Survey \citep{Howard10_Science} used 5 yr of Keck-HIRES data collected from 166 Sun-like stars to measure the occurrence of planets with orbital periods less than 50 days, ranging from super-Earths to gas giants, and found both an abundance of planets within 10 day orbits and a mass function that increases with decreasing mass for close-in planets. The APF-50 Survey combined 5 yr of high-cadence Automated Planet Finder data on a sample of 50 bright, nearby stars with 20 yr of Keck-HIRES data to constrain the mass function of super-Earths and sub-Neptunes, and discovered several planets of both varieties \citep{Fulton16}.

We constructed an aggregate survey from these distinct RV surveys, known hereafter as the California Legacy Survey (CLS), in order to measure exoplanet occurrence, particularly for planets with long orbital periods. We selected every star in the CPS catalog that was observed as part of an occurrence survey, added 31 CPS stars that satisfied our stellar selection criteria (described below), and regularly observed these stars using the Keck and UCO-Lick observatories. The California Legacy Survey contains 103,991 RVs, and reaches observational baselines beyond three decades. We wrote an automated planet search pipeline to systematically recover all planets that are detectable in the CLS and to measure the search completeness of each star's RV time series. We can use these completeness contours to calculate exoplanet occurrence rates with respect to planetary and host-star properties \citep[e.g.][]{Cumming08, Howard10_Science}.

In this paper, we present the CLS stellar sample and the \knownplanets\ known exoplanets orbiting these stars, as well as \newplanets\ newly discovered and vetted exoplanets and substellar companions. In Section 2, we describe our methodology for stellar selection. In Section 3, we describe the RVs measured for this survey. In Section 4, we describe our methods for computing the stellar properties of our sample. In Section 5, we describe the methods by which we search for exoplanets in the RVs, confirm their planetary status, and characterize their orbits. In Section 6, we present the catalog of known exoplanets, and describe in detail each of the new exoplanet candidates. In Section 7, we discuss the significance of our catalog, and conclude with plans for future work.

\section{Stellar Sample Selection}

Our goal for this study was to construct a sample of RV-observed FGKM stars and their associated planets, in order to provide a stellar and planetary catalog for occurrence studies. We want a survey that is quantifiably complete in some way, such as being volume- or magnitude-limited, so that we can perform unbiased occurrence measurements. One way to do this would be to observe every HD star within our desired range of stellar parameters, with the same cadence and a thirty year baseline. Given the constraints of finite observing time and instrumental magnitude limits, this is not possible. More importantly, there is no achievable, Platonic ideal of a quantifiably complete survey. However, we can approximate one by selecting CPS-observed stars that were originally chosen without bias toward a higher- or lower-than-average likelihood of hosting planets. Multiple CPS surveys, including the Keck Planet Search and Eta-Earth Survey, performed their stellar selection with these criteria.

We began with the Keck Planet Search sample, so that we can make direct comparisons to their results. We then supplemented those 585 stars with 135 stars that were not originally included as part of that sample, but they have since been observed by the CPS team and satisfy a set of criteria intended to ensure survey quality and statistical rigor for planet occurrence measurements. We selected these criteria to ensure data quality, both of individual measurements and stellar datasets, and proper stellar selection, without bias toward known or likely planet hosts, which would skew our occurrence measurements. We included CPS-observed stars that have at least 20 total RVs and at least 10 High Resolution Echelle Spectrometer (HIRES) RVs collected after the HIRES CCD upgrade in 2004, to guarantee enough RVs for well-constrained Keplerian fits, and have an observational baseline of at least 8 yr, which is the maximum baseline of the \cite{Cumming08} sample at the time of publication. All stars in the Keck Planet Search sample pass these criteria, since we have collected more than 10 new HIRES RVs for each of them since 2004.

In order to ensure proper stellar selection, we did not include CPS-observed stars that were chosen for surveys that deliberately selected known planet hosts, metal-rich stars, or non-main-sequence stars, since these surveys would bias planet occurrence measurements. We excluded stars that were observed as part of the ``N2K'' and ``M2K'' surveys, which targeted metal rich stars to search for gas giants \citep{Fischer05, Apps10}. We excluded all massive stars that were observed as part of a search for planets orbiting subgiants \citep{Johnson11}, since that survey used a particular observing strategy geared solely toward detecting giant planets. We excluded all young stars that were selected for CPS observing based on photometric IR excess, since such stars were selected for an increased probability of planet occurrence \citep{Hillenbrand15}. We excluded all stars from the ``Friends of Hot Jupiters'' surveys, which targeted known planet hosts \citep{Knutson14}. For the same reason, we excluded all stars that were observed as part of \emph{Kepler}, K2, TrES, HAT, WASP, or KELT transiting planet surveys \citep{Bakos02, Alonso04, Pollacco06, Pepper07, Borucki16}.

\begin{figure}[ht!]
\includegraphics[width = 0.49\textwidth]{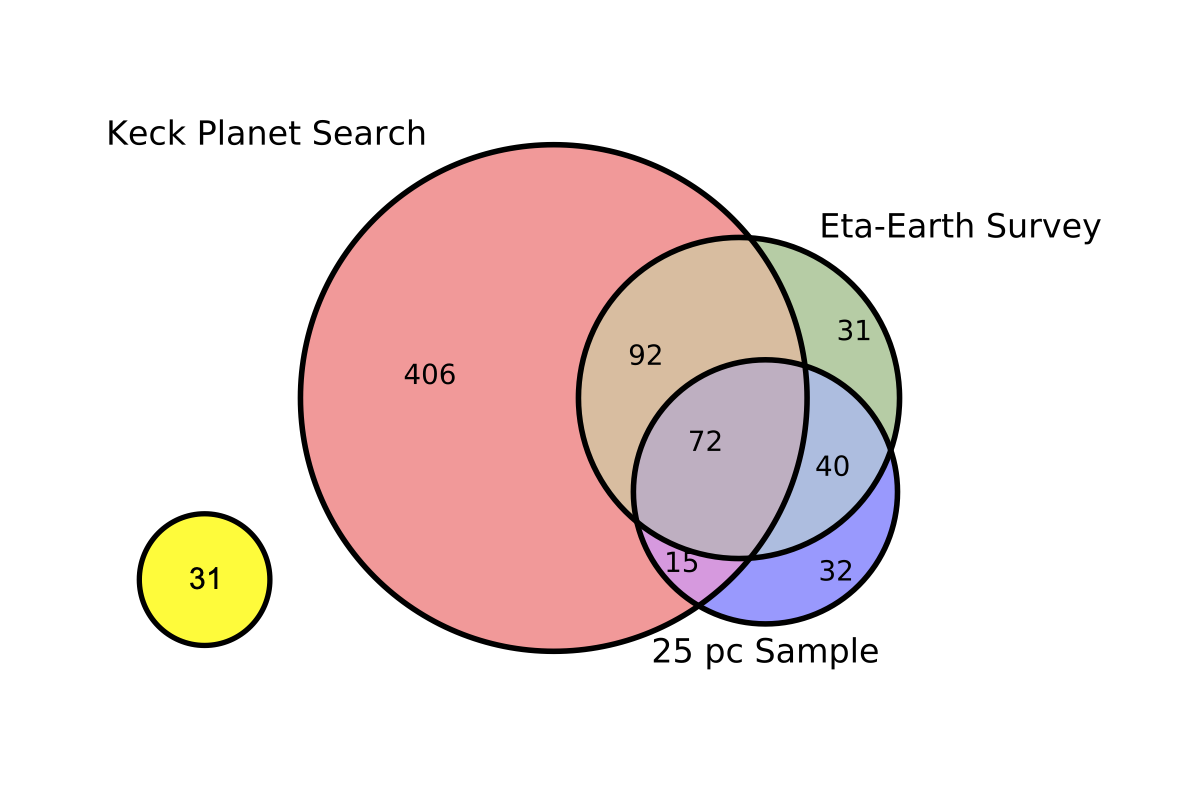}
\caption{Venn diagram showing the overlap between the stars in the Keck Planet Search sample \citep{Cumming08}, the Eta-Earth sample \citep{Howard10}, and a 25 pc northern hemisphere volume-limited survey (Hirsch et al. in prep). 31 stars in the California Legacy Survey do not belong to the union of these three surveys.}
\label{fig:venn}
\end{figure}

This selection process left us with \val{nstars-legacy} stars. Figure \ref{fig:venn} shows the entire CLS samples as a Venn diagram, illustrating the overlap of the \cite{Cumming08} sample with the Eta-Earth \citep{Howard10_Science} and 25 pc northern hemisphere volume-limited \citep{Hirsch20} samples. The 25 pc sample includes 255 G  and early K dwarfs with apparent $V$ magnitudes ranging from $V\approx 3$ to $V\approx 9$. These stars have a median temperature of 5360 K and a median mass of 0.86 $M_{\odot}$. The median number and duration of RV observations for this sample was 71 RVs spanning 21 yr, while the minimum number and duration of observations in the sample was 20 RVs spanning 3 yr.  The architects of all three of these surveys designed them for planet occurrence studies. Therefore, they did not construct these catalogs by selecting on properties known to correlate or anticorrelate with planet occurrence. There are only 31 stars in the California Legacy Survey that do not belong to any of these three surveys but do still pass of our selection criteria. This survey has no hard constraints on distance, apparent magnitude, or color, as seen in Figure \ref{fig:stellar_hists}.

\section{Observations}

\subsection{Keck-HIRES}

HIRES \citep{Vogt94} has been in operation on the Keck I Telescope since 1994 and has been used to measure stellar RVs via the Doppler technique since 1996 \citep{Cumming08}.  This technique relies on measuring the Doppler shift of starlight relative to a reference spectrum of molecular iodine, which is at rest in the observatory frame \citep{Butler1996}.  We consistently set up HIRES with the same wavelength format on the CCDs for each observation and followed other standard procedures of CPS \cite{Howard10}. With the iodine technique, starlight passes through a glass cell of iodine gas heated to 50$^\circ$ C, imprinting thousands of molecular absorption lines onto the stellar spectrum, which act as a wavelength reference. We also collected an iodine-free ``template'' spectrum for each star.  This spectrum is naturally convolved with the instrumental point spread function (PSF) and is sampled at the resolving power of HIRES ($R$ = 55,000--86,000, depending on the width of the decker used).  These spectra are deconvolved using PSF measurements from spectra of featureless, rapidly rotating B stars with the iodine cell in the light path. The final, deconvolved intrinsic stellar spectra serve as ingredients in a forward-modeling procedure from which we measure relative Doppler shifts of each iodine-in spectrum of a given star \citep{Valenti95}. We also used this process to compute uncertainties on the Doppler shifts.  The uncertainty for each measurement is the standard error on the mean of the RVs for 700 segments of each spectrum (each 2\,\AA wide) run through the Doppler pipeline.  We distinguish between ``pre-upgrade'' RVs (1996--2004; $\sim$3 \ms\ uncertainties) and ``post-upgrade'' RVs (2004--present; $\sim$1 \ms\ uncertainties).  In 2004, HIRES was upgraded with a new CCD and other optical improvements.  We account in the time series modeling for different RVs zero points ($\gamma$) for data from the two different eras.

The RVs reported here stem from HIRES observations with a long history.  The RVs from 1996 to 2004 are based on HIRES spectra acquired by the California \& Carnegie Planet Search (CCPS) collaboration and were reported in \cite{Cumming08}.  CCPS continued to observe these stars, but split into two separate collaborations: CPS and the Lick-Carnegie Exoplanet Survey (LCES). This paper principally reports results from 41,804 CPS and CCPS HIRES spectra that were obtained and analyzed by our team during 1996--2020.  In addition, we have included RVs computed by our pipeline for 7530 spectra of CLS stars taken by LCES during 2008--2014.  These HIRES spectra were acquired with the same instrumental setup as the CPS spectra and are publicly available in the Keck Observatory Archive. \cite{Butler2017} separately published RVs based on the same HIRES observations from CCPS, CPS, and LCES for the 1996--2014 time span. The LCES and CPS Doppler pipelines diverged in $\sim$2007. \cite{Tal-Or19} uncovered the 2004 zero-point offset, which we model with two independent offsets. They also claimed two second-order systematics in the LCES 2017 dataset: a long-term drift of order $<$ 1 \ms, and a correlation between stellar RVs and time of night with respect to midnight. They estimated the long-term drift by averaging the zero points of three RV-quiet stars on each night, where possible. However, by our estimates, even the quietest stars exhibit 1--2 \ms jitter in HIRES time series. Averaging the zero points of three such stars will likely yield a scatter of 1 \ms across many nights. Additionally, they did not remove planet RV signals from their data before estimating the linear correlation between RV and time of night, and it is unclear how they derived the uncertainty in that correlation.

\subsection{Automated Planet Finder}

The APF-Levy spectrograph is a robotic telescope near the summit of Mt. Hamilton, designed to find and characterize exoplanets with high-cadence Doppler spectroscopy \citep{Vogt14, Radovan14}. The facility consists of a 2.4-m telescope and the Levy Spectrometer, which has been optimized for optical Doppler shift measurements. The Doppler pipeline that was developed for Keck-HIRES also extracts RV measurements from APF spectra. Most of the APF data in the California Legacy Survey was collected as part of the APF-50 Survey \citep{Fulton17}, the stellar sample of which was drawn entirely from the Eta Earth sample. These two surveys have slightly different selection criteria. While both surveys have a distance cut $d < 25$ pc and luminosity cut $M_V < 3$, Eta-Earth cuts on apparent magnitude $V < 11$, whereas APF-50 has $V < 7$; Eta-Earth cuts on chromospheric activity $\mathrm{log}R'_{\mathrm{HK}} < -4.7$, whereas APF-50 has $\mathrm{log}R'_{\mathrm{HK}} < -4.95$; and Eta-Earth cuts on declination $> -30^{\circ}$, whereas APF-50 has declination $> -10^{\circ}$. These stricter cuts were made to ensure higher data quality for the high-cadence APF survey.

\subsection{Lick-Hamilton}

The Hamilton Spectrograph is a high-resolution echelle spectrometer, attached to the 3 m Shane telescope on Mt. Hamilton. Beginning in 1987, and ending in 2011 with a catastrophic iodine cell failure, the Lick Planet Search program \citep{Fischer14} monitored 387 bright FGKM dwarfs to search for and characterize giant exoplanets. This was one of the first surveys to produce precise RVs via Doppler spectroscopy with iodine cell calibration, and yielded RVs with precision in the range 3--10 \ms. The Lick Planet Search overlaps heavily with the Keck Planet Search and other CPS surveys, since these surveys drew from the same bright-star catalogs.

\subsection{Activity Indices}

For each HIRES and APF spectrum from which we measure radial velocities, we also measure the strength of emission in the cores of the \caii\ lines (S-values) following the techniques of \cite{Isaacson10} and \cite{Robertson14}. There is a small, arbitrary offset between the HIRES and APF activity indices. We adopted uniform S-value uncertainties with values of 0.002 and 0.004 for HIRES and APF respectively. We provide activity indices along with our RV measurements. Missing values are the result of sky contamination and/or low SNR.

\subsection{APT Photometry}

We collected long-term photometric observations of the subset of our sample that were included in the APF-50 survey \citep{Fulton17}, in order to search for evidence of rotation-induced stellar activity. We collected these measurements with Tennessee State University's Automated Photometric Telescopes (APTs) at Fairborn Observatory as part of a long-term program to study stellar magnetic activity cycles \citep{Lockwood13}. Most stars have photometric datasets spanning 15 -- 23 yr. The APTs are equipped with photomultiplier tubes that measure the flux in the Stromgren $b$ and $y$ bands relative to three comparison stars. We combined the differential $b$ and $y$ measurements into a single $(b+y)/2$ ``passband'' then converted the differential magnitudes into a relative flux normalized to 1.0. The precision in relative flux is typically between 0.001 and 0.0015. Further details of the observing strategy and data reduction pipeline are available in \cite{Henry99, Eaton03, Henry13}. We make the photometric data available as a machine-readable table.

\subsection{Observational Statistics}

We examined the range of observing cadences and observational baselines within the CLS sample, to determine whether stars without known planets were observed with strategies that differed significantly from those for stars with known planets. Figure 2 shows the distribution of number of observations and observational baselines for three groups of stars: the entire sample, the stars around which we detected planets, and the star around which we did not detect planets. Each of these three samples has a median baseline of 21 yr. Stars with detected planet have a median of 74 observations, compared to 35 observations for stars without detected planets and 41 observations for the entire CLS sample. A factor of two in number of observations will have a small but measurable impact on planet detectability of a given data set -- and therefore on its search completeness contours.

\begin{figure}[ht!]
\includegraphics[width = 0.49\textwidth]{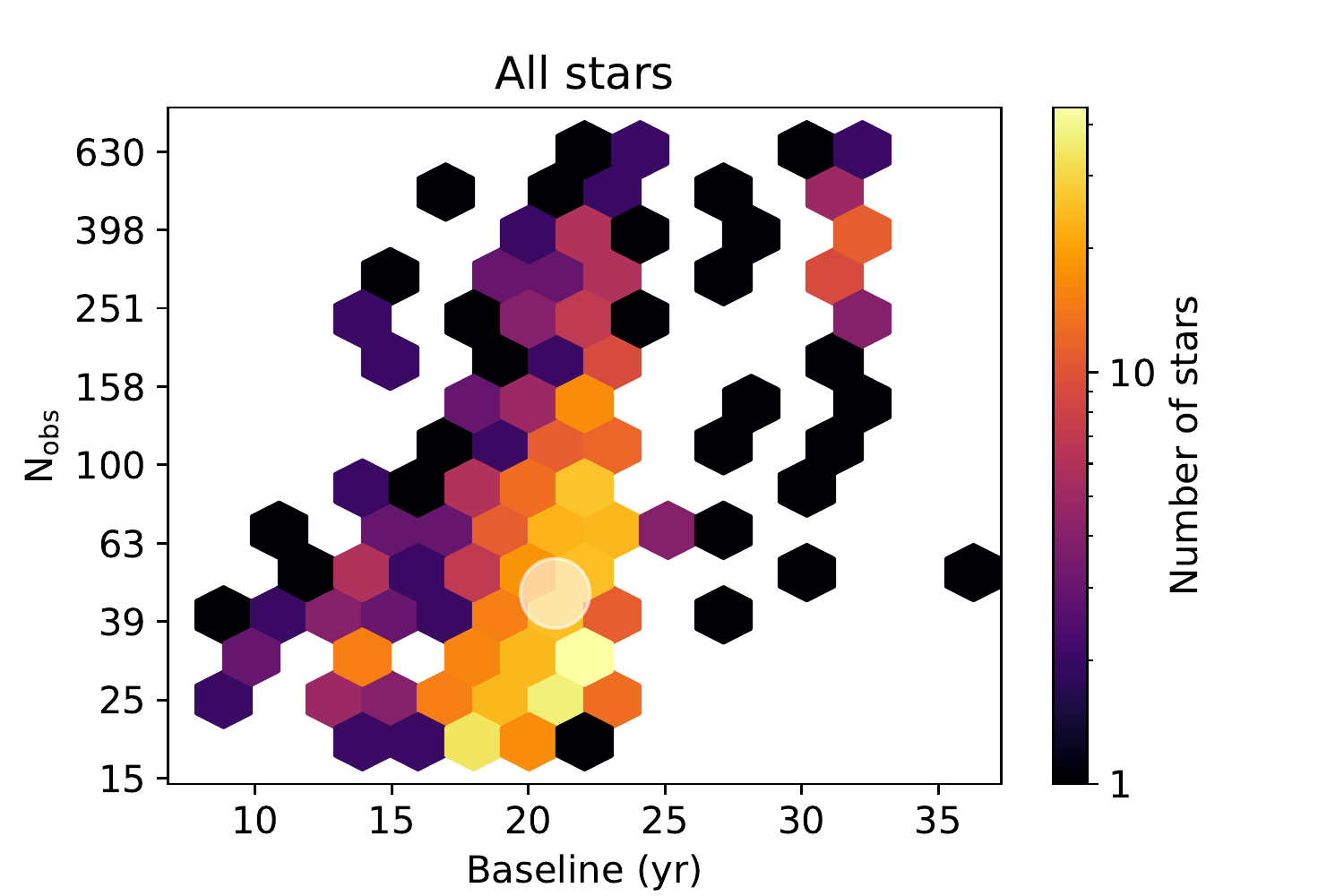} \\
\includegraphics[width = 0.49\textwidth]{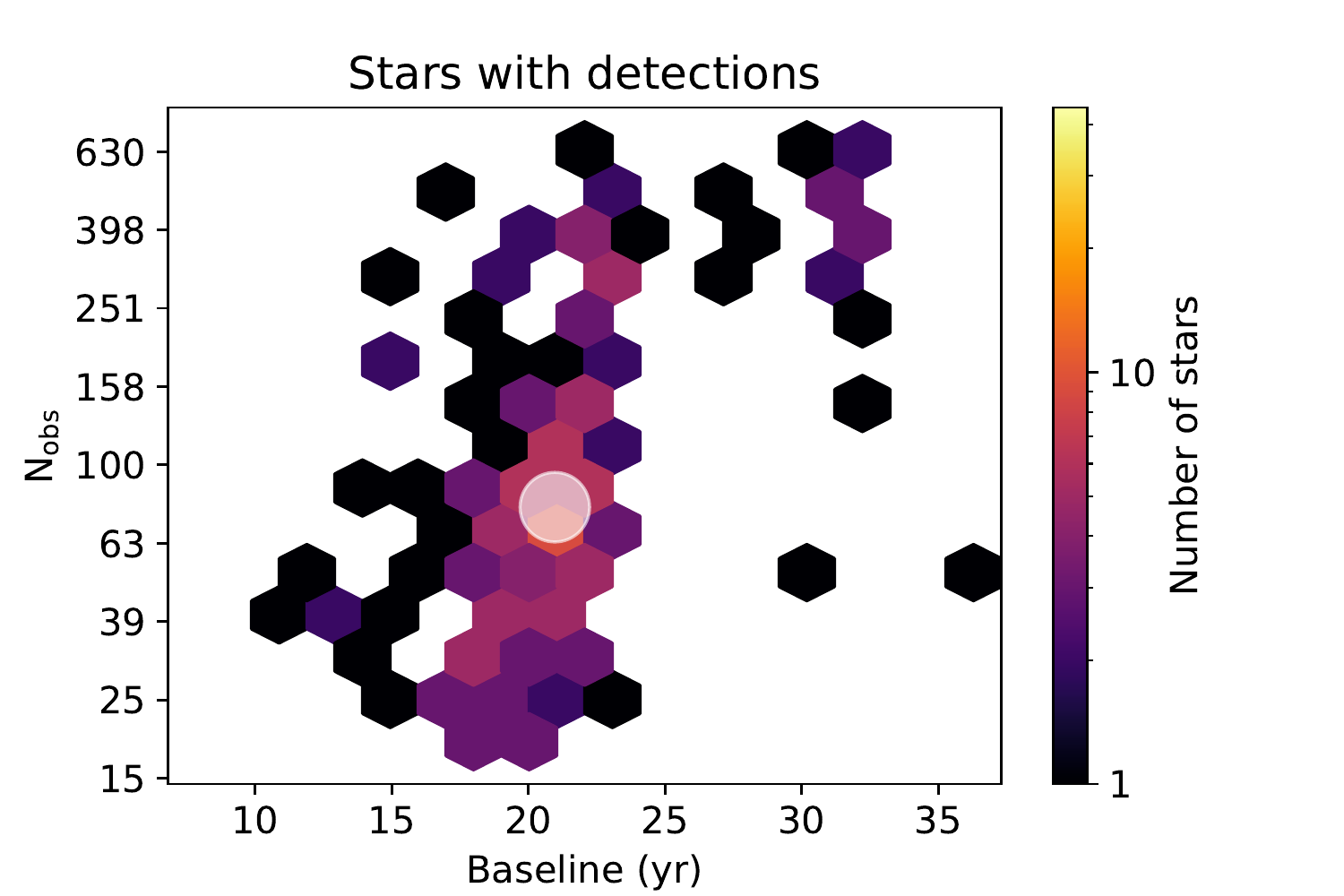} \\
\includegraphics[width = 0.49\textwidth]{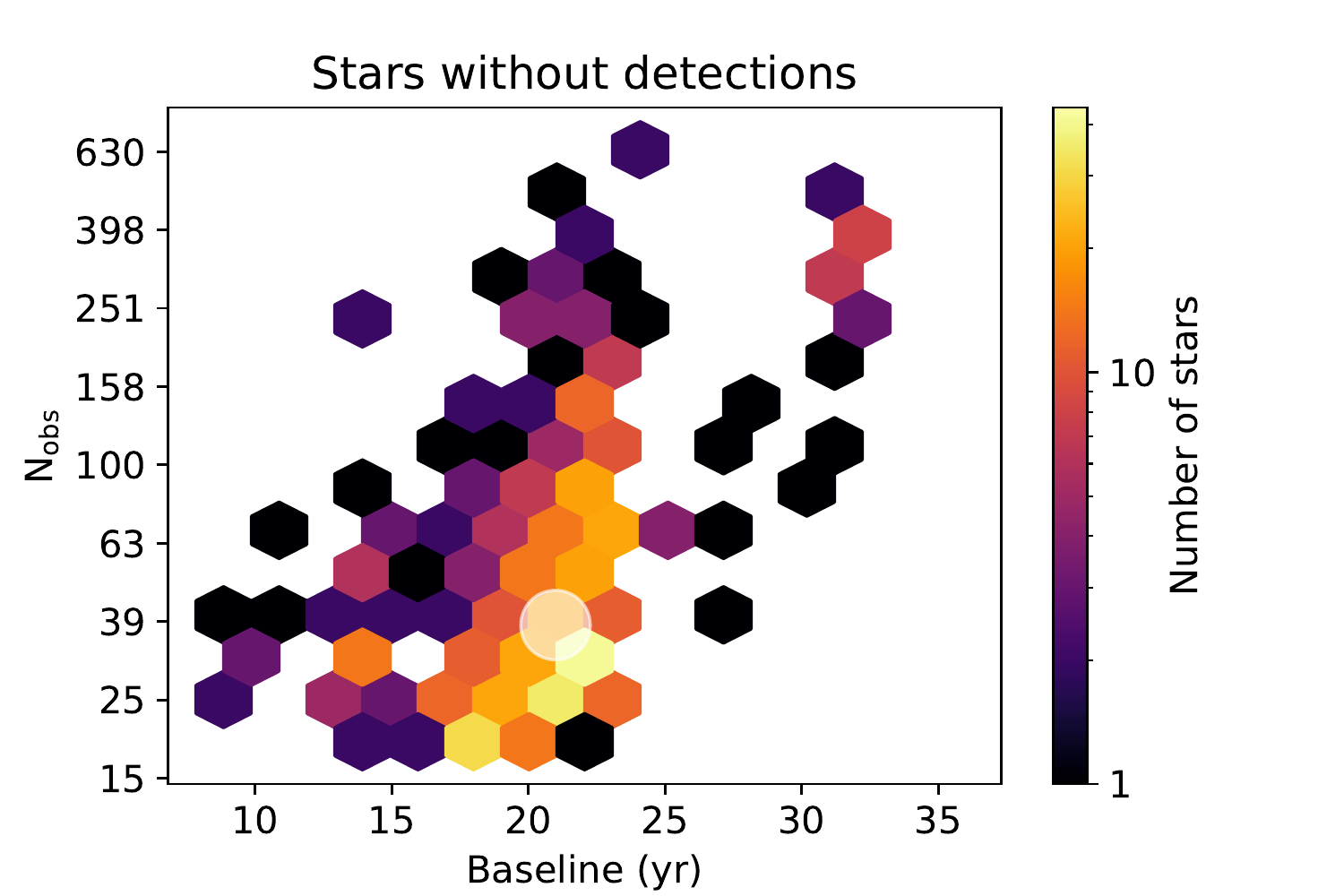} \\
\caption{Distributions of observational baseline versus number of observations. The top panel shows these statistics for all stars in the CLS sample; the center panel shows stars around which we detect planets; the bottom panel shows stars around which we do not detect planets. Median baseline and number of observations for each sample are overplotted as translucent circles.}
\end{figure}

\section{Stellar properties}

We derived stellar properties for our sample by applying the \texttt{SpecMatch} \citep{Petigura15} and \texttt{Isoclassify} \citep{Huber17} software packages to the template Keck-HIRES spectra of our stars. \texttt{Specmatch} takes an optical stellar spectrum as input, and by interpolating over a grid of template spectra with known associated stellar properties, returns three spectral properties and uncertainties. For stars hotter than 4700 K, we interpolated over synthetic spectra to derive spectral parameters \citep{Petigura15}. For stars below this threshold, we interpolated over real spectra of cool stars with well-characterized stellar properties, since synthetic spectral models are unreliable below this temperature \citep{Yee17}.

\texttt{Specmatch} produces metallicity, effective temperature, and surface gravity when interpolating over synthetic spectra; it produces metallicity, effective temperature, and radius when interpolating over empirical spectra. \texttt{Isoclassify} takes effective temperature, metallicity, and surface gravity as spectral parameter inputs, and uses isochrone models and multinest Bayesian sampling \citep{Buchner16} to produce estimates and uncertainties of physical parameters, in particular stellar mass. For stars cooler than 4700 K, we passed \texttt{Isoclassify} a wide Gaussian input prior on surface gravity, since temperature and metallicity strongly constrain the masses of cool, main-sequence stars \citep{Johnson17}.

Almost all stars in the California Legacy Survey have both Gaia-measured parallaxes \citep{GaiaDR1, GaiaDR2, Lindegren18} and apparent $K$-band magnitudes. For stars with both of these measurements available, we pass them and their uncertainties into \texttt{Isoclassify}  as additional inputs, since taken together, they constrain stellar luminosity and therefore place tighter constraints on stellar mass. \texttt{Isoclassify}  also returns more precise estimates of stellar radius when provided with parallax and apparent magnitude. With the inclusion of this luminosity constraint, the median precision of our stellar mass measurements is 3.6$\%$.

In Table 2 in Appendix B, we report stellar mass, radius, surface gravity, effective temperature, and metallicity for a subsection of the CLS sample. We make this table available for the entire sample in machine-readable format, with additional columns including $V$-band magnitude and Gaia parallax. Figure 3 is a visualization of these stellar properties, while Figure 4 shows individual histograms for mass, metallicity, and effective temperature, as well as for the following observational properties: parallax-inverse distance, $V$, and $B - V$.

\begin{figure*}[ht!]
\begin{center}
\includegraphics[width = 0.99\textwidth]{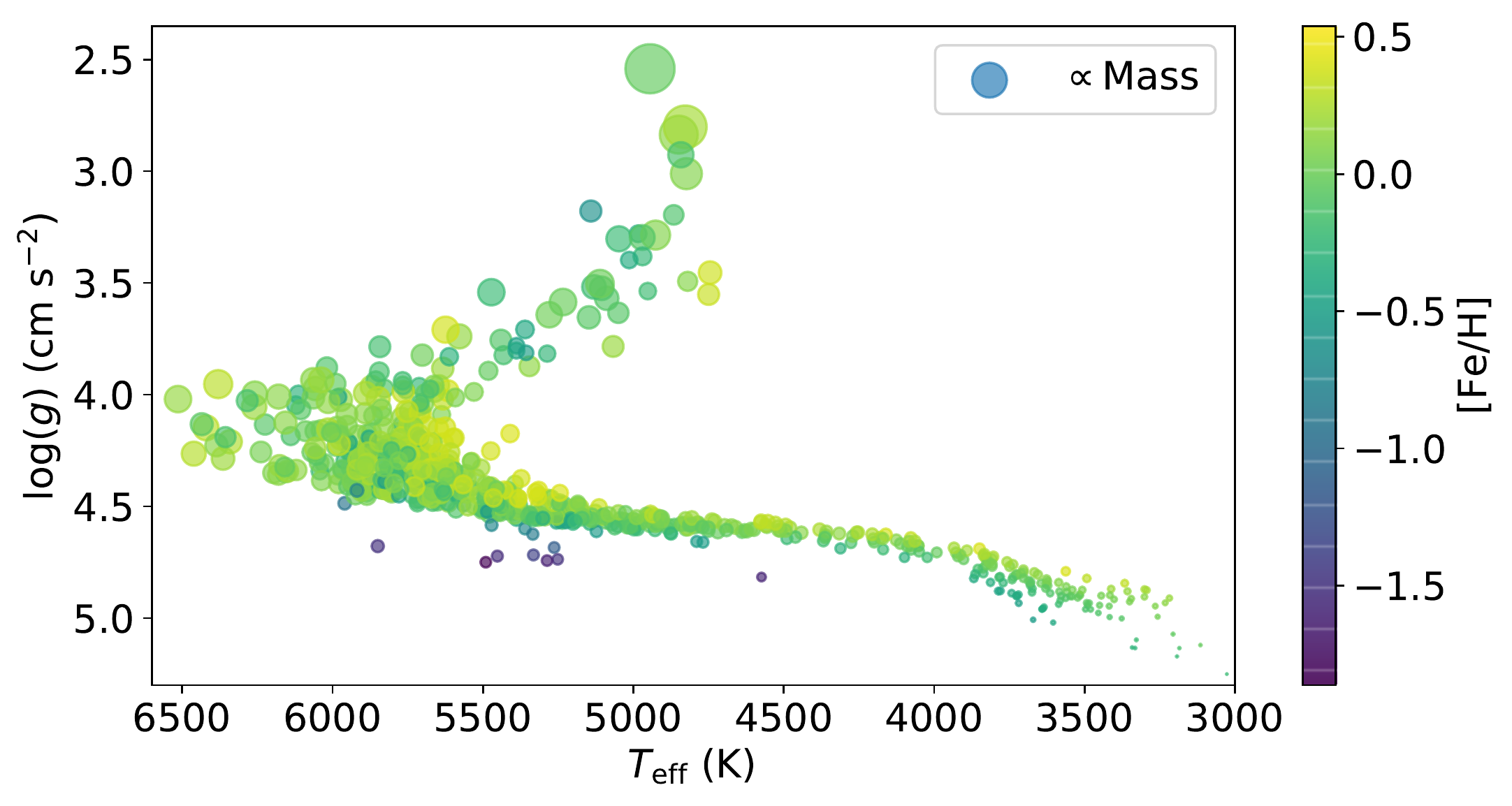} \\
\caption{Stellar property measurements of the California Legacy Survey, in effective temperature, surface gravity, and mass. The sample consists of stars spanning spectral types F, G, K, and M, some of which have evolved off of the main sequence. Most stars have metallicities within 0.4 dex of Solar metallicity, with the exception of a small handful of extremely metal-poor stars, which lie below the main sequence on this plot.}
\label{fig:hr}
\end{center}
\end{figure*}

\begin{figure*}[ht!]
\begin{center}
\includegraphics[width = 0.49\textwidth]{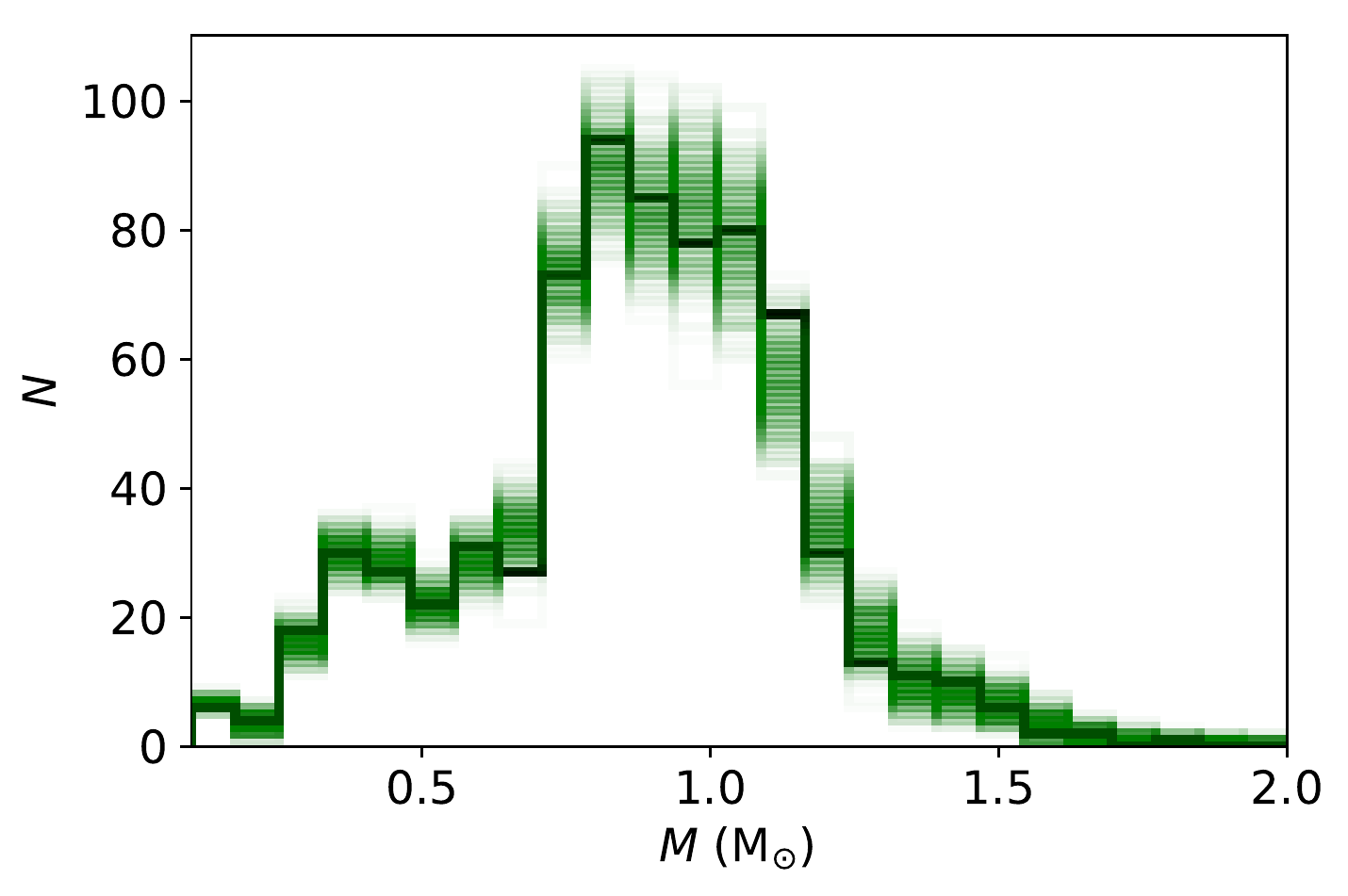}
\includegraphics[width = 0.49\textwidth]{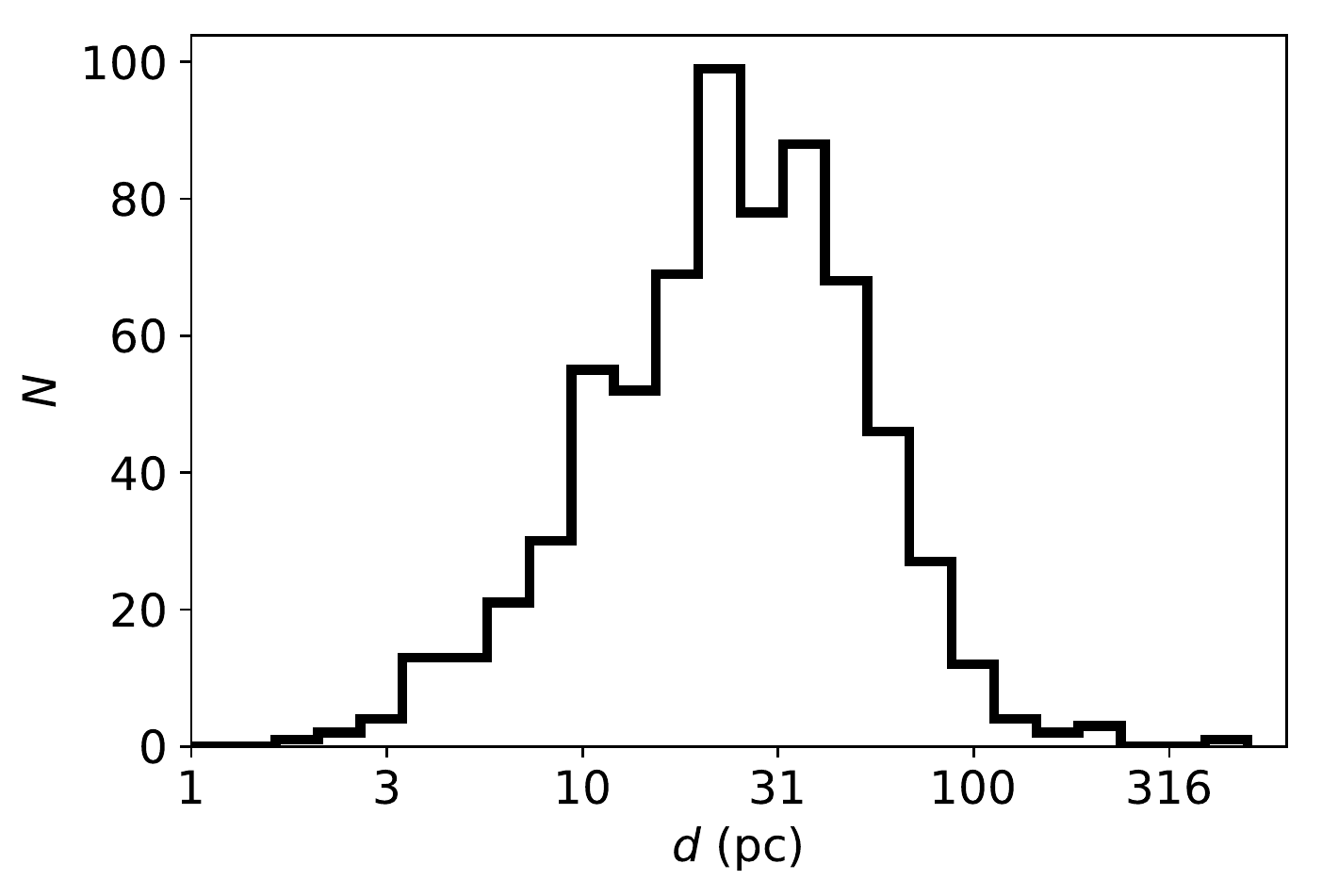} \\

\includegraphics[width = 0.49\textwidth]{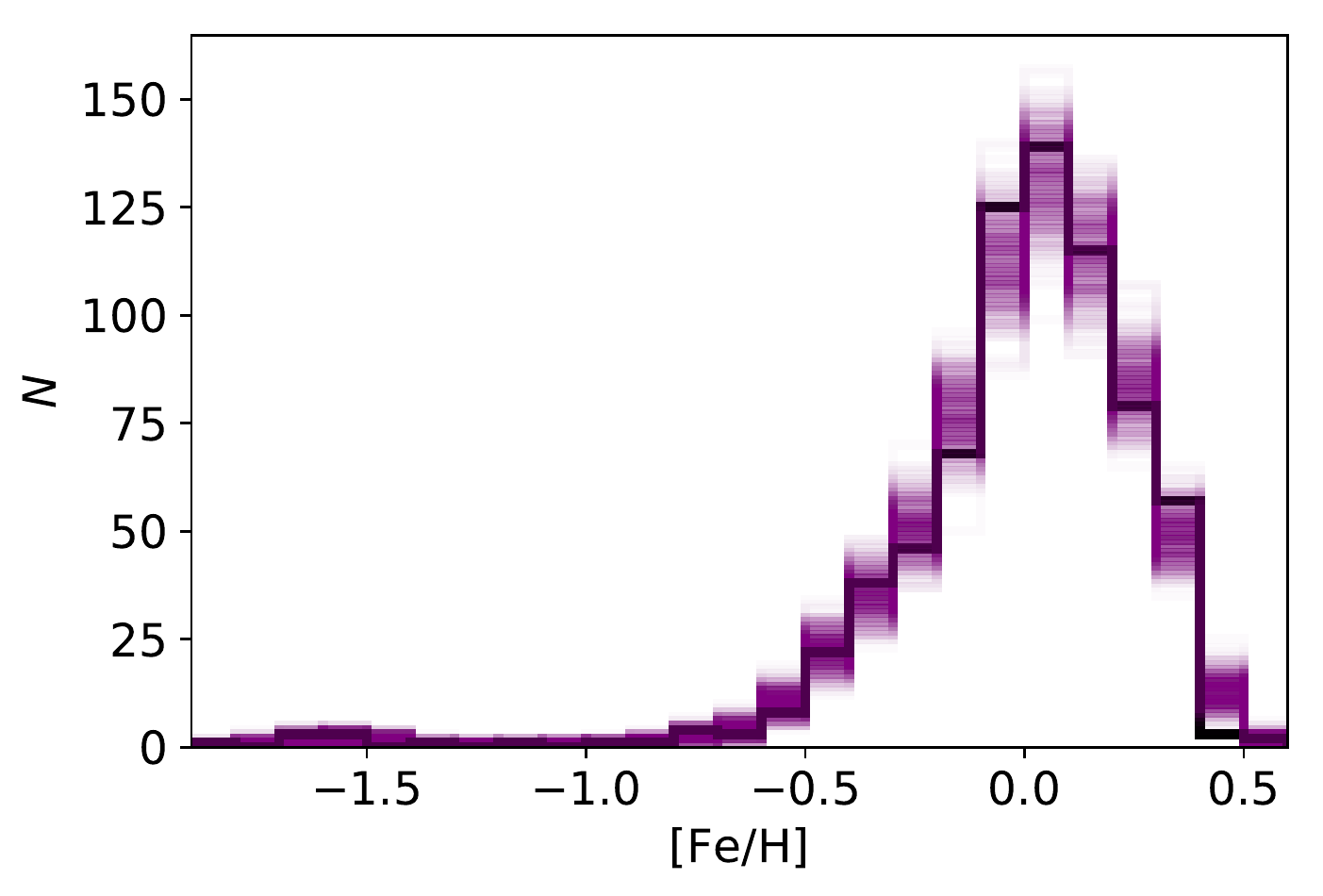}
\includegraphics[width = 0.49\textwidth]{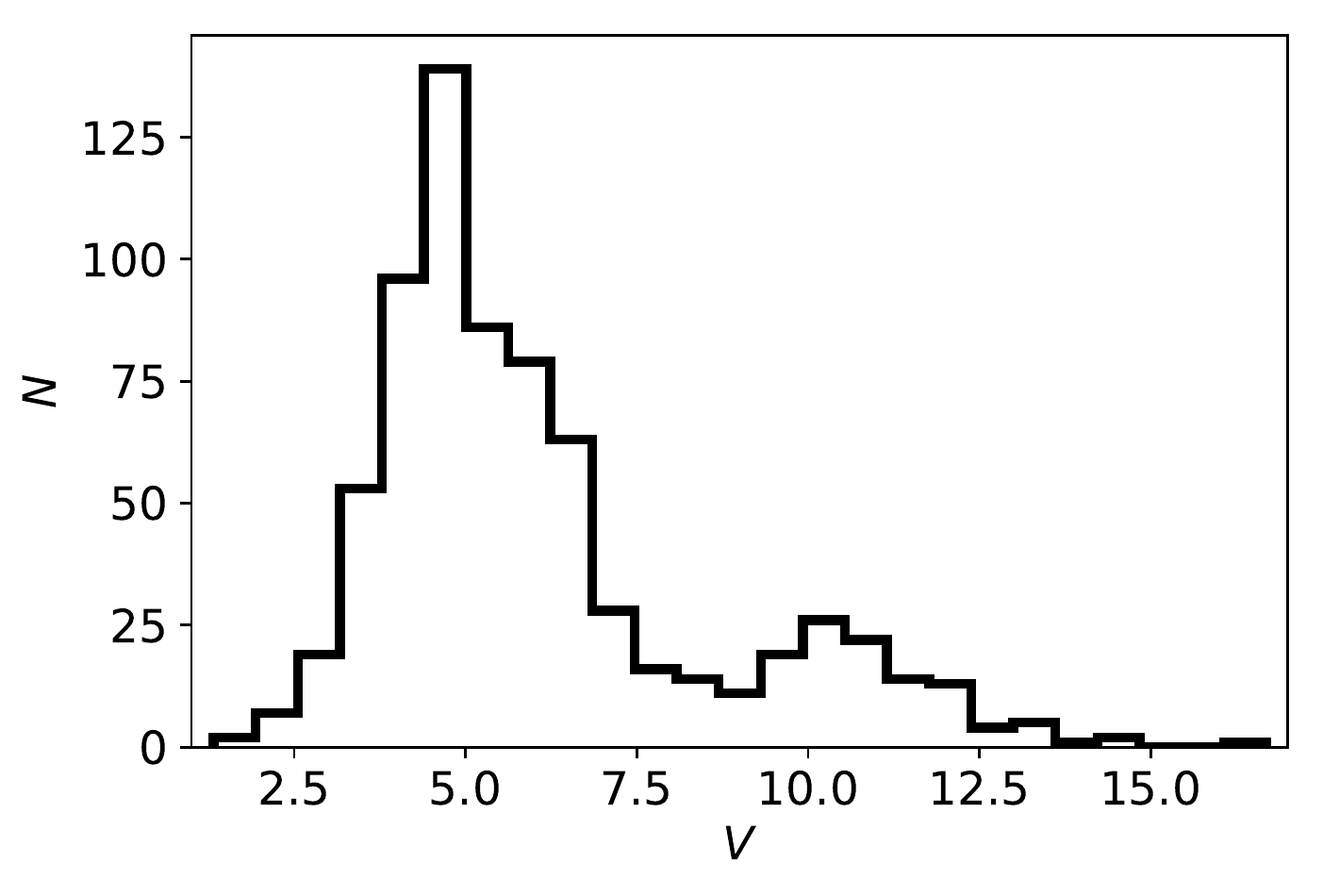} \\

\includegraphics[width = 0.49\textwidth]{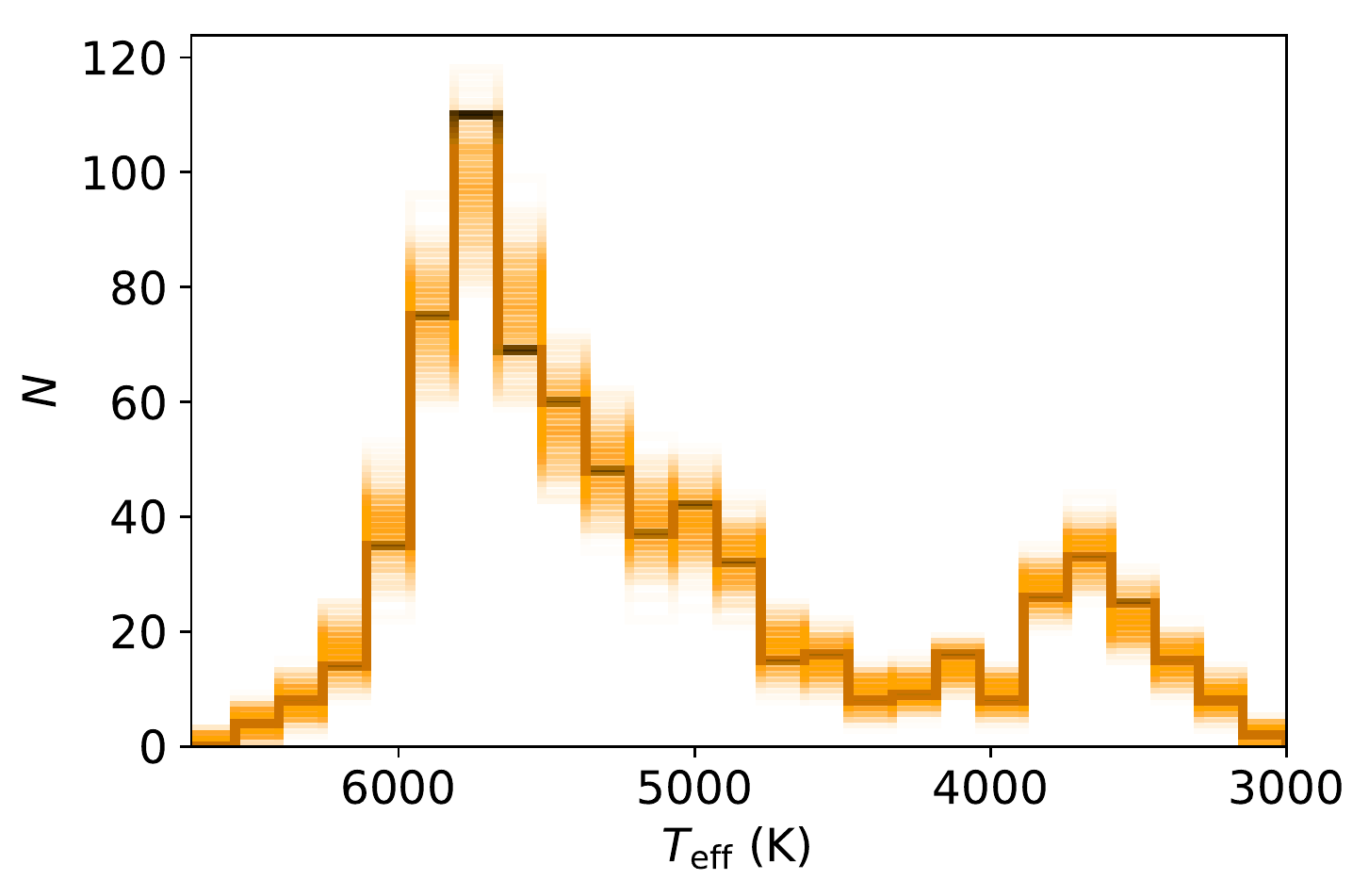}
\includegraphics[width = 0.49\textwidth]{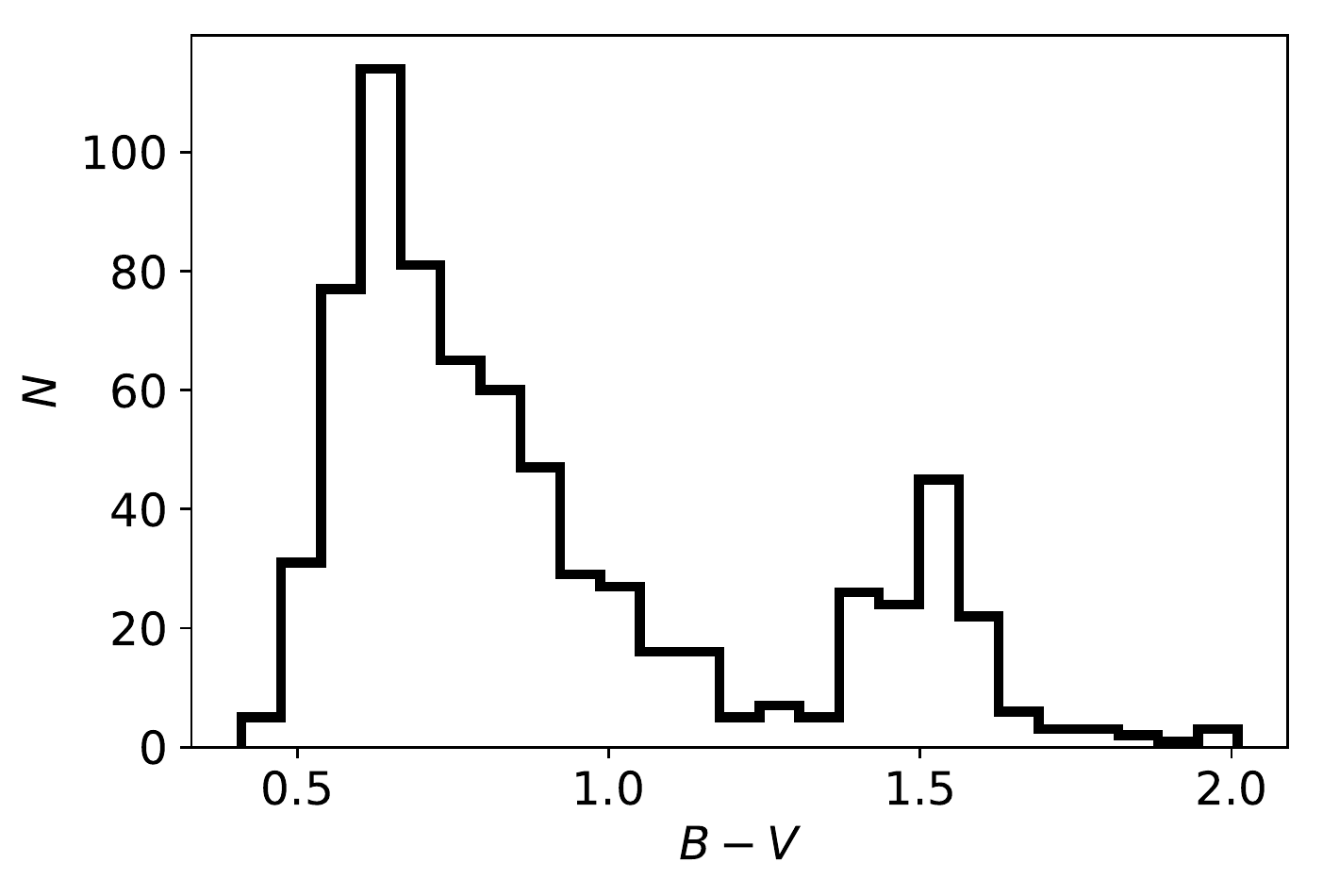} \\

\caption{Stellar parameter distributions. The left column shows  mass, metallicity, and effective temperature, while the right column shows parallax-inferred distance, $V$-band magnitude, and $B - V$ color. Black lines are histograms of the stellar parameter median values. For the left column, colored lines are 500 histograms per panel, with parameters redrawn from normal distributions with width equal to their individual measurement uncertainties. We left these redrawn parameter histograms out for the plots in the right column because distance, magnitude, and color have uncertainties that are smaller than the chosen bin size.}
\label{fig:stellar_hists}
\end{center}
\end{figure*}

\section{Planet Catalog Methods}

\subsection{Planet Search}

\begin{figure*}[ht!]
\begin{center}
\includegraphics[width=0.7\textwidth]{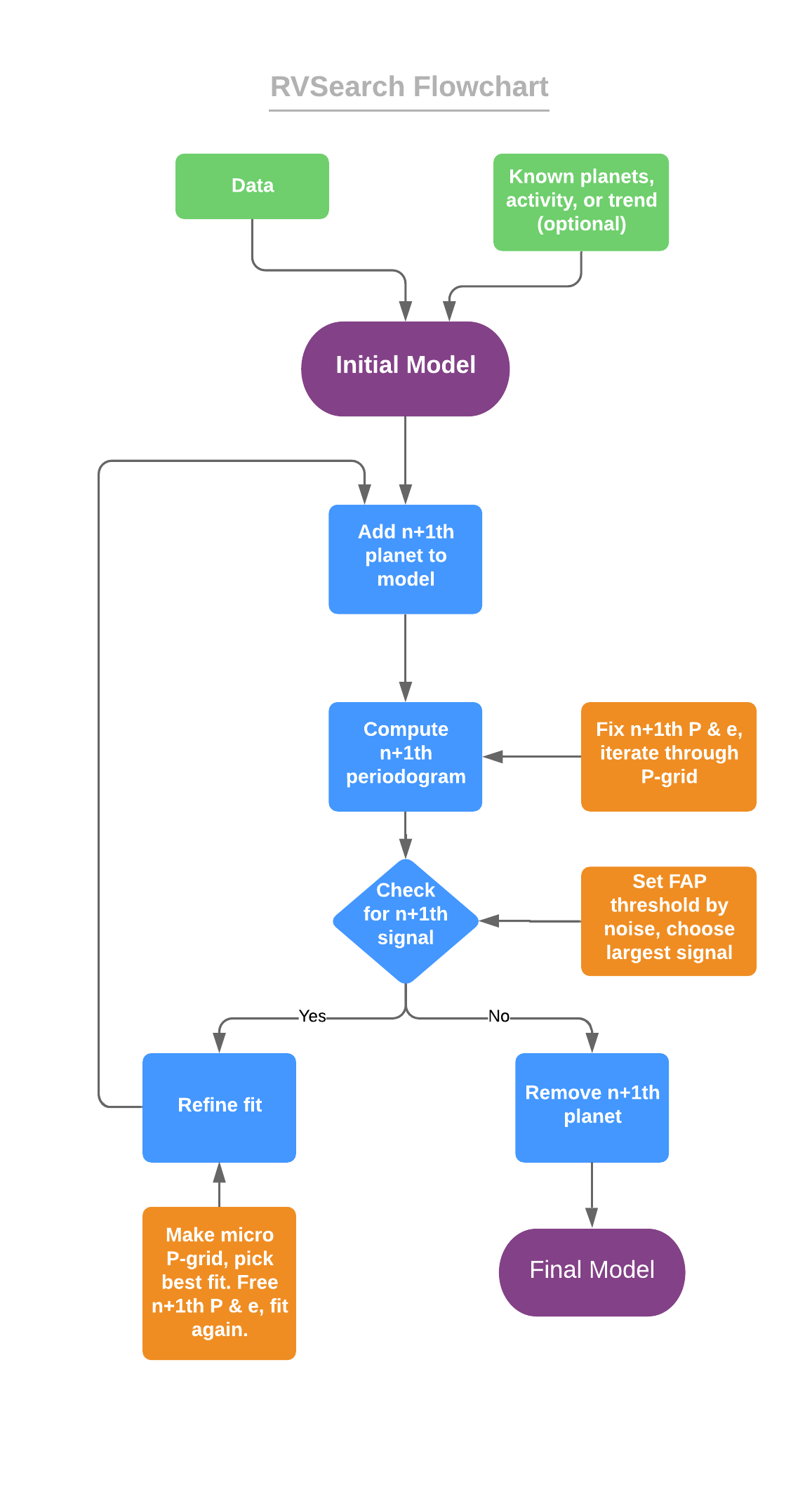}
\caption{Search algorithm flowchart.}
\label{fig:search_flowchart}
\end{center}
\end{figure*}

\begin{figure*}[ht!]
\begin{center}
\includegraphics[width=0.8\textwidth]{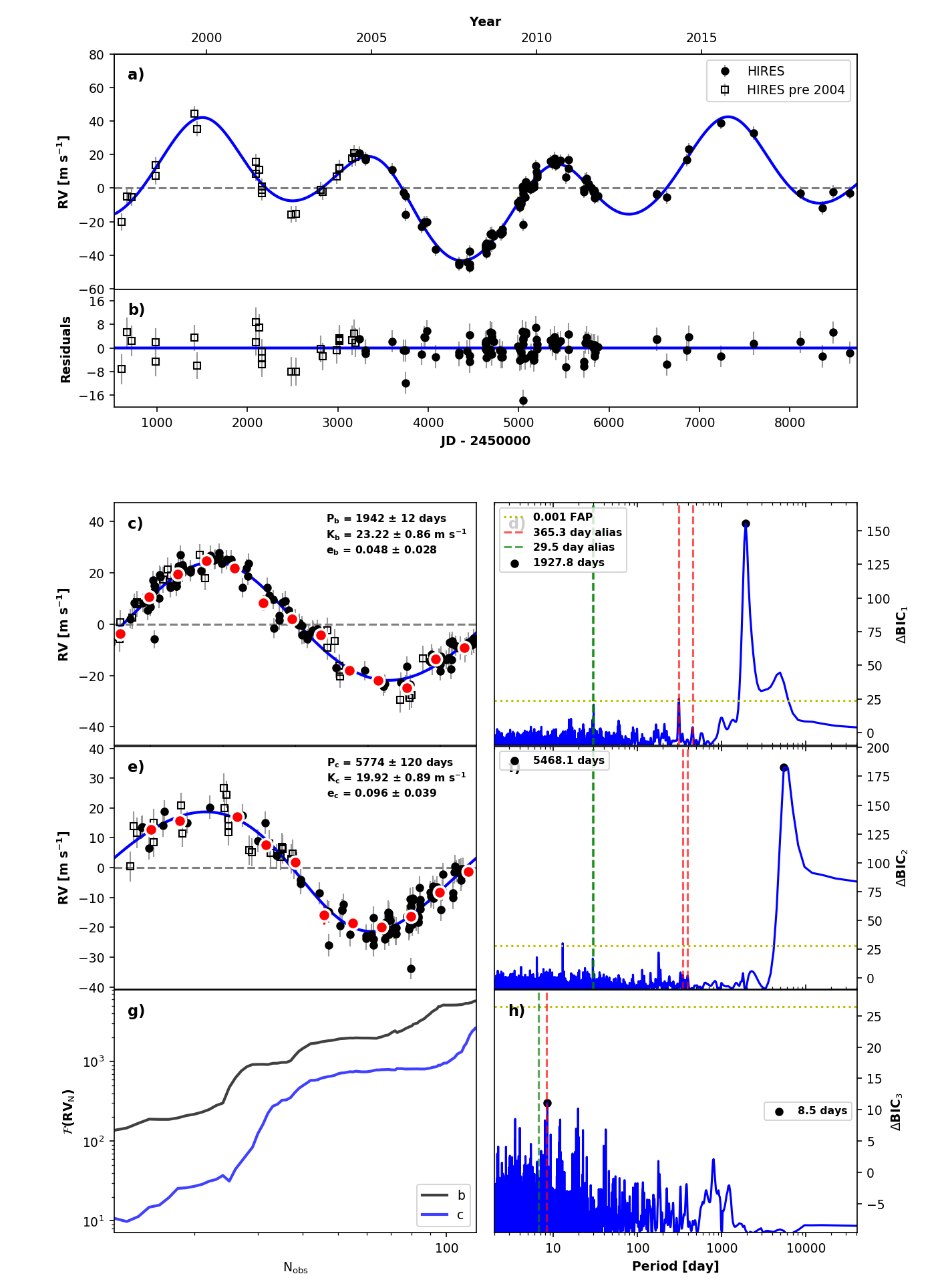}
\caption{Example \texttt{RVSearch} summary plot, for the known two-planet system HIP 109388. Panel a) shows the total model plotted over the radial velocity time series, while panel b) shows the model residuals. Each successive row shows a phase-folded signal discovered by \texttt{RVSearch} on the left, and the associated periodogram on the right. The final row shows the running periodograms of each signal, generated with Lomb--Scargle power, on the left, and the final periodogram on the right.}
\label{fig:hip109388_summary}
\end{center}
\end{figure*}

We developed an iterative approach to a search for periodic signals in RV data in order to generate the CLS planet catalog. We outline this algorithm, which we developed as the open-source Python package \texttt{RVSearch} and have made public alongside the publication of this paper. Figure \ref{fig:search_flowchart} is a flowchart that lays out each step of the algorithm, and Figure \ref{fig:hip109388_summary} is a visualization of an example \texttt{RVSearch} output, where the top two panels show the final model, and each successive row shows an iterative search for each signal in the model. First, we provide an initial model, from which the iterative search begins. This initial model contains an RV data set, and a likelihood function. The natural logarithm of the latter is defined as

\begin{equation}
\textrm{ln}(\mathcal{L}) = -\frac{1}{2}\sum_i\left[\frac{(v_i - m(t_i) - \gamma_D)^2}{\sigma_i^2} + \textrm{ln}(2\pi \sigma_i^2 )\right], \\
\end{equation}

\noindent where $i$ is the measurement index, $v_i$ is the $i$th RV measurement, $\gamma_D$ is the offset of the instrumental dataset from which the $i$th measurement is drawn, and $\sigma_i^2$ is the quadrature sum of the instrumental error and the stellar jitter term of the $i$th measurement's instrumental data set. Here, $m(t_i)$ is the model RV at time $t_i$, defined as

\begin{equation}
m(t) = \sum_n\textrm{K}(t | K_n, P_n, e_n, \omega_n, t_{c n}) + \dot{\gamma}(t - t_0) + \ddot{\gamma}(t - t_0)^2, \\
\end{equation}

 where $n$ is a given Keplerian orbit in the model, $\textrm{K}(t | K, P, e, \omega, t_c)$ is the Keplerian orbit RV signature at time $t$ given RV amplitude $K$, period $P$, eccentricity $e$, argument of periastron $\omega$, and time of inferior conjunction $t_c$, $\dot{\gamma}$ is a linear trend term, $\ddot{\gamma}$ is a quadratic trend term, and $t_0$ is a reference time, which we defined as the median time of observation.

We used \texttt{RadVel} \citep{Fulton18} to fit Keplerian orbits. The initial likelihood model contains either a one-planet Keplerian model with undefined orbital parameters, or a predefined model including trend/curvature terms and/or Keplerian terms associated with known orbital companions. We defaulted to performing a blind search starting with the undefined single-planet model, and we only supply a predefined model if there is evidence for a highly eccentric companion whose period is misidentified by our search algorithm. Several highly eccentric stellar binaries satisfy this criterion, as do two planets: HD 120066 b \citep{Blunt19}, and HD 80606 b \citep{Wittenmyer07}.

Before beginning a blind search, \texttt{RVSearch} determines whether the data merits a trend with curvature, a linear trend, or no trend. It does this by fitting each of these three models to the data, then performing a goodness-of-fit test to decide which model is favored. We measured the Bayesian Information Criterion (BIC) for each of the three models, and computed the \dBIC\ between each model. RVSearch selects the linear model if it has \dBIC\ = 5 with respect to the flat model, and the quadratic model if it has \dBIC\ = 5 with respect to the linear model. We did not perform this test on datasets that contain eccentric companions with orbital periods greater than the data's observational baseline, since such datasets would be better fit with a long-period Keplerian orbit than with linear and parabolic trends. The Bayesian information criterion is defined as
\begin{equation}
\textrm{BIC} = k\textrm{ln}(n_{\textrm{obs}}) - 2\textrm{ln}(\mathcal{L}), \\
\end{equation}
\noindent where $n_\textrm{obs}$ is the number of observations, $k$ is the number of free model parameters, and $\textrm{ln}(\mathcal{L})$ is the log-likelihood of the model in question.

Once we provide an initial model, \texttt{RVSearch} defines an orbital period grid over which to search, with sampling such that the difference in frequency between adjacent grid points is $\frac{1}{2\pi \tau}$, where $\tau$ is the observational baseline. We chose this grid spacing in accordance with \cite{Horne86}, who state that, in frequency space, a Lomb--Scargle periodogram has a minimum peak width of ~$\frac{1}{2\pi \tau}$. For each dataset, we searched for periodicity between two days and five times the observational baseline. Searching out to five times the baseline only adds a few more points to the period grid, and it allows for the possibility of  recovering highly eccentric, ultra-long-period planet candidates with best-fit orbital period.

The search algorithm then computes a goodness-of-fit periodogram by iterating through the period grid and fitting a sinusoid with a fixed period to the data. We measure goodness-of-fit as the \dBIC\ at each grid point between the best-fit, $n$+1-planet model with the given fixed period, and the $n$-planet fit to the data (this is the zero-planet model for the first planet search).

After constructing a \dBIC\ periodogram, the algorithm performs a linear fit to a log-scale histogram of the periodogram power values. The algorithm then extrapolates a \dBIC\ detection threshold corresponding to an empirical false-alarm probability of 0.1$\%$, meaning that, according to the power-law fit, only 0.1$\%$ of periodogram values are expected to fall beyond this threshold. This process follows the detection methodology outlined in \cite{Howard16}.

If a periodic signal exceeds this detection threshold, \texttt{RVSearch} refines the fit of the corresponding Keplerian orbit by performing a maximum \textit{a posteriori} (MAP) fit with all model parameters free, including eccentricity, and records the BIC of that best-fit model. \texttt{RVSearch} includes two hard-bound priors, which constrain $K > 0$ and $0 <= e < 1$. The algorithm then adds another planet to the RV model and conducts another grid search, leaving all parameters of the known Keplerian orbits free so that they might converge to a more optimal solution. In the case of the search for the second planet in a system, the goodness of fit is defined as the difference between the BIC of the best-fit one-planet model and the BIC of the two-planet model at each fixed period in the grid. \texttt{RVSearch} once again sets a detection threshold in the manner described above, and this iterative search continues until it returns a nondetection.

This iterative periodogram search is superior to a Lomb--Scargle residual subtraction search in two key ways. First, this process fits for the instrument-specific parameters of each dataset, stellar jitter and RV-offset, as free parameters throughout the search. Second, by leaving the known model parameters free while searching for each successive planet, we allow the solutions for the already discovered planets to reach better max-likelihood solutions that only become evident with the inclusion of another planet in the model.

Note that our search and model comparison process is not Bayesian; we do not use priors to inform our model selection, and we do not sample posteriors, beyond a grid search in period space, until we settle upon a final model. We use the BIC as our model comparison metric because it incorporates the number of free parameters as a penalty on more complex models, which, in our case, corresponds to models with additional planets.

We make \texttt{RVSearch} publicly available alongside this paper via a GitHub repository. See the \href{https://california-planet-search.github.io/rvsearch/}{RVSearch website} for installation and use instructions.

\clearpage

\subsection{Search Completeness}

We characterized the search completeness of each individual dataset, and of the entire survey, by running injection-recovery tests. Once \texttt{RVSearch} completed an iterative search of a dataset, it injected synthetic planets into the data and ran one more search iteration to determine whether it recovers these synthetic planets in that particular dataset. We ran 3000 injection tests for each star. We drew the injected planet period and \msini\ from log-uniform distributions, and drew eccentricity from the beta distribution described in \cite{Kipping13}, which was fit to a population of RV-observed planets.

\begin{figure*}[ht!]
\begin{center}
\includegraphics[width = 0.49\textwidth]{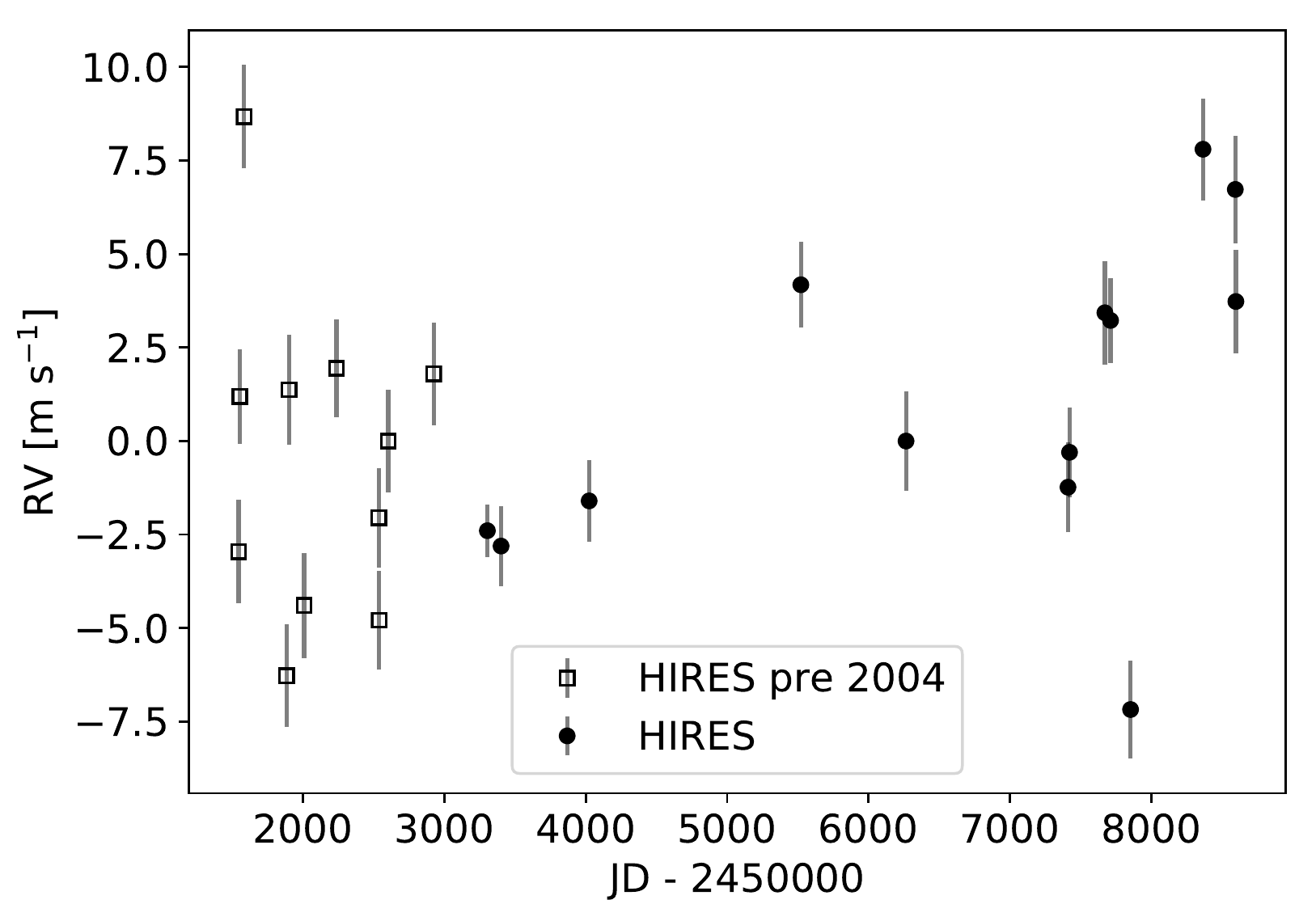}
\includegraphics[width = 0.49\textwidth]{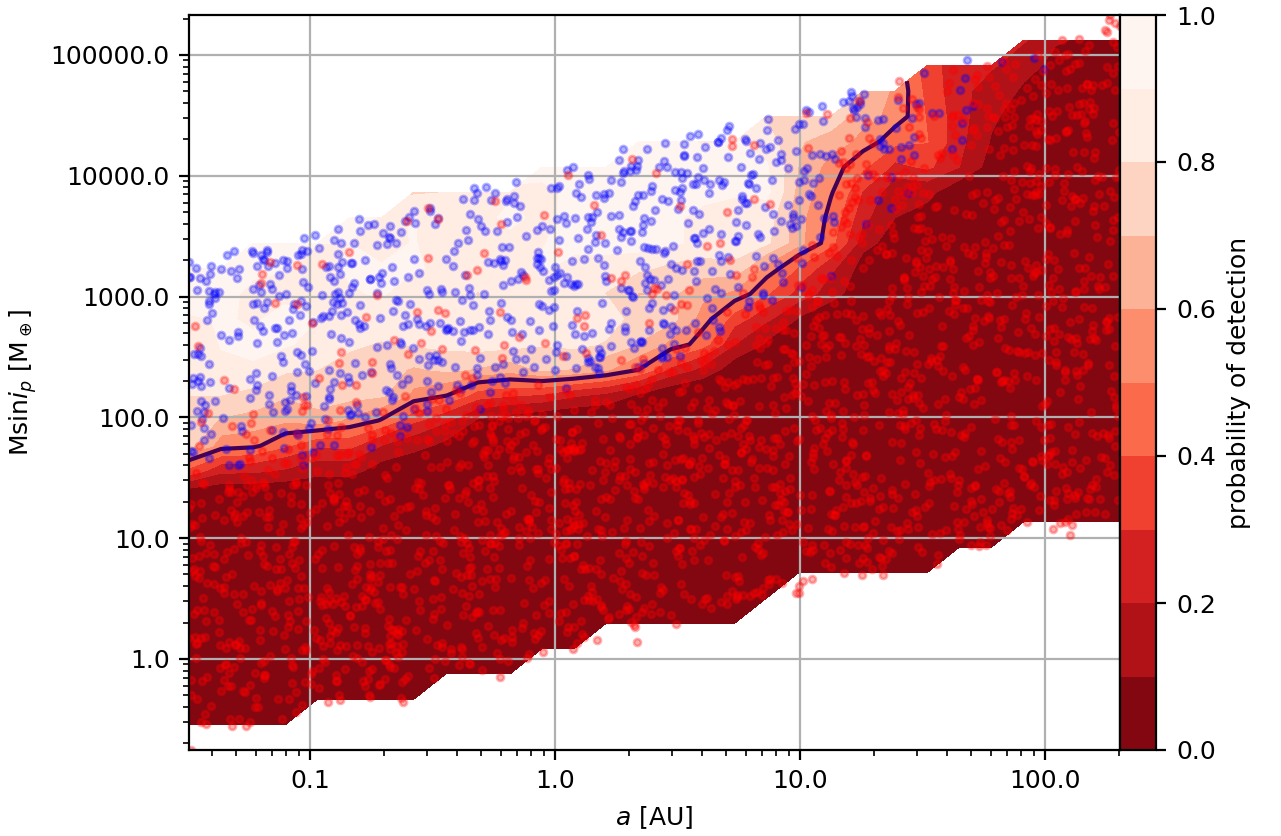} \\

\includegraphics[width = 0.49\textwidth]{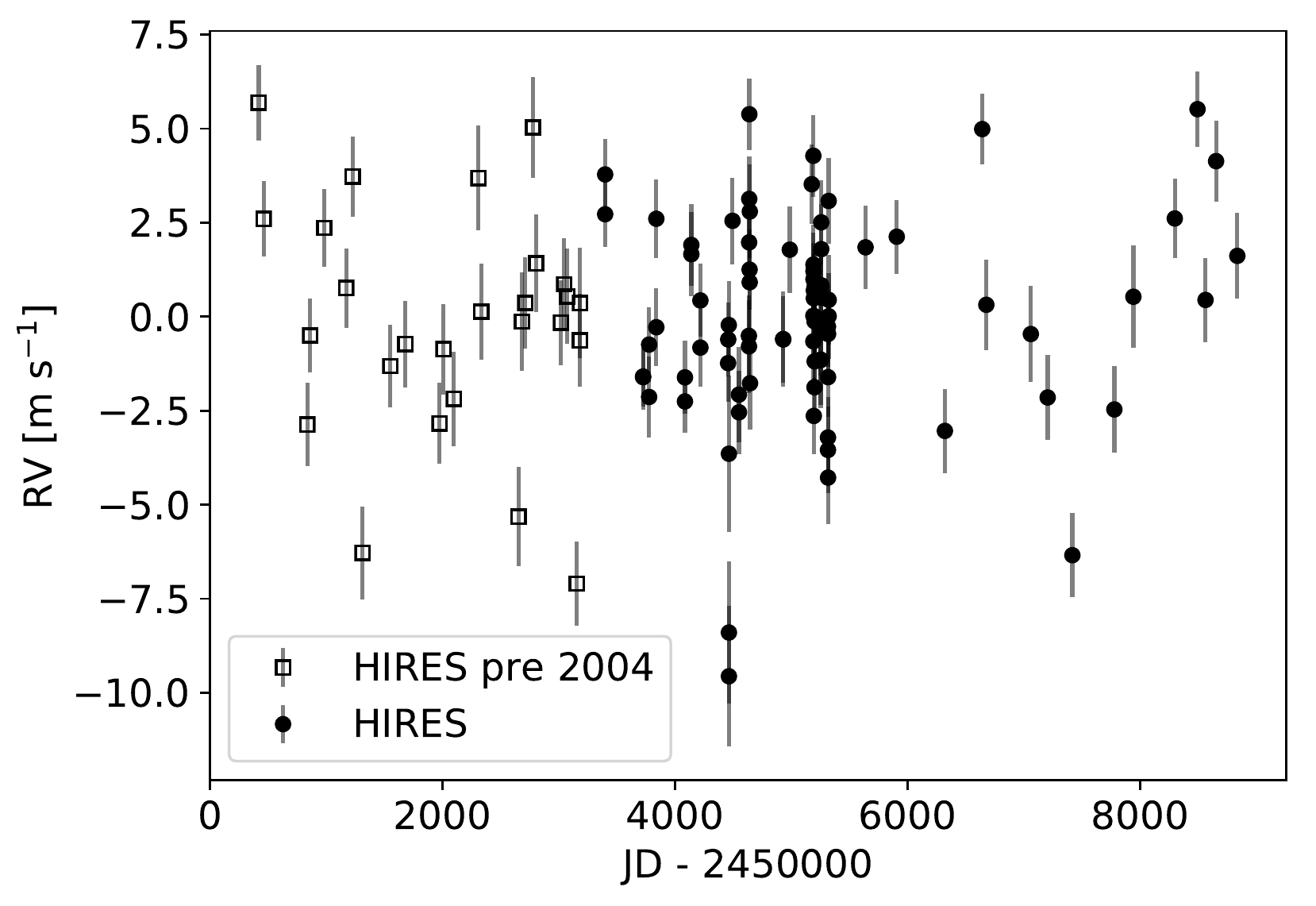}
\includegraphics[width = 0.49\textwidth]{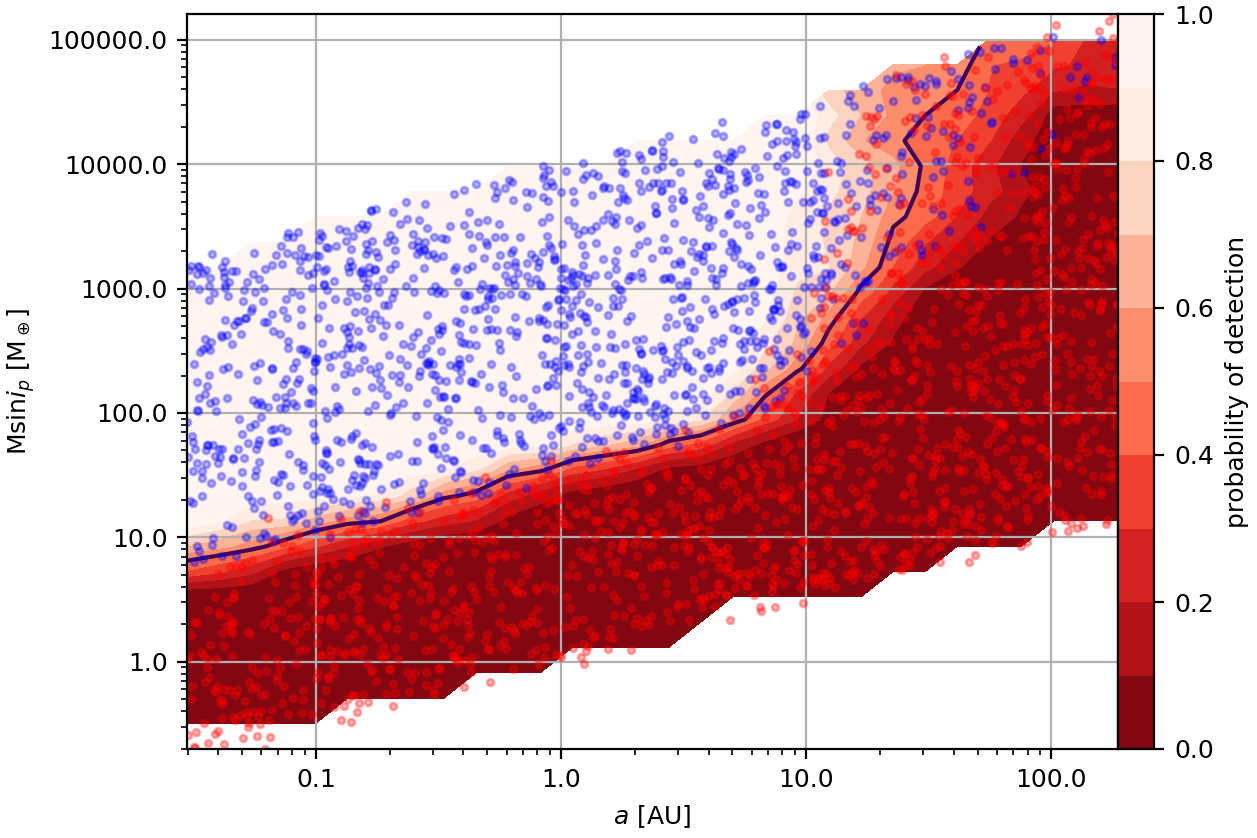} \\

\includegraphics[width = 0.49\textwidth]{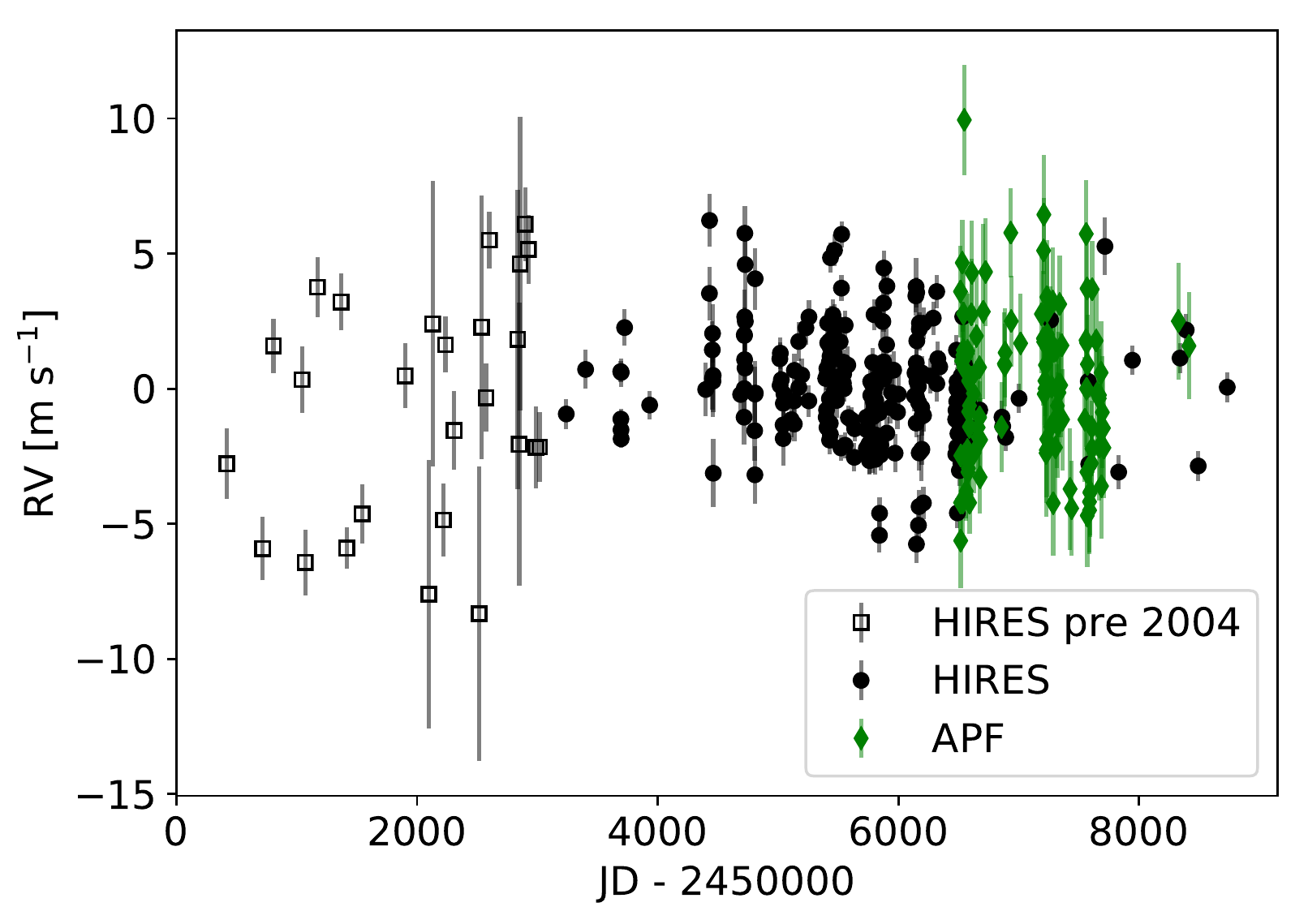}
\includegraphics[width = 0.49\textwidth]{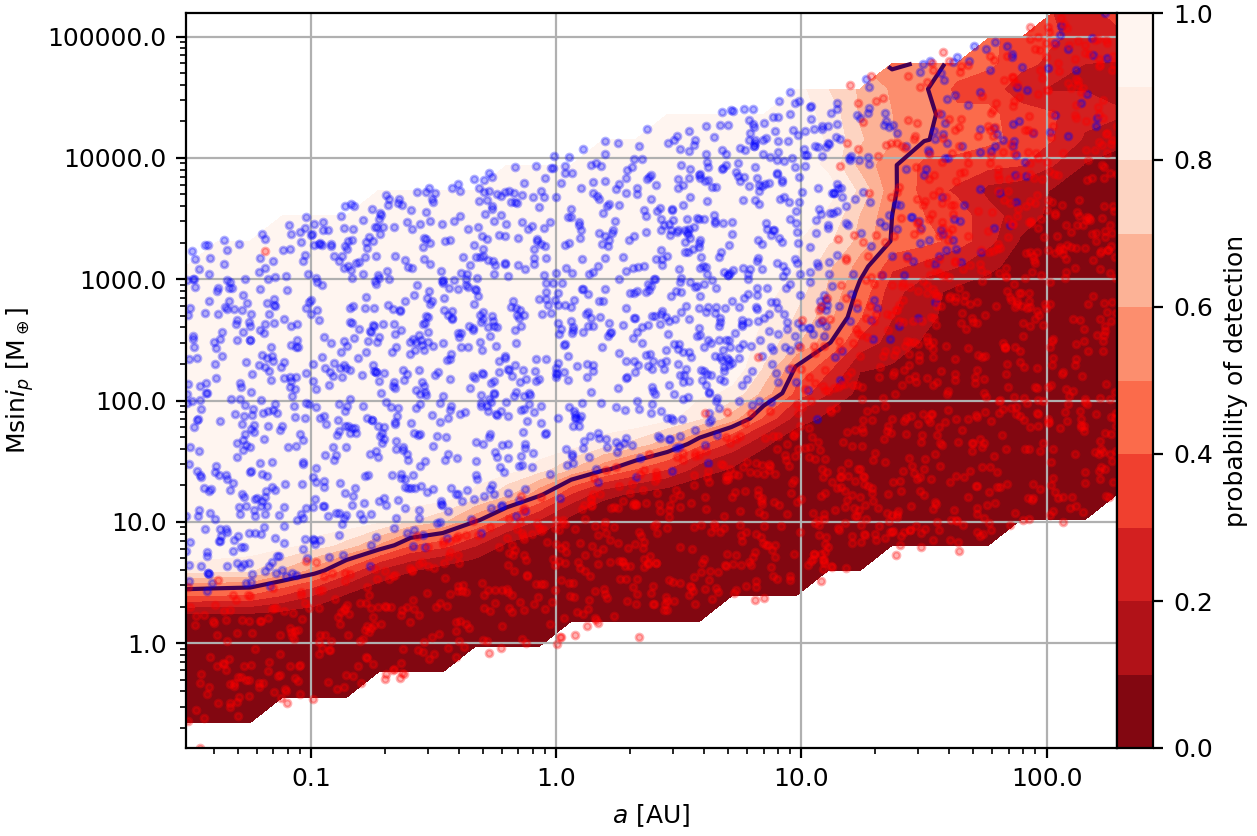} \\

\caption{RVs and completeness contours for three datasets with similar baselines, median measurement errors, and stellar jitter. The left column plots RVs with respect to time, while the right column plots injected signals in the \msini\ and $a$ plane, where blue dots are recovered injections and red dots are not. The right column also shows detection probability contours, with 50$\%$ plotted as a solid black line. From top to bottom, we show RVs and contours for HD 44420, for which we have 24 RVs; HD 97343, for which we have 94 RVs; and HD 12051, for which we have 372 RVs.}
\label{fig:injections}
\end{center}
\end{figure*}

We used the results of these injection tests to compute search completeness for each individual dataset, and report constant  \msini\ / $a$ contours of detection probability. Figure \ref{fig:injections} shows examples of these contours and the corresponding RVs for three different stars, all early G-type: one with 25 observations, one with 94, and one with 372. We make the 10th, 50th, and 90th percentile completeness contours for each individual star available in machine-readable format.

\subsection{Model posteriors}

Once \texttt{RVSearch} returned max-likelihood estimates of the orbital model parameters for a given dataset, we sampled the model posterior using affine-invariant sampling, implemented via \texttt{emcee} and \texttt{RadVel} \citep{Foreman-Mackey13, Fulton18}. We sampled using the orbital parameter basis $\{$ $\mathrm{log}P\ K\ t_c\ \sqrt{e}\mathrm{sin}\omega\ \sqrt{e}\mathrm{cos}\omega$ $\}$. We placed uniform priors on all fitting parameters, with hard bounds such that $K > 0$ and $0 \leq e < 1$. We fit in $\mathrm{log}P$ space to efficiently sample orbits with periods longer than our observational baseline, and in $\sqrt{e}\mathrm{sin}\omega$ and $\sqrt{e}\mathrm{cos}\omega$ to minimize bias toward higher eccentricities \citep{Lucy71}. We reported parameter estimates and uncertainties as the median and $\pm$1$\sigma$ intervals.

If a dataset is so poorly constrained by a Keplerian model that \texttt{emcee}'s affine-invariant sampler cannot efficiently sample the posterior distribution, we instead used a rejection sampling algorithm to estimate the posterior. In these cases, we used \texttt{TheJoker} \citep{Price-Whelan17}, a modified MCMC algorithm designed to sample Keplerian orbital fits to sparse radial velocity measurements. We chose a flat prior on $\mathrm{log}P$, with a minimum at the observing baseline and a maximum at twenty times the observing baseline. We drew orbital eccentricity from a beta prior weighted toward zero, as modeled in \cite{Kipping13}, in order to downweight orbits with arbitrarily high eccentricity, which can be viable fits to sparse or otherwise underconstraining RV data sets.

\subsection{False-positive vetting}

We performed a series of tests to vet each planet candidate discovered by our search pipeline. The following subsections each detail one test we perform to rule out one way in which a signal might be a false-positive. We also represent this process with a flowchart in Figure \ref{fig:vetting_flowchart}, and include a table of all false-positive signals recovered by \texttt{RVSearch} in Table 6.

\subsubsection{Stellar activity, magnetic/long-period}
Many main-sequence stars, particularly F- and G-type, have magnetic activity cycles on timescales of several to tens of years. These fluctuations in activity can cause changes in the core depths of stellar Calcium H \& K lines, which manifest as apparent RV shifts \citep{Isaacson10}. To evaluate whether stellar activity may be the cause of a signal recovered by our search pipeline, we measure the linear correlation between the RV signature of that signal and a measured stellar activity metric--in our case, S-values. We computed S-values for both post-upgrade HIRES and APF data by measuring the core flux of Calcium H \& K lines.

If we found a periodic signal in the S-value data that has a similar period and phase similar to one of the Keplerian terms in our RV model, we searched for correlations between our RV model and S-values. If we found one periodic signal in an RV  dataset, we measured its correlation with stellar activity simply as the linear correlations between the RVs of each instrument and their associated S-values. If we found multiple periodic signals, then for each signal, we subtracted the associated RV models of all other signals from the data, and measured the correlations between these residuals and the S-values. A significant linear correlation between a signal's RV residuals and the associated S-values does not necessarily mean that this signal is caused by stellar activity, even when these signals also have the same period and phase, but we took it as sufficient evidence to remove such signals from our catalog of confirmed planets.

It is important to note that our approach to vetting our planet candidates is systematic but not exhaustive, particularly with respect to stellar activity. One might use activity metrics beyond S-values and photometry, such as H$\alpha$ line modulation. Furthermore, there are more sophisticated ways to deal with activity than searching for linear correlations with RVs. For instance, one might actively model stellar activity during the search process, using a Gaussian Process \citep{Haywood14} or some other correlated noise model. Such techniques might improve the accuracy of our planet candidate parameters and catalog selection, but require case-by-case analysis for each stellar system, as activity modeling is sometimes unwarranted or even counterproductive, e.g., for low-activity stars or confirmed planets that have periods similar to their host star's activity cycle. We chose to perform uniform, after-the-fact vetting for our catalog, and invite others to perform more sophisticated modeling for individual systems of interest.

\subsubsection{Stellar activity, rotation/short-period}

We only detected planet candidates that are low-amplitude and short-period enough to possibly be stellar rotation false positives in sustained, high-cadence datasets. Almost all CLS datasets that satisfy this criteria were collected as part of the APF-50 survey. We collected APT photometry of all APF-50 stars, which we can use to search for evidence of stellar rotation with moving-average smoothing and periodogram analysis. If we find strong evidence for rotation in APT photometry, or spectral S-value measurements, we discount planet candidates with periods close to the apparent rotation timescale or its harmonics.

\subsubsection{Yearly alias}

When we find a signal with a period of a year or an integer fraction of a year, we investigate whether it is an alias of long-period power, or a systematic that is correlated with the barycentric velocity at the time of observation or Doppler fitting parameters. We do this by recomputing the associated RVs using a different template observation. When another template observation was unavailable, we were able to take one using Keck-HIRES during collaborator observing nights. Templates taken in poor observing conditions or when barycentric velocity with respect to the observed star is high can produce systematic errors in the Doppler code. If a search of this new dataset returns a nondetection, or detection at a significantly different period, we conclude that this signal is an alias. Figure 8 shows the presence of yearly alias power in our survey, seen in a stack of the the final nondetection periodograms of all CLS stars.

\begin{figure}[ht!]
\begin{center}
\includegraphics[width=0.5\textwidth]{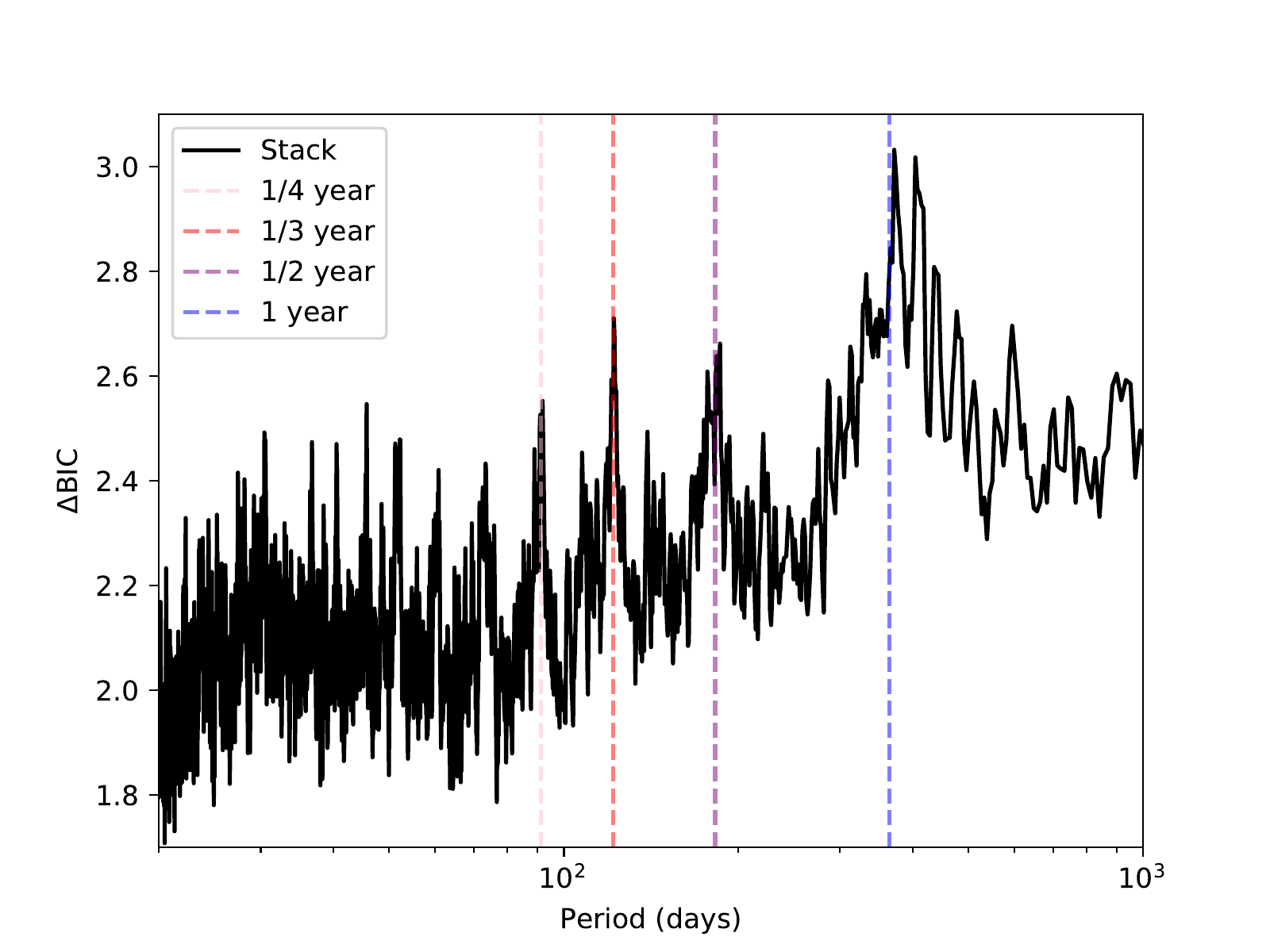}
\caption{Stack of all final nondetection periodograms in the CLS planet search, linearly interpolated to the same period grid. A broad peak around 1 year is evident, as well as narrow peaks at 1/2-year, 1/3-year, and 1/4-year.}
\label{fig:stacked_periodogram}
\end{center}
\end{figure}

\begin{figure*}[ht!]
\begin{center}
\includegraphics[width=0.8\textwidth]{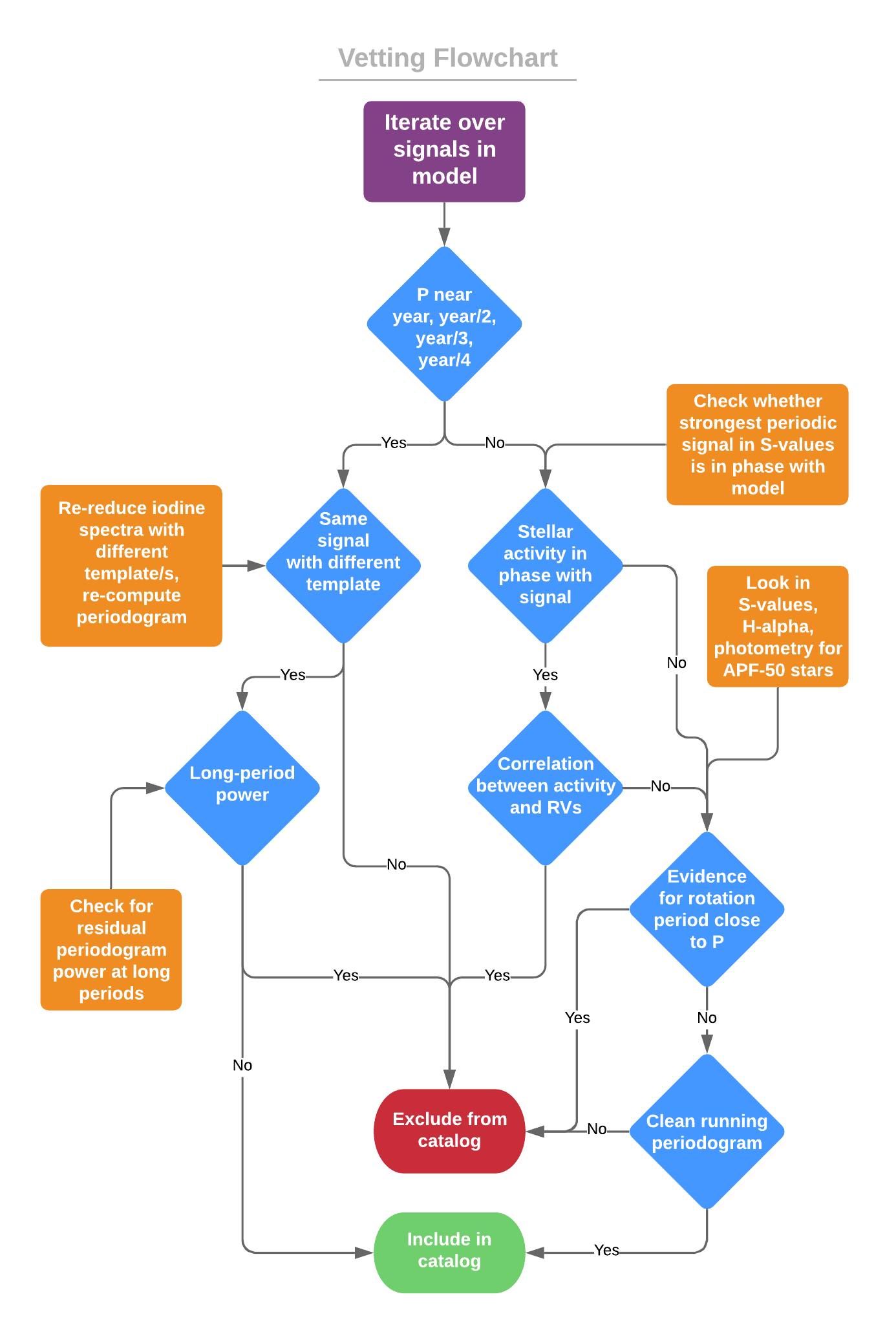}
\caption{Candidate vetting flowchart.}
\label{fig:vetting_flowchart}
\end{center}
\end{figure*}

\clearpage

\section{Planet and Stellar/Substellar Companion Catalog}

We present orbital solutions for the known planets, substellar companions, and stellar binaries that \texttt{RVSearch} has recovered in the California Legacy Survey. As mentioned in Section 5.1, where appropriate, we modeled long-period companions with linear or parabolic trends. We included in the appendix portions of the tables associated with each class of object: one for planets, one for stellar and substellar companions that are best modeled by Keplerian orbits, and one for stars with linear or parabolic RV trends. We also present 14 newly confirmed or significantly revised exoplanets and substellar companions. We list them and their orbital parameters in Table 1, and include individual notes on each system in Appendix A. Figure \ref{fig:catalog} shows all recovered planets in our survey, and distinguishes between known planets and new discoveries.

\begin{figure*}[ht!]
\begin{center}
\includegraphics[width = \textwidth]{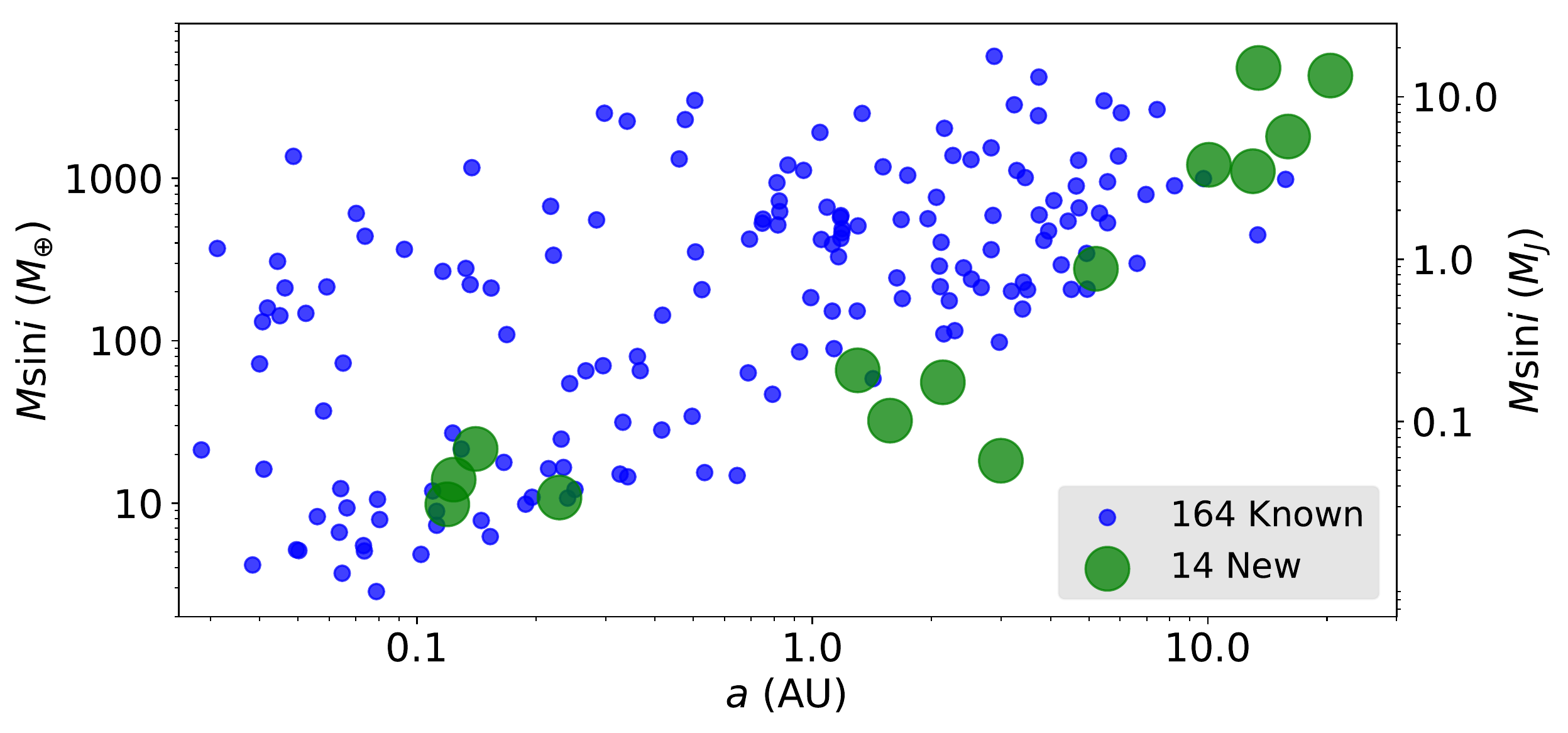}
\caption{Scatterplot of best-fit \msini\ and semi-major axis values for planets in the CLS catalog. Blue dots represent known planets, while green circles represent newly discovered planets and planets with significantly revised orbits.}
\label{fig:catalog}
\end{center}
\end{figure*}

\begin{longtable*}{llrrrrrr}
\caption{Discovered or Revised Planets and Substellar Companions} \\
\toprule
\midrule

CPS Name & Lit. Name & \msini\ [\mjup] (\mearth) & $a$ [AU] & $e$ & $P$ [d] & $t_c - 2450000$ [d] & $\omega$ \\
\toprule
HD 107148 c & HD 107148 c & $0.0626^{+0.0097}_{-0.0098}$ ($19.9^{+3.1}_{-3.1}$) & $0.1406^{+0.0018}_{-0.0018}$ & $0.34^{+0.13}_{-0.16}$ & $18.3267^{+0.0022}_{-0.0024}$ & $5203.4^{+1.4}_{-1.3}$ & $-0.78^{+0.67}_{-0.42}$ \\
HD 136925 b & HD 136925 b & $0.84^{+0.078}_{-0.074}$ ($267^{+25}_{-24}$) & $5.13^{+0.12}_{-0.11}$ & $0.103^{+0.094}_{-0.070}$ & $4540^{+160}_{-140}$ & $-310^{+280}_{-330}$ & $0.42^{+0.69}_{-1.30}$ \\ 
HD 141004 b & HIP 77257 b & $0.0428^{+0.0047}_{-0.0045}$ ($13.6^{+1.5}_{-1.4}$) & $0.1238^{+0.002}_{-0.002}$ & $0.16^{+0.11}_{-0.10}$ & $15.5083^{+0.0016}_{-0.0018}$ & $6704.33^{+0.55}_{-0.49}$ & $0.34^{+0.55}_{-0.74}$ \\
HD 145675 c & 14 Her c & $5.8^{+1.4}_{-1.0}$ & $16.4^{+9.3}_{-4.3}$ & $0.45^{+0.17}_{-0.15}$ & $25000^{+24000}_{-9200}$ & $4680^{+440}_{-310}$ & $-0.13^{+0.14}_{-0.18}$ \\
HD 156668 b & HD 156668 b & $0.0991^{+0.0079}_{-0.0077}$ ($31.5^{+2.5}_{-2.5}$) & $1.57^{+0.017}_{-0.017}$ & $0.089^{+0.084}_{-0.061}$ & $811.3^{+5.2}_{-5.3}$ & $5890^{+25}_{-19}$ & $-0.19^{+0.92}_{-0.74}$ \\
HD 164922 e & HD 164922 e & $0.0331^{+0.0031}_{-0.0031}$ ($10.52^{+0.99}_{-0.97}$) & $0.2292^{+0.0026}_{-0.0027}$ & $0.086^{+0.083}_{-0.060}$ & $41.763^{+0.012}_{-0.012}$ & $5256.6^{+1.2}_{-1.0}$ & $-0.3^{+1.30}_{-0.87}$ \\
HD 168009 b & HIP 89474 b & $0.03^{+0.0038}_{-0.0037}$ ($9.5^{+1.2}_{-1.2}$) & $0.1192^{+0.0017}_{-0.0018}$ & $0.121^{+0.110}_{-0.082}$ & $15.1479^{+0.0035}_{-0.0037}$ & $5201.56^{+0.82}_{-0.77}$ & $0.23^{+0.8}_{-1.1}$ \\ 
HD 213472 b & HD 213472 b & $3.48^{+1.10}_{-0.59}$ & $13.0^{+5.7}_{-2.6}$ & $0.53^{+0.120}_{-0.085}$ & $16700^{+12000}_{-4800}$ & $9580^{+190}_{-160}$ & $0.65^{+0.24}_{-0.25}$ \\
HD 24040 c & HD 24040 c & $0.201^{+0.027}_{-0.027}$ ($63.9^{+8.6}_{-8.6}$) & $1.3^{+0.021}_{-0.021}$ & $0.11^{+0.120}_{-0.079}$ & $515.4^{+2.2}_{-2.5}$ & $4984^{+19}_{-17}$ & $-0.3^{+1.30}_{-0.87}$ \\
HD 26161 b & HD 26161 b & $13.5^{+8.5}_{-3.7}$ & $20.4^{+7.9}_{-4.9}$ & $0.82^{+0.061}_{-0.050}$ & $32000^{+21000}_{-10000}$ & $10540^{+450}_{-280}$ & $-0.07^{+0.13}_{-0.12}$ \\
HD 3765 b & HIP 3206 b & $0.173^{+0.014}_{-0.013}$ ($54.8^{+4.3}_{-4.2}$) & $2.108^{+0.032}_{-0.033}$ & $0.298^{+0.078}_{-0.071}$ & $1211^{+15}_{-16}$ & $5607^{+52}_{-54}$ & $-1.1^{+2.60}_{-0.32}$ \\
HD 66428 c & HD 66428 c & $27^{+22}_{-17}$ & $23.0^{+19.0}_{-7.6}$ & $0.32^{+0.23}_{-0.16}$ & $39000^{+56000}_{-18000}$ & $-4100^{+4600}_{-4500}$ & $0.5^{+0.65}_{-1.40}$ \\
HD 68988 c & HD 68988 c & $15.0^{+2.8}_{-1.5}$ & $13.2^{+5.3}_{-2.0}$ & $0.45^{+0.130}_{-0.081}$ & $16100^{+11000}_{-3500}$ & $1660^{+110}_{-180}$ & $-0.13^{+0.037}_{-0.043}$ \\
HD 95735 c & GJ 411 c & $0.0568^{+0.0091}_{-0.0083}$ ($18.0^{+2.9}_{-2.6}$) & $3.1^{+0.13}_{-0.11}$ & $0.14^{+0.160}_{-0.095}$ & $3190^{+200}_{-170}$ & $7440^{+150}_{-150}$ & $0.1^{+1.1}_{-1.3}$ \\
\bottomrule
\end{longtable*}

\section{Discussion} \label{sec:discussion}

Through the use of high-cadence APF observations and long-baseline HIRES observations, we have expanded the population of known exoplanets along the current mass and semi-major axis boundary of detectability, as seen in Figure \ref{fig:catalog}. We recovered 43 planets with \msini\ $<$ 30 \mearth\ , including four new discoveries within 1 AU. In a future paper in the California Legacy Survey series, we will leverage the decades-long-baseline datasets in which these planets were discovered, in order to constrain the probability that a host of a small planet also hosts an outer companion, as explored in \cite{Bryan19} and \cite{Zhu18}. We will also directly place a lower limit on the conditional occurrence of inner small planets given the presence of an outer gas giant.

\begin{figure}[ht!]
\includegraphics[width = 0.49\textwidth]{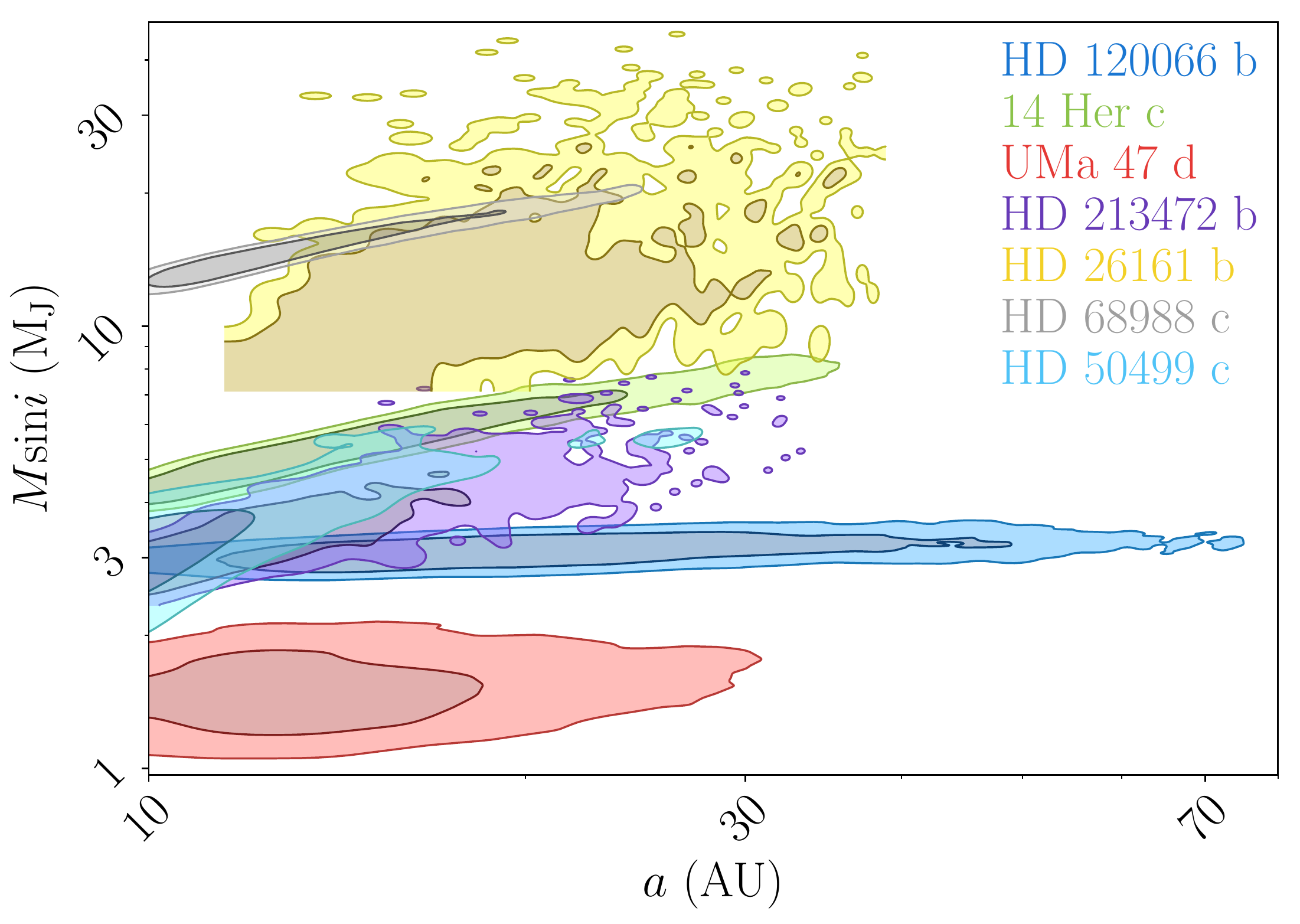}
\caption{Contours (1- and 2-$\sigma$) of \msini\ and semi-major axes for planets in the CLS sample whose semi-major axis posteriors extend beyond 10 AU. Contours for HD 26161 b have hard cutoffs due to sparsity below 7 \mjup\ and 12 AU; these limits come from the data's baseline and RV increase to date.}
\label{fig:cold_contours}
\end{figure}

\begin{figure*}[ht!]
\begin{center}
\includegraphics[width = \textwidth]{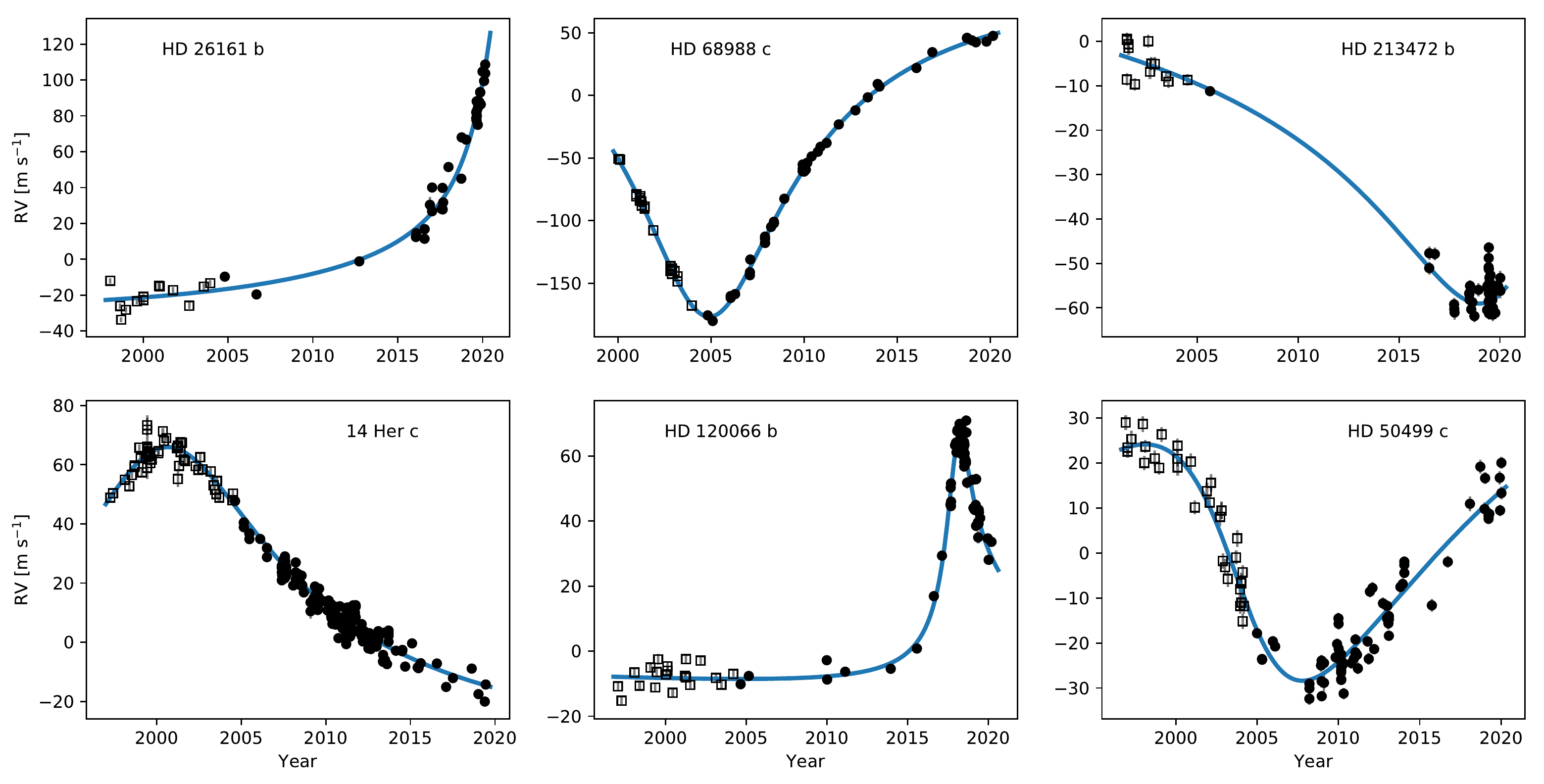}
\caption{Orbit gallery for six of the coldest companions in our survey. We plot RV data and Keplerian model versus year, and subtract off the model signatures of inner companions and stellar activity. We did not include UMa 47 d, seen in Figure \ref{fig:cold_contours}, in this plot, because its detection relied on early Lick-Hamilton RVs, and we wanted to showcase HIRES RV measurements from the past twenty-four years.}
\label{fig:gallery}
\end{center}
\end{figure*}

In addition to expanding the population of small planets with measured \msini\ , we discovered or revised the orbits of ten planets with orbital separations greater than 1 AU, six of them beyond 4 AU. We represent the model posteriors for the coldest of these planets in Figure \ref{fig:cold_contours}, and show a gallery of some of their orbits in Figure \ref{fig:gallery}. These discoveries include two new detections with incomplete orbits, HD 213472 b and HD 26161 b.  Details are provided in Appendices A.3 and A.14. Using HIRES to extend the observational baseline of our survey by another decade will tighten our \msini\ and orbital parameter constraints for these planets, and may reveal more cold companions beyond 10 AU.

In a future paper in the CLS series, we will use our sample of long-period planets and completeness contours to measure the mass--period planet occurrence distribution out to 10 AU, extending beyond the Keck Planet Search's limit of 5 AU \citep{Cumming08} and the 9 AU limit of \cite{Wittenmyer20}. This will provide novel constraints on planet occurrence beyond the water ice line, resolve the discrepancy between the results of \cite{Fernandes19} and those of \cite{Wittenmyer20}, and provide new insight into planet formation across protoplanetary disks.

\begin{figure}[ht!]
\includegraphics[width = 0.49\textwidth]{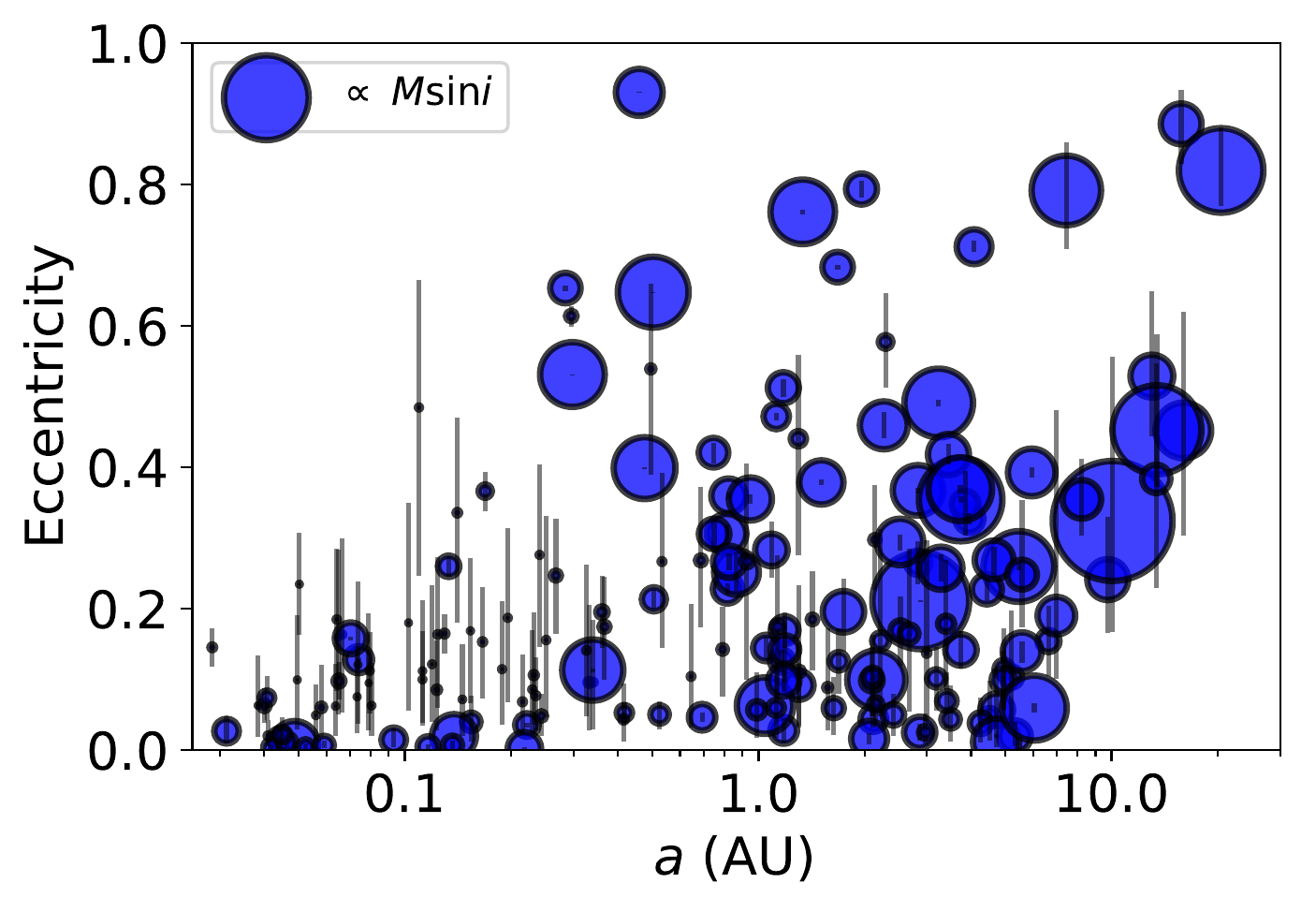}
\caption{\msini\ , $a$, and eccentricity of the CLS sample. Eccentricity is plotted in medians and 68$\%$ confidence intervals, while scatter size is proportional to \msini\ posterior mode.}
\label{fig:axis_eccentricity}
\end{figure}

Figure \ref{fig:axis_eccentricity} is a visualization of the eccentricities of all planets in the California Legacy Survey. In future work, we will quantify the eccentricity distribution of gas giants in our sample and its dependence on planet mass and multiplicity, as well as the eccentricity distributions of brown dwarfs and other substellar companions, in order to clarify possible formation pathways. We will extend the wide-orbit population comparisons of \cite{Bowler20} to our sample of planets and brown dwarfs within 20 AU of their hosts. We will also explore the eccentricity distribution of gas giants beyond 7 AU. As Figures \ref{fig:gallery} and \ref{fig:axis_eccentricity} show, all planets recovered beyond 7 AU are eccentric with significance $e > 2\sigma_e$. This may be a selection effect, as the median baseline of observations in our sample is 21 yr, which corresponds to a semi-major axis of 7.6 AU for a planet orbiting a solar-mass star. It is possible that planets with orbital periods beyond our observational baselines are more easily detectable if they are eccentric. We can use injection-recovery tests to determine whether there is a detection bias toward eccentric planets beyond observational baselines. If this phenomenon is not a selection effect, it might imply that most giant planets beyond 7 AU have undergone a scattering event or otherwise been excited to high eccentricity. Taken together, these studies will leverage this decades-long observational undertaking to provide new insights into planet formation and evolution.

\acknowledgments
L.J.R.\ led the construction of this paper, including finalizing the stellar sample, running the Keplerian search, assessing planet candidates and generating the planet catalog, generating most of the figures, and writing this manuscript. The RVsearch pipeline was developed by L.J.R., B.J.F., and L.A.H., with assistance from A.W.H., H.T.I., E.A.P., and C.M.D.
B.J.F., A.W.H., L.A.H., H.T.I., E.A.P., and I.A.S.\ assisted L.J.R.\ in vetting the planet candidates and insuring the integrity of the RVs and the planet catalog.
A.W.H., G.W.M.\ (though 2015), D.A.F., and J.T.W.\ provided leadership and funding to CPS and CCPS.
L.J.R., B.J.F., L.A.H., H.T.I., A.W.H., S.C.B., E.A.P., A.B., A.C., J.R.C., I.J.M.C., P.A.D., D.A.F., M.K., G.W.M., R.A.R., L.M.W., and J.T.W.\ contributed significantly to the Doppler observations.
H.T.I., A.W.H., B.J.F., and G.W.M.\ executed and refined the Doppler pipeline that produced the RVs reported here.
G.W.H.\ contributed photometry and analysis that were used to rule out stellar activity signals.
I.A.S.\ provided similar analysis of activity based on a suite of indicators.
B.J.F., A.W.H., E.A.P., L.M.W., R.A.R., and H.T.I.\ created an internal data visualization system (``Jump'') that was integral to this project.
L.J.R., B.J.F., A.W.H., L.A.H., H.T.I., E.A.P., H.A.K., S.R.K., P.A.D., and L.M.W.\ contributed to the discussion section and structure of this paper, as well as the strategy of this paper and successors in the CLS series.

We thank Jay Anderson, G\'asp\'ar Bakos, Mike Bottom, John Brewer, Christian Clanton, Jason Curtis, Fei Dai, Steven Giacalone, Sam Grunblatt, Michelle Hill, Lynne Hillenbrand, Rebecca Jensen-Clem, John A.\ Johnson, Chris McCarthy, Sean Mills, Teo Mo\v{c}nik, Ben Montet, Jack Moriarty, Tim Morton, Phil Muirhead, Sebastian Pineda, Nikolai Piskunov, Eugenio Rivera, Julien Spronck, Jonathan Swift, Guillermo Torres, Jeff Valenti, Sharon Wang, Josh Winn, Judah van Zandt, Ming Zhao, and others who contributed to the observations and analysis reported here.  We acknowledge R.\ P.\ Butler and S.\ S.\ Vogt for many years of contributing to this dataset. This research has made use of the Keck Observatory Archive (KOA), which is operated by the W. M. Keck Observatory and the NASA Exoplanet Science Institute (NExScI), under contract with the National Aeronautics and Space Administration.  We acknowledge RVs stemming from HIRES data in KOA with principal investigators from the LCES collaboration (S.\ S.\ Vogt, R.\ P.\ Butler, and N.\ Haghighipour).
We gratefully acknowledge the efforts and dedication of the Keck Observatory staff for support of HIRES and remote observing. We are grateful to the time assignment committees of the Caltech, the University of California, the University of Hawaii, NASA, and NOAO for their generous allocations of observing time. Without their long-term commitment to radial velocity monitoring, these planets would likely remain unknown.

We thank Ken and Gloria Levy, who supported the construction of the Levy Spectrometer on the Automated Planet Finder, which was used heavily for this research. We thank the University of California and Google for supporting Lick Observatory, and the UCO staff as well as UCO director Claire Max for their dedicated work scheduling and operating the telescopes of Lick Observatory. G.W.H.\ acknowledges long-term support from NASA, NSF, Tennessee State University, and the State of Tennessee through its Centers of Excellence program. A.W.H.\ acknowledges NSF grant 1753582. H.A.K. acknowledges NSF grant 1555095. P.D.\ gratefully acknowledges support from a National Science Foundation (NSF) Astronomy \& Astrophysics Postdoctoral Fellowship under award AST-1903811.

This work has made use of data from the European Space Agency (ESA) mission {\it Gaia} (\url{https://www.cosmos.esa.int/gaia}), processed by the {\it Gaia} Data Processing and Analysis Consortium (DPAC, \url{https://www.cosmos.esa.int/web/gaia/dpac/consortium}). Funding for the DPAC has been provided by national institutions, in particular the institutions participating in the {\it Gaia} Multilateral Agreement.

Finally, we recognize and acknowledge the cultural role and reverence that the summit of Maunakea has within the indigenous Hawaiian community. We are deeply grateful to have the opportunity to conduct observations from this mountain.

\facilities{Keck:I (HIRES), Automated Planet Finder (Levy), Lick (Hamilton)}

\software{All code, plots, tables, and data used in this paper are available at \url{github.com/leerosenthalj/CLSI}. Data and tables, including the full stellar catalog with $\{$ $M$, $R$, $T_{\mathrm{eff}}$, log$g$, [Fe/H] $\}$, as well as APT photometry, are also available in the associated .tar.gz file available through ApJ. \texttt{RVSearch} is available at \url{github.com/California-Planet-Search/rvsearch}. This research makes use of GNU Parallel \citep{Tange11}. We made use of the following publicly available Python modules:
          \texttt{pandas} \citep{pandas},
          \texttt{numpy/scipy} \citep{numpy/scipy},
          \texttt{emcee}    \citep{Foreman-Mackey13},
          \texttt{Specmatch}  \citep{Petigura15, Yee17},
          \texttt{Isoclassify}  \citep{Huber17},
          \texttt{TheJoker}   \citep{Price-Whelan17},
          \texttt{RadVel}   \citep{Fulton18},
          \texttt{RVSearch} (this work).
          }
\newpage

\bibliographystyle{aasjournal}
\bibliography{legacy_catalog}{}

\appendix

In our appendices, we have included individual notes on each planet discovery reported in this paper; complete tables of recovered planets, Keplerian-resolved stellar binaries, and substellar companions in the California Legacy Survey; signals that \texttt{RVSearch} recovered and we determined to be false-positives; linear and parabolic RV trends; and excerpts from the stellar sample and RV dataset, which are available in their entirety in machine-readable format.

\section{Individual Discoveries and Revised Orbits}

\subsection{HD 3765}

\begin{figure*}[ht!]
\begin{center}
\includegraphics[width=0.99\textwidth]{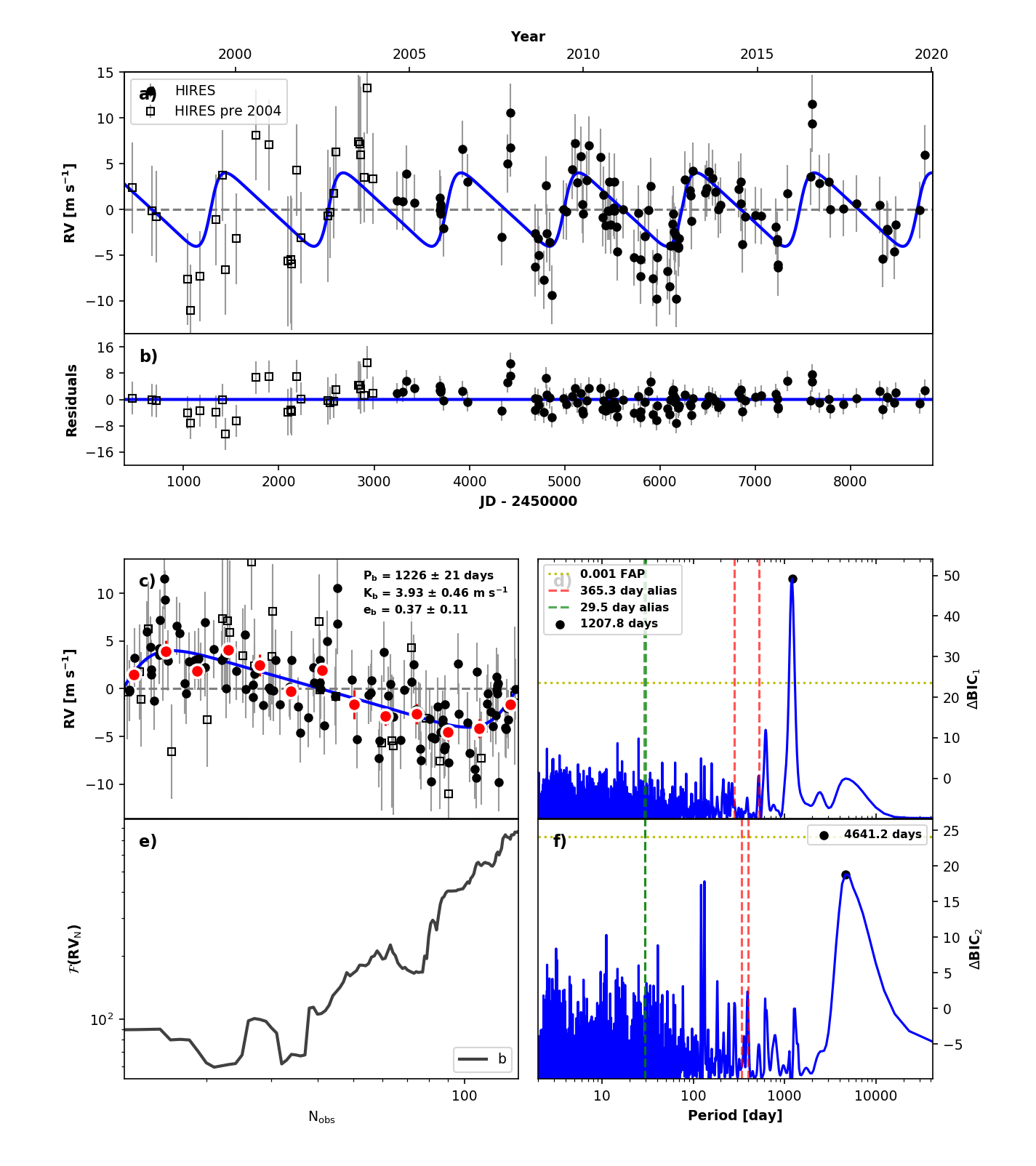}
\caption{\texttt{RVSearch} summary plot for HD 3765.  See Figure \ref{fig:hip109388_summary} for plot description.}
\label{fig:3765_summary}
\end{center}
\end{figure*}

\begin{figure*}[ht!]
\begin{center}
\includegraphics[width=0.99\textwidth]{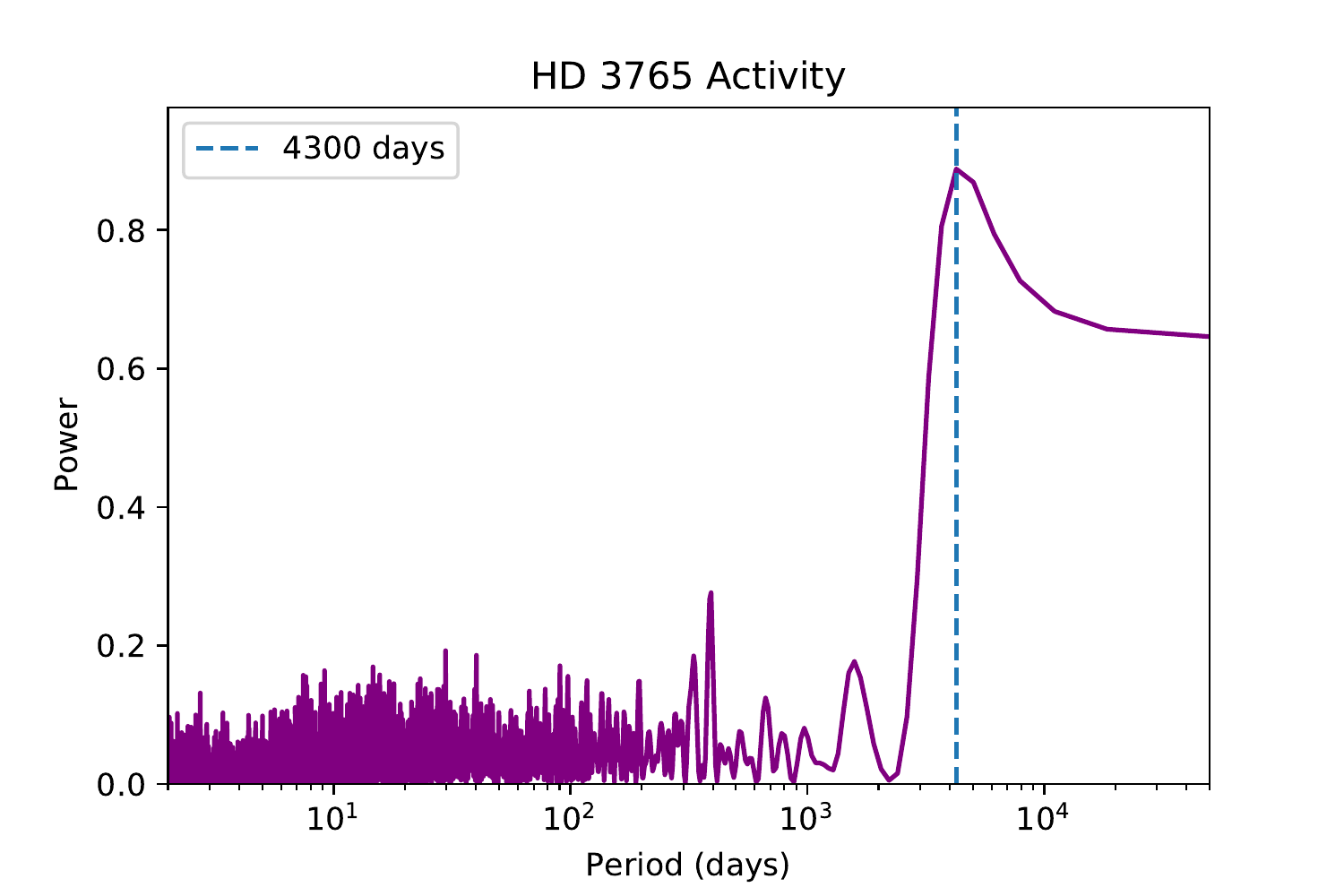}
\caption{Lomb-Scargle periodogram of HIRES S-values for HD 3765. Significant power at and beyond 4,300 days.}
\label{fig:3765_svals}
\end{center}
\end{figure*}

\begin{figure*}[ht!]
\begin{center}
\includegraphics[width=0.99\textwidth]{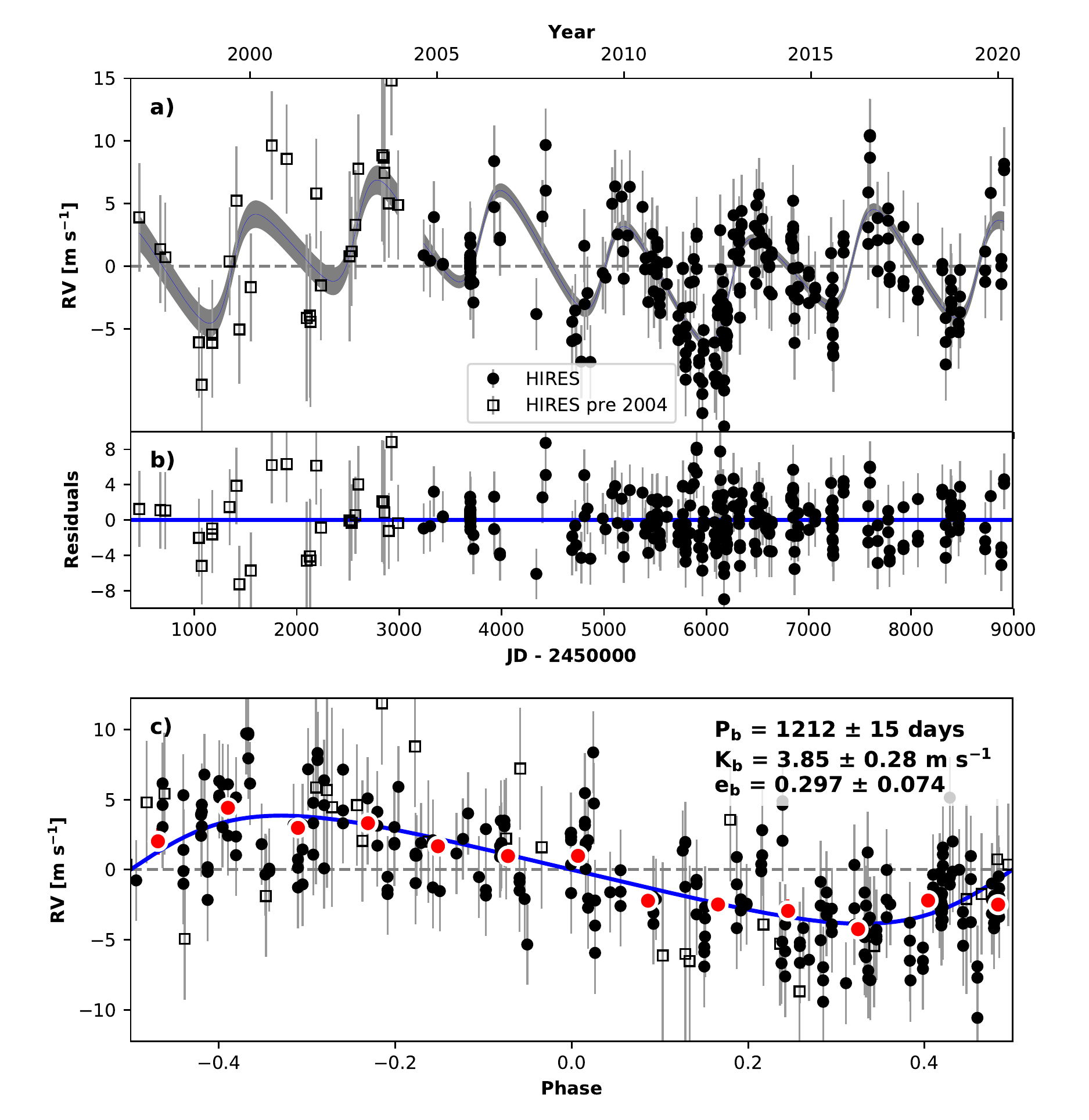}
\caption{\texttt{RadVel} model orbital plot for HD 3765, including a Gaussian process with a squared-exponential kernel. The grey shaded curve represents the 68$\%$ interval for the Gaussian process RV signature.}
\label{fig:3765_gp}
\end{center}
\end{figure*}

\begin{figure*}[ht!]
\begin{center}
\includegraphics[width=\textwidth]{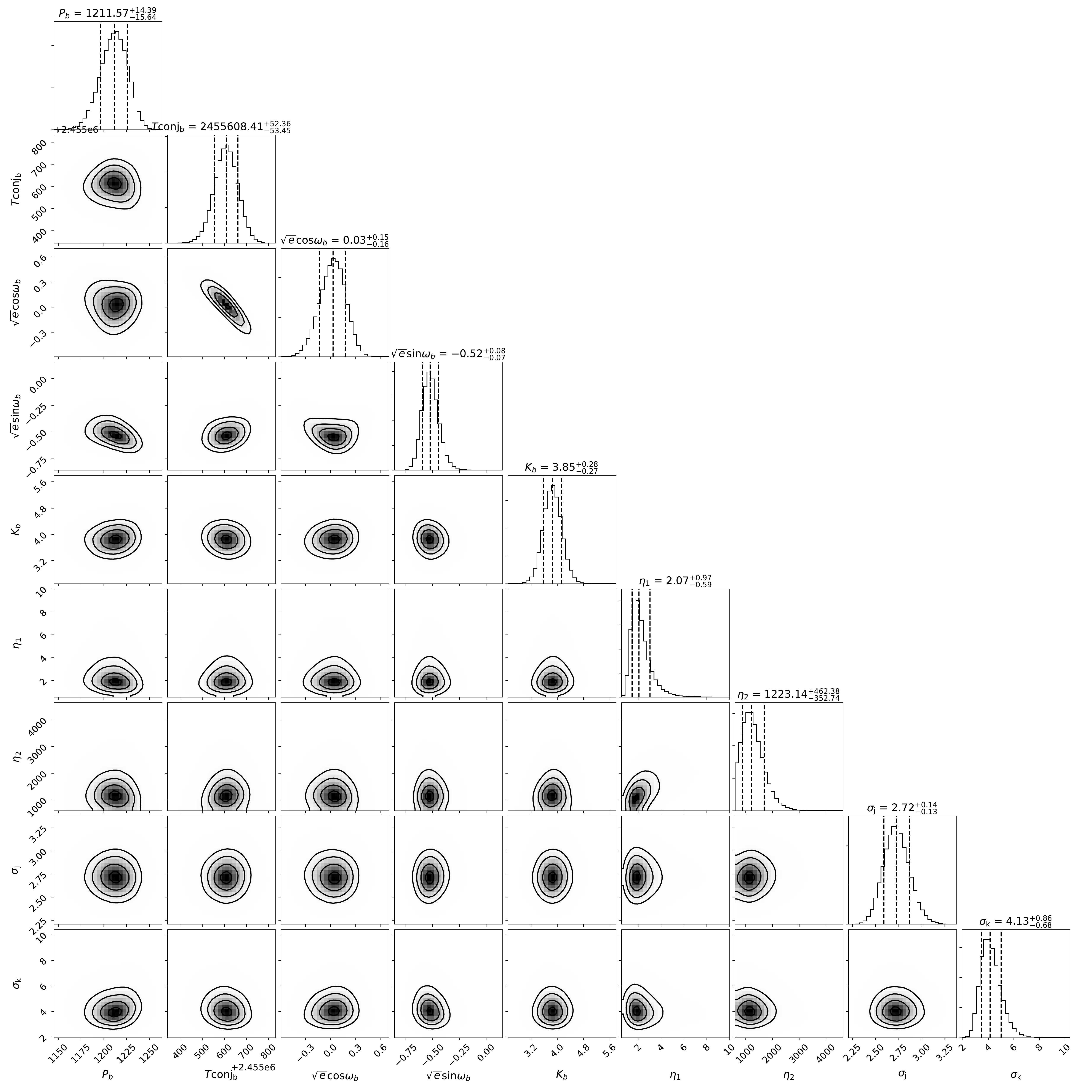}
\caption{Orbital and Gaussian process parameter posteriors for HD 3765. $\eta_1$ is the GP amplitude, while $\eta_2$ is the GP exponential decay timescale.}
\label{fig:3765_corner}
\end{center}
\end{figure*}

HD 3765 is a K2 dwarf at a distance of 17.9 pc \citep{GaiaDR2}. Figure \ref{fig:3765_summary} shows the \texttt{RVSearch} results for this star. We recovered a signal with a period of 3.36 yr. Table 1 reports all planet parameters. There is significant periodicity in the S-value time series, but concentrated around a period of 12 yr. Furthermore, we find no correlation between the RVs and S-values. Figure \ref{fig:3765_svals} shows a Lomb--Scargle periodogram of the S-value time series. Thus, we label this signal as a confirmed planet, with \msini\ = $0.173 \pm 0.014$ \mjup\ and $a = 2.108 \pm 0.033$ AU. The magnetic activity cycle is too weak for \texttt{RVSearch} to recover, but is evident in the best-fit RV residuals. We used RadVel to model this activity cycle with a squared-exponential Gaussian process, and report MCMC-generated posteriors for both orbital and Gaussian process parameters in Figure \ref{fig:3765_gp} and Figure \ref{fig:3765_corner}.

\clearpage

\subsection{HD 24040}

\begin{figure*}[ht!]
\begin{center}
\includegraphics[width=0.6\textwidth]{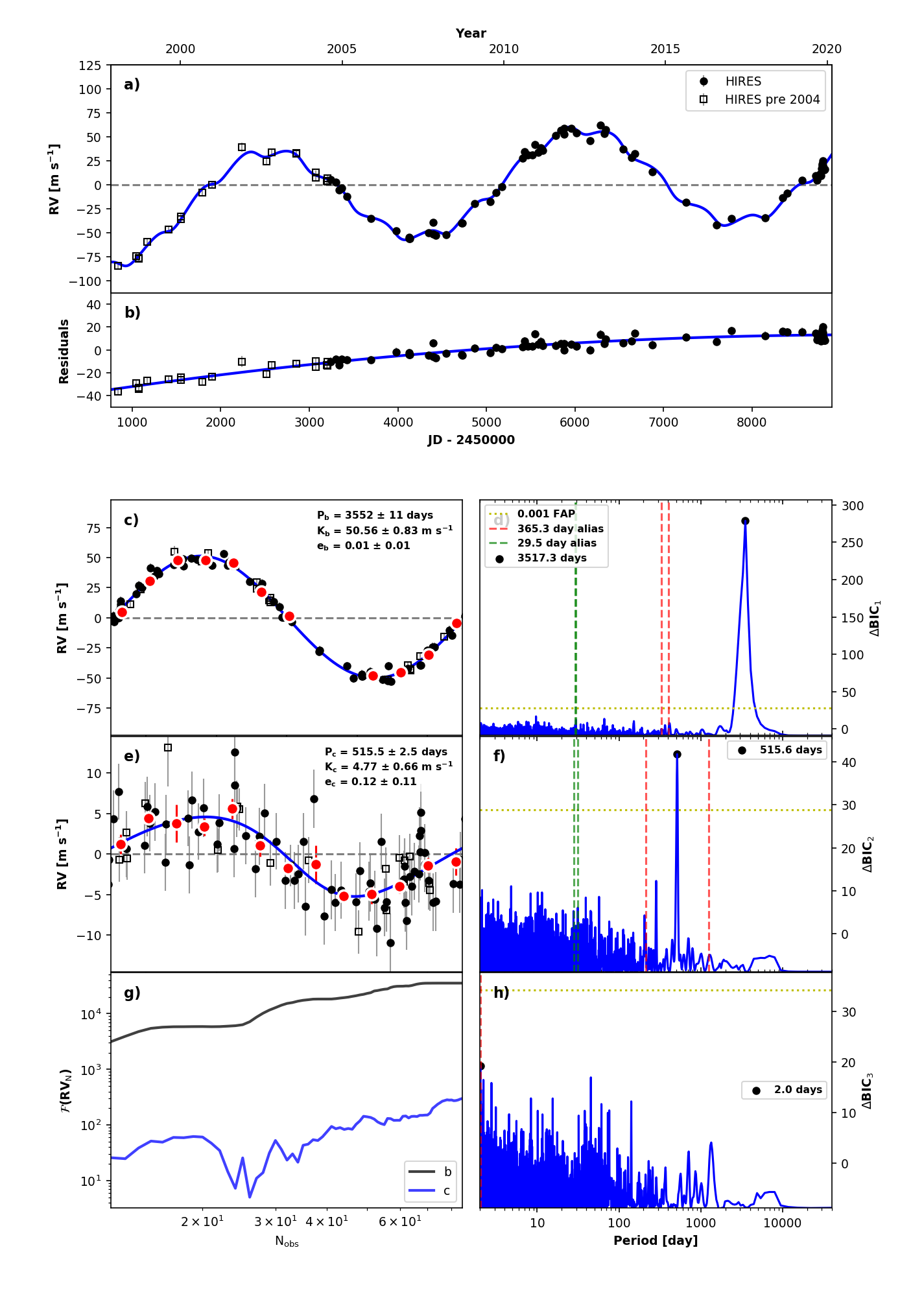}
\caption{\texttt{RVSearch} summary plot for HD 24040.  See Figure \ref{fig:hip109388_summary} for plot description.}
\label{fig:24040_summary}
\end{center}
\end{figure*}

\begin{figure*}[ht!]
\begin{center}
\includegraphics[width=0.99\textwidth]{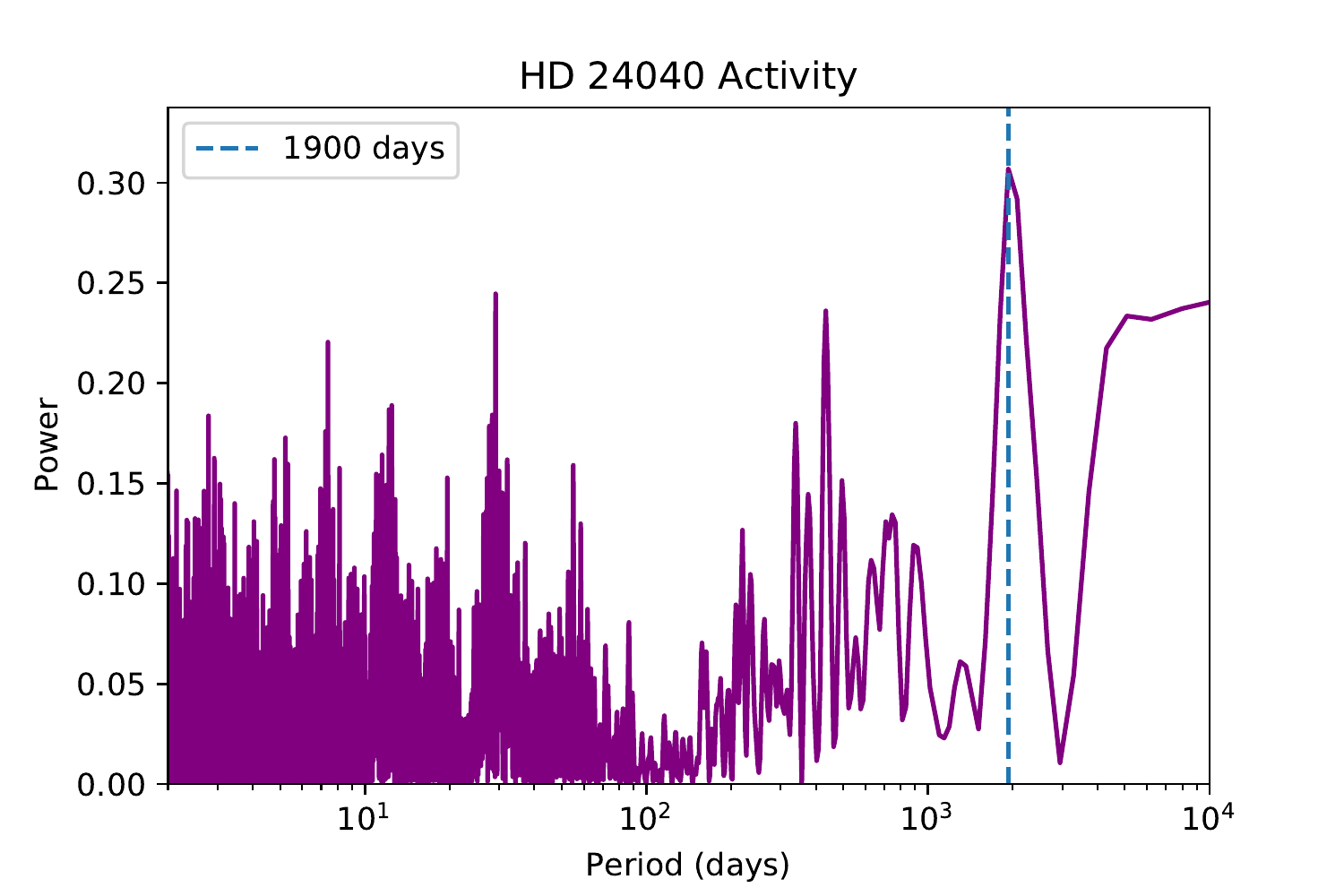}
\caption{Lomb-Scargle periodogram of HIRES S-values for HD 24040. No periods show power that is statistically significant.}
\label{fig:24040_svals}
\end{center}
\end{figure*}

HD 24040 is a G1 dwarf at a distance of 46.7 pc. Figure \ref{fig:24040_summary} shows the \texttt{RVSearch} results for this star. It hosts a known gas giant \citep{Wright07, Feng15} with a semi-major axis that we measured as $a = 4.72 \pm 0.18$ AU, an orbital period of $9.53 \pm 10^{-4}$ yr, and a minimum mass \msini\ = $4.09 \pm 0.22$ \mjup. We have extended the observational baseline of our HIRES measurements to 21.7 yr, constrained the long-term trend and curvature of the RVs, and discovered a new exoplanet, a sub-Saturn (\msini\ = $0.201 \pm 0.027$ \mjup) on a 1.4 yr orbit ($a = 1.30 \pm 0.021$ AU) that is consistent with circular.  The S-values are uncorrelated with the the RVs of both planet signals, after removing the long-term trend. Figure \ref{fig:24040_svals} shows a Lomb--Scargle periodogram of the S-value time series. Table 1 reports all planet parameters.

In addition to the newly detected sub-Saturn, we further constrained the known linear trend in the RVs and found evidence for a curvature term as well. \texttt{RVSearch} detected a curvature term with model preference \dBIC $>$ 10 over a purely linear trend. We measured the linear trend to be $0.00581 \pm 0.00044$ \msd, and the curvature to be $-6.6 \times 10^{-7} \pm 1.2 \times 10^{-7}$ \msdd, a 5.5$\sigma$ detection. The trend and curvature parameters are slightly correlated in the posterior, but neither is correlated with any of the Keplerian orbital parameters in the model. Therefore, we kept the curvature term that \texttt{RVSearch} selected in our model. This long-term trend is low-amplitude enough that it may be caused by another planet in the system, orbiting beyond 30 AU. Gaia astrometry or another two decades of RVs may provide further constraints on this object.

\clearpage

\subsection{HD 26161}

\begin{figure*}[ht!]
\begin{center}
\includegraphics[width=\textwidth]{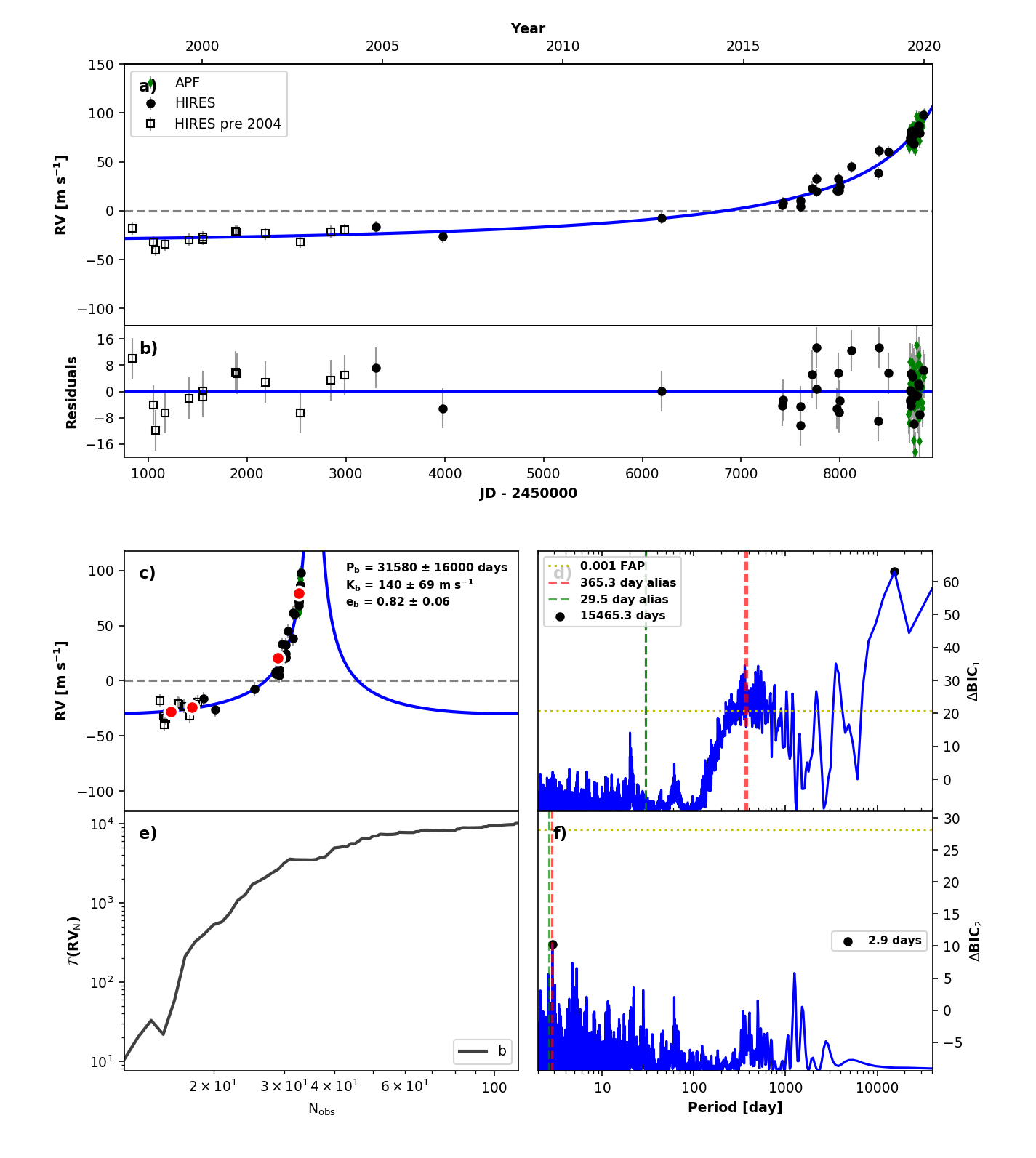}
\caption{\texttt{RVSearch} summary plot for HD 26161.  See Figure \ref{fig:hip109388_summary} for plot description.}
\label{fig:26161_summary}
\end{center}
\end{figure*}

\begin{figure*}[ht!]
\begin{center}
\includegraphics[width=\textwidth]{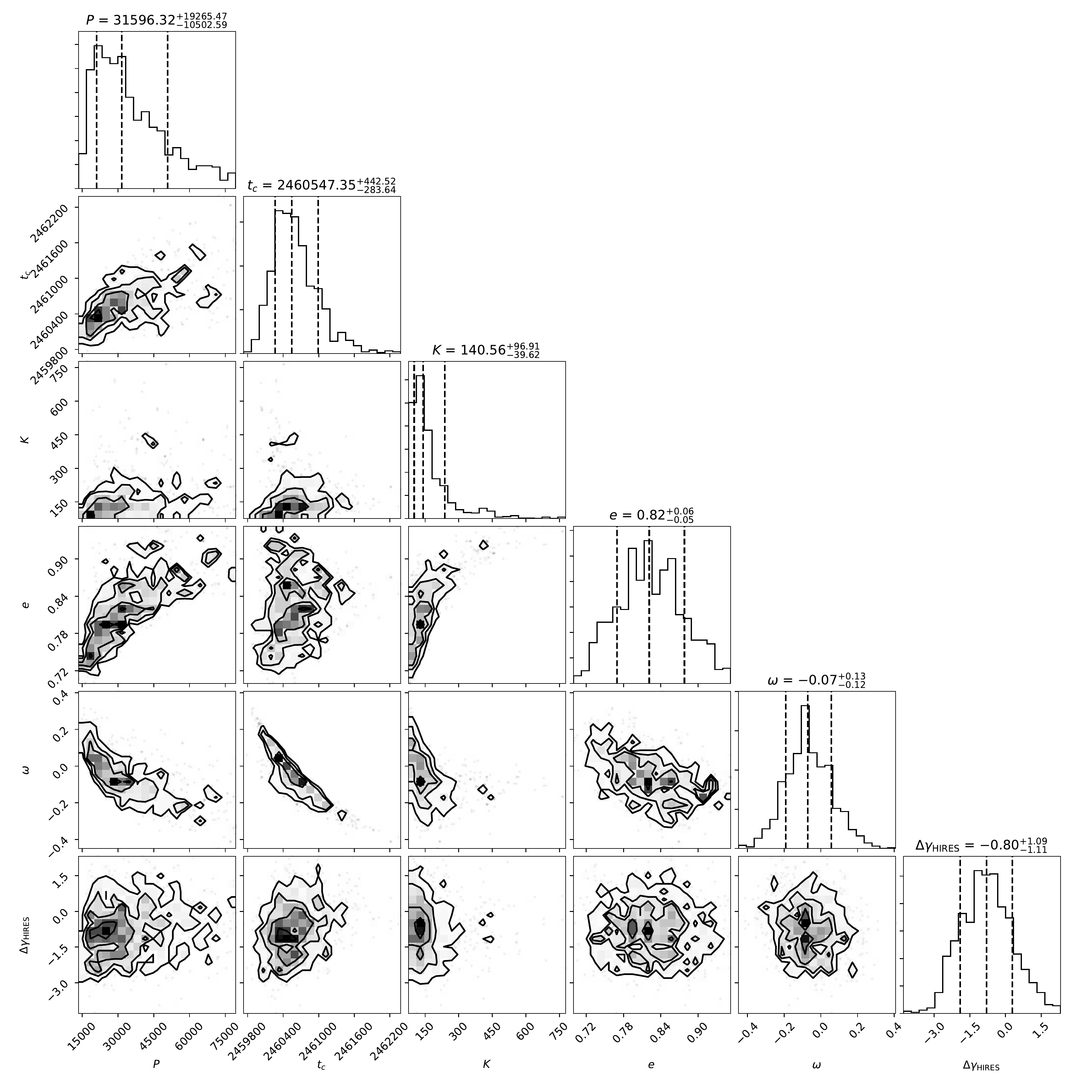}
\caption{Rejection sampling posterior for HD 26161.}
\label{fig:26161_corner}
\end{center}
\end{figure*}

HD 26161 is a G0 dwarf located at a distance of 50.0 pc. Figure \ref{fig:26161_summary} shows the \texttt{RVSearch} results for this star. Our RVs are consistent with a long-period, eccentric companion, and \texttt{RVSearch} detected this long-period signal. Due to the sparseness of the data and the fractional orbital coverage, traditional MCMC methods fail to return a well-sampled model posterior. Since the data underconstrains our model, we used \texttt{TheJoker} to sample the posterior, which is consistent with an extremely long-period gas giant with minimum mass \msini\ = $13.5\mathrm{\substack{+8.5 \\ -3.7}}$ \mjup, semi-major axis $a = 20.4\mathrm{\substack{+7.9 \\ -4.9}}$ AU, and eccentricity $e=0.82\mathrm{\substack{+0.06 \\ -0.05}}$. Table 1 reports current estimates of all orbital parameters, and Figure \ref{fig:26161_corner} shows their posterior distributions. A Keplerian model is significantly preferred over a quadratic trend, with \dBIC $>$ 15.

The Simbad stellar catalog designates HD 26161 as a stellar multiple. We used Gaia to identify a binary companion with similar parallax and within 60 arcseconds. This companion has an effective temperature identified from Gaia colors of 4053 K, and a projected separation of 562 AU. A stellar companion that is currently separated from its primary by more than 560 AU could not cause a change in RV of 100 \ms over 4 yr. This curve is far more likely caused by an inner planetary or substellar companion approaching periastron.

\begin{figure*}[ht!]
\begin{center}
\includegraphics[width=\textwidth]{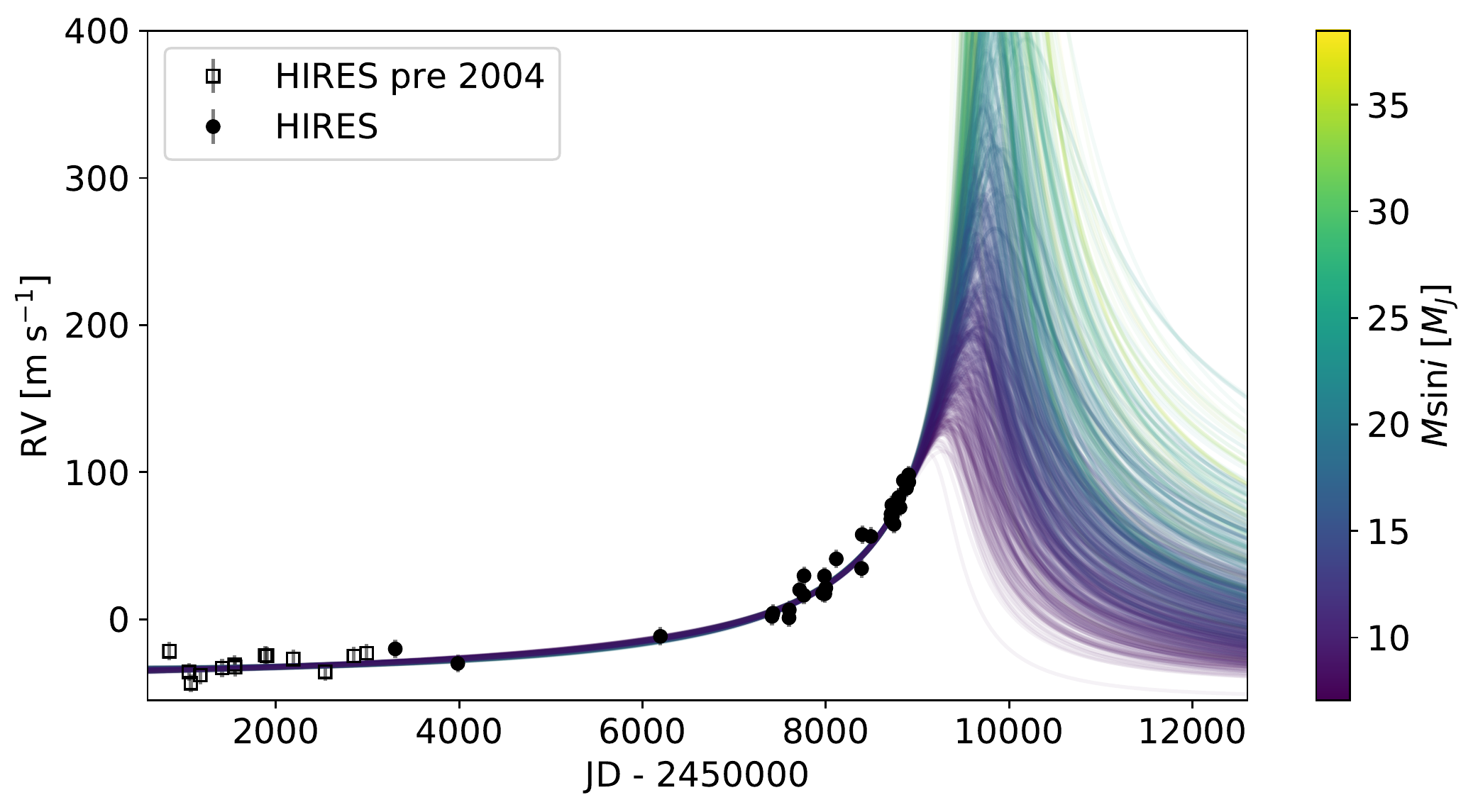}
\caption{Possible orbits for HD 26161 b. RV curves are drawn from the rejection sampling posterior generated with \texttt{TheJoker}. The color of each orbit drawn from the posterior scales with \msini\ .}
\label{fig:26161_orbits}
\end{center}
\end{figure*}

Figure \ref{fig:26161_orbits} shows a sample of possible orbits for HD 26161 b, drawn from our rejection sampling posteriors and projected over the next decade. We will continue to monitor HD 26161 with HIRES at moderate cadence, and have begun observing this star with APF. As we gather more data during the approach to periastron, we can tighten our constraints on the minimum mass, eccentricity, and orbital separation of HD 26161 b.

\clearpage

\subsection{HD 66428}

\begin{figure*}[ht!]
\begin{center}
\includegraphics[width=0.6\textwidth]{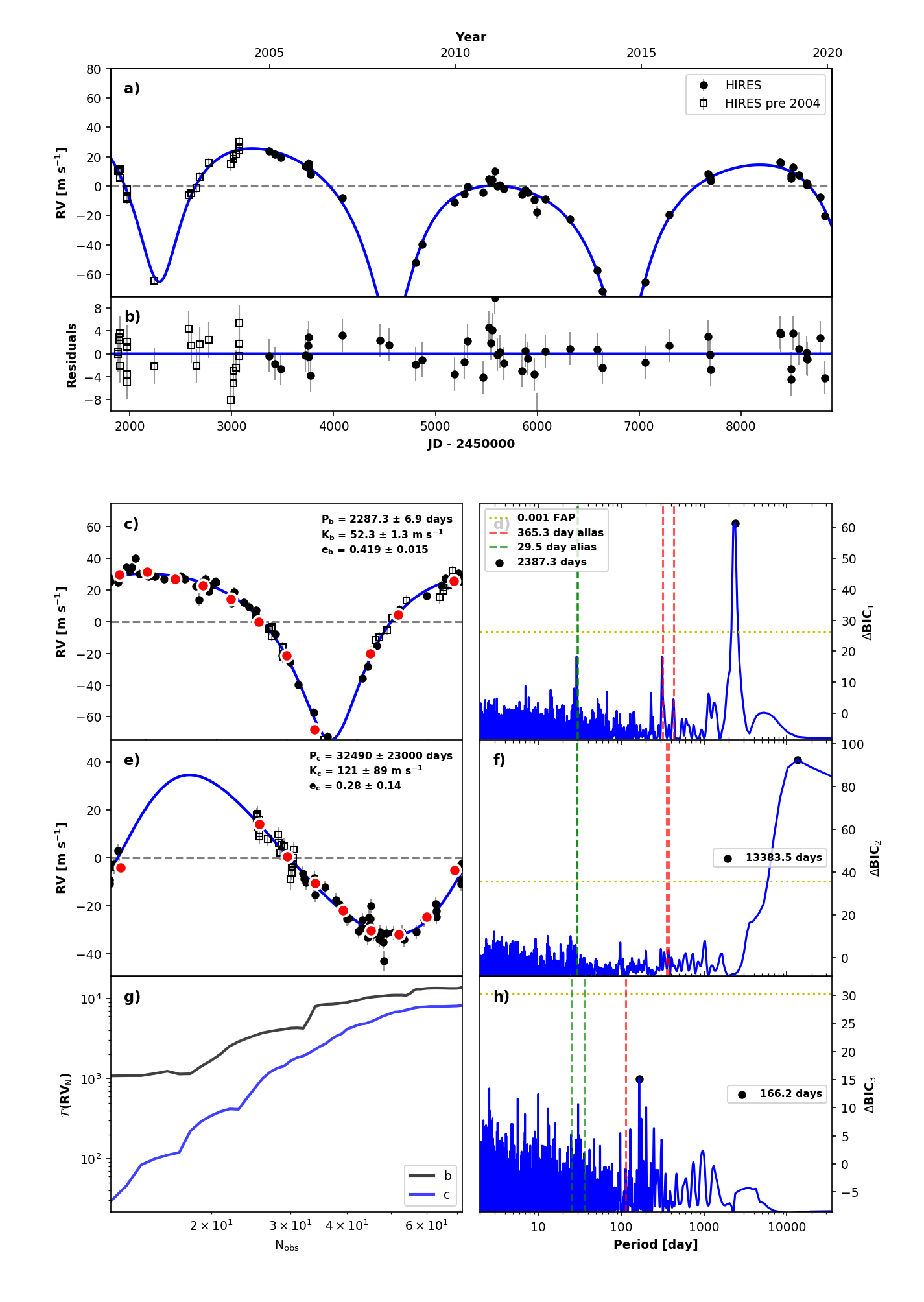}
\caption{\texttt{RVSearch} summary plot for HD 66428.  See Figure \ref{fig:hip109388_summary} for plot description.}
\label{fig:66428_summary}
\end{center}
\end{figure*}

HD 66428 is a G8 dwarf found at a distance of 53.4 pc. Figure \ref{fig:66428_summary} shows the \texttt{RVSearch} results for this star. This system has one well-constrained cold Jupiter \citep{Butler06} and an outer companion candidate first characterized in \cite{Bryan16} as a linear trend. With four more years of HIRES data, we now see curvature in the RVs and a clear detection in \texttt{RVSearch}, and can place constraints on this outer candidate's orbit with a Keplerian model. The Keplerian orbit for the outer candidate is preferred to a parabolic trend with \dBIC $>$ 30. A maximum likelihood fit gives an orbital period of $P = 36.4$ yr. However, since we have only observed a partially resolved orbit so far, the orbit posterior in period space is wide and asymmetric. MCMC sampling produces $P = 88\mathrm{\substack{+153 \\ -49}}$ yr. Table 1 reports current estimates of all orbital parameters.

The model parameters are \msini\ = $27\mathrm{\substack{+22 \\ -17}}$ \mjup, $a = 23.0\mathrm{\substack{+19.0 \\ -7.6}}$ AU, and $e = 0.31\mathrm{\substack{+0.13 \\ -0.13}}$. This orbital companion could be a massive gas giant or a low-mass star, if we only consider constraints from RV modeling. However, \cite{Bryan16} used NIRC2 Adaptive-Optics images to place upper bounds on the mass and semi-major axis of an outer companion, at a time when it only presented as a linear trend in HIRES RVs. They found an upper bound of $\approx$100 \mjup\ on mass, not just \msini, and an upper bound of $\approx$150 AU on $a$. We will continue to monitor this star with HIRES to further constrain the mass and orbit of HD 66428 c.

\clearpage

\subsection{HD 68988}

\begin{figure*}[ht!]
\begin{center}
\includegraphics[width=0.6\textwidth]{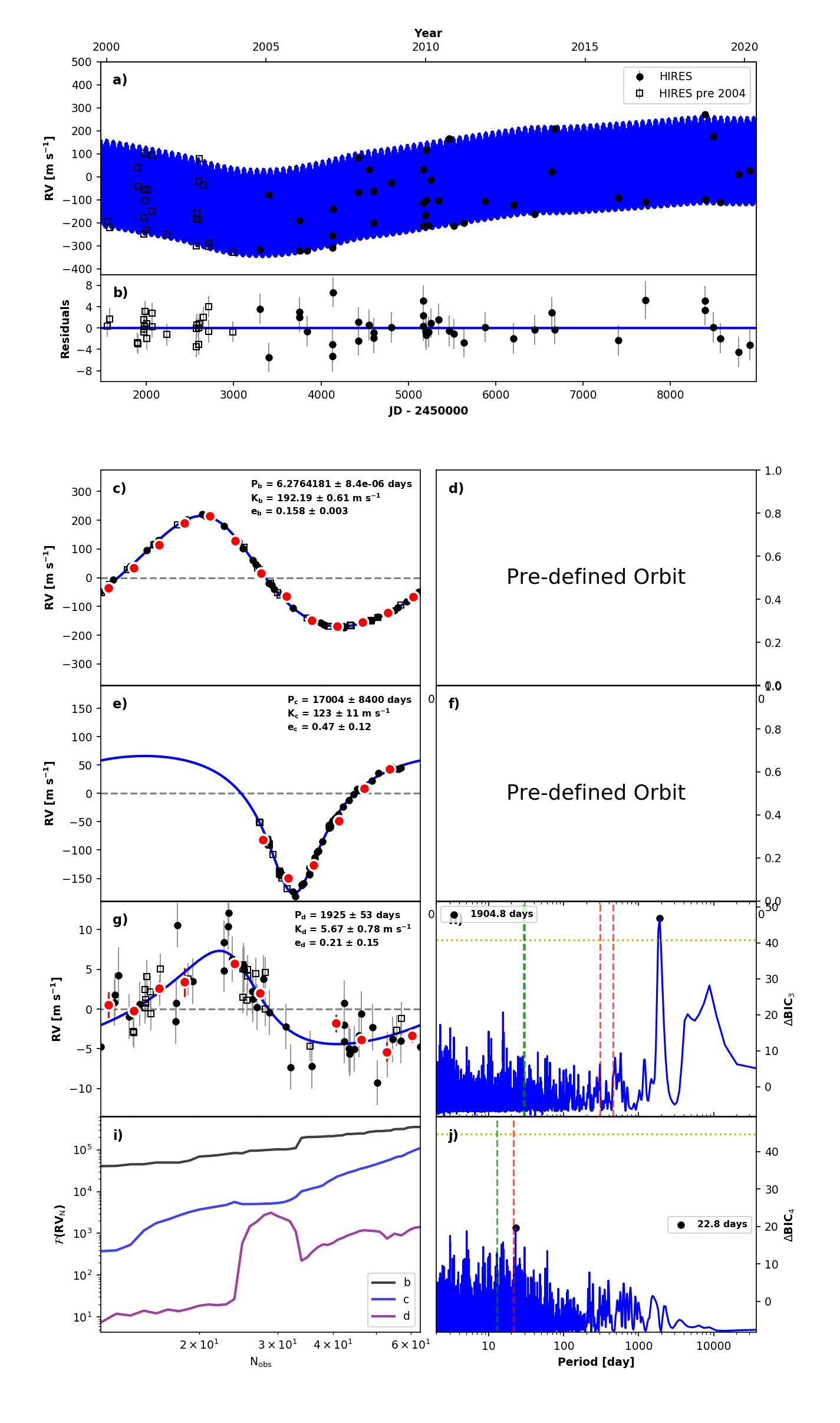}
\caption{\texttt{RVSearch} summary plot for HD 68988.  See Figure \ref{fig:hip109388_summary} for plot description.}
\label{fig:68988_summary}
\end{center}
\end{figure*}

HD 68988 is a G0 dwarf found at a distance of 61 pc. Figure \ref{fig:68988_summary} shows the \texttt{RVSearch} results for this star. This system has one well-constrained hot Jupiter \citep{Vogt02} and an outer companion candidate that was first characterized in \cite{Bryan16} as a partially resolved Keplerian orbit. With four more years of HIRES data, we can place tighter constraints on this outer candidate's orbit. A maximum likelihood fit gives an orbital period of 49.2 yr.  However, since we have only observed a partially resolved orbit so far, the orbit posterior is wide and asymmetric in period space. MCMC sampling produces $P = 61\mathrm{\substack{+28 \\ -20}}$ yr. 
The model parameters are \msini\ = $17.6\mathrm{\substack{+2.4 \\ -2.5}}$ \mjup, $a = 16.5\mathrm{\substack{+4.8 \\ -3.8}}$ AU, and $e = 0.53\mathrm{\substack{+0.13 \\ -0.09}}$. Table 1 reports all companion parameters.

\texttt{RVSearch} detects a third periodic signal, with $P$ = 1900 days, that has the same period and phase as the peak period in the S-value time series. This signal also has a low RV amplitude, ~6 m s$^{-1}$.  Therefore, we designated this signal as a false-positive corresponding to stellar activity. 

\clearpage

\subsection{HD 95735}

\begin{figure*}[ht!]
\begin{center}
\includegraphics[width=0.6\textwidth]{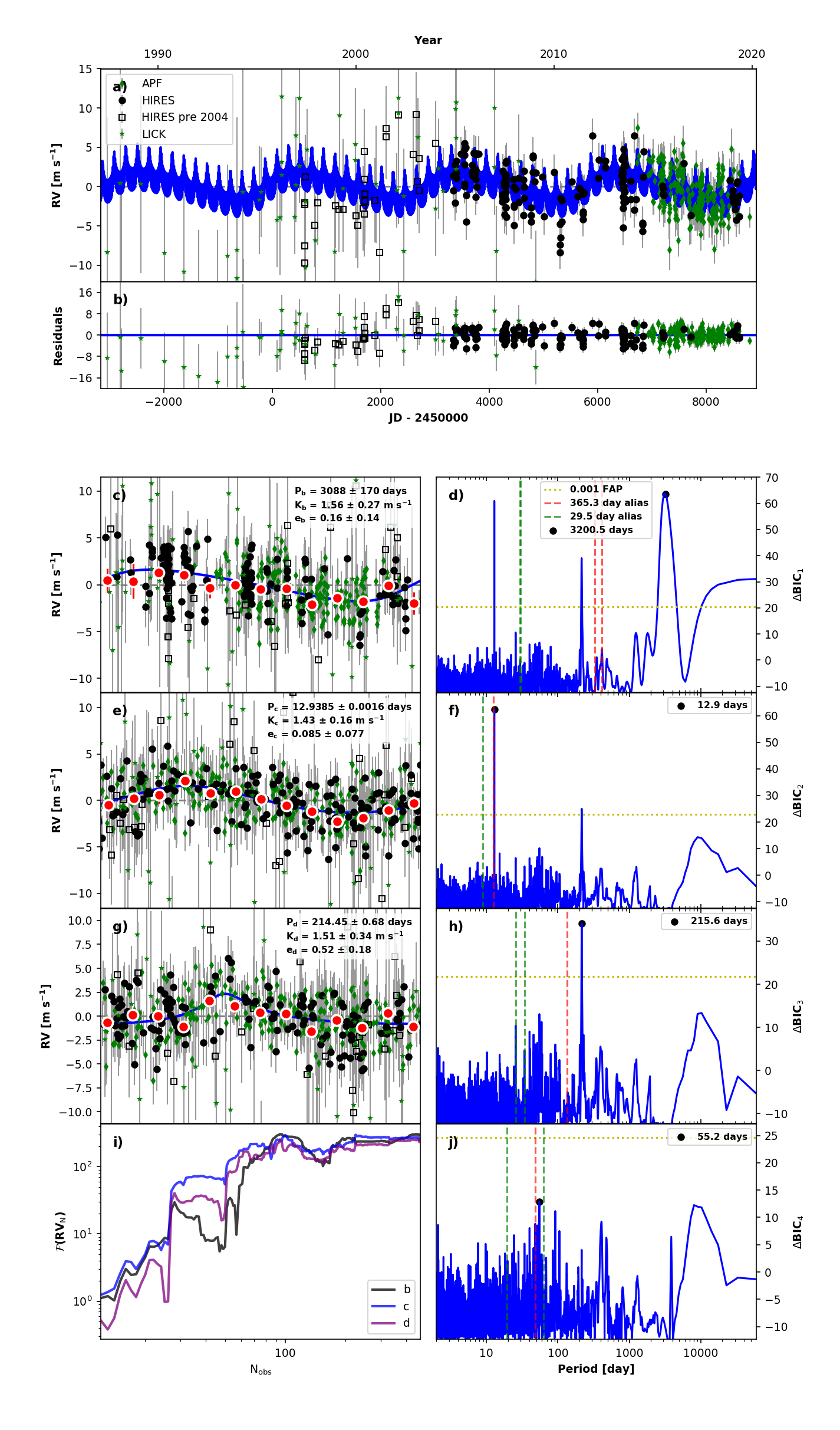}
\caption{\texttt{RVSearch} summary plot for HD 95735.  See Figure \ref{fig:hip109388_summary} for plot description.}
\label{fig:95735_summary}
\end{center}
\end{figure*}

HD 95735 (GJ 411) is an M2 dwarf found at a distance of 2.55 pc. Figure \ref{fig:95735_summary} shows the \texttt{RVSearch} results for this star. This system has one known short-period super-Earth, with \msini\ = 3.53 \mearth\ and an orbital period of 12.9 days. Our detection of this planet was driven by high-cadence APF data. This planet was first reported by \cite{Diaz19}, who also noted long-period power in their SOPHIE RV data, but they did not have a sufficiently long baseline or the activity metrics necessary to determine the origin of this power. With our HIRES post-upgrade and APF observations, we have an observational baseline of 14 yr, allowing us to confirm this long-period signal as a planet with \msini\ = $24.7 \pm 3.6$ \mearth\ and an orbital period $P = 8.46$ yr. Table 1 reports all planet parameters. Since GJ 411 is a cool M dwarf, the Lick-Hamilton and HIRES pre-upgrade data are not reliable, because those detectors are not sufficiently high-resolution to capture a cool M dwarf's dense spectral lines \citep{Fischer14}.

There is a long-period trend in the HIRES S-value time series, with significant power at and beyond 25 yr, but no significant power near the orbital period of the outer candidate. Therefore, we included this candidate in our catalog as a new planet candidate, to be verified and constrained with several more years of HIRES observations.

\texttt{RVSearch} also recovered a highly eccentric, 216 day signal, but this signal correlates with APF systematics. Therefore, we labeled it as a false positive. This systematic remained when we applied \texttt{RVSearch} only to the HIRES post-upgrade and APF data and left out the problematic pre-upgrade and Lick data.

\begin{figure*}[ht!]
\begin{center}
\includegraphics[width=0.99\textwidth]{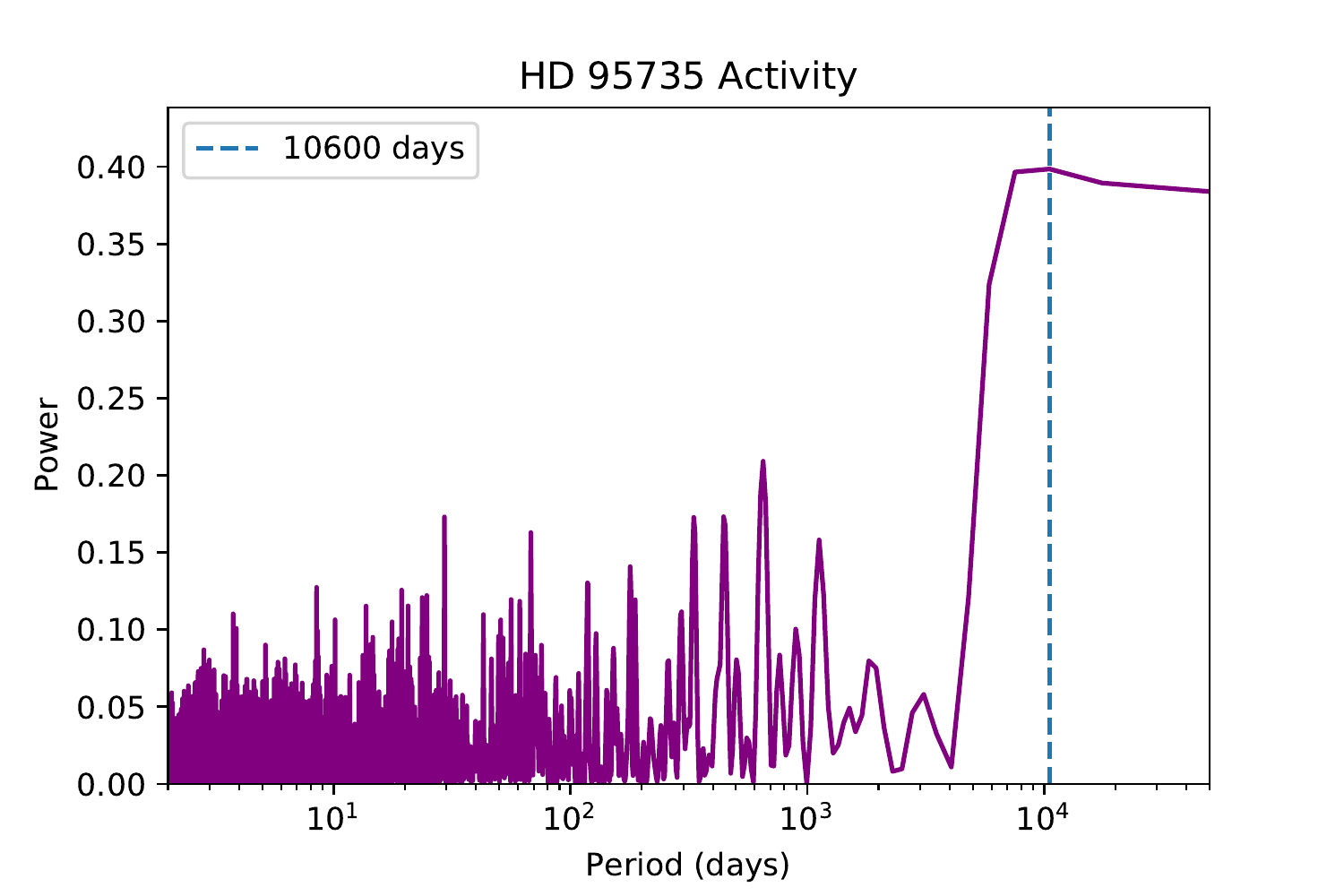}
\caption{Lomb-Scargle periodogram of HIRES S-values for HD 95735. There is evidence for an activity cycle longer than 10,000 days, but no significant power near the period of our 3,000-day planet candidate.}
\label{fig:95735_svals}
\end{center}
\end{figure*}


\clearpage

\subsection{HD 107148}

\begin{figure*}[ht!]
\begin{center}
\includegraphics[width=0.99\textwidth]{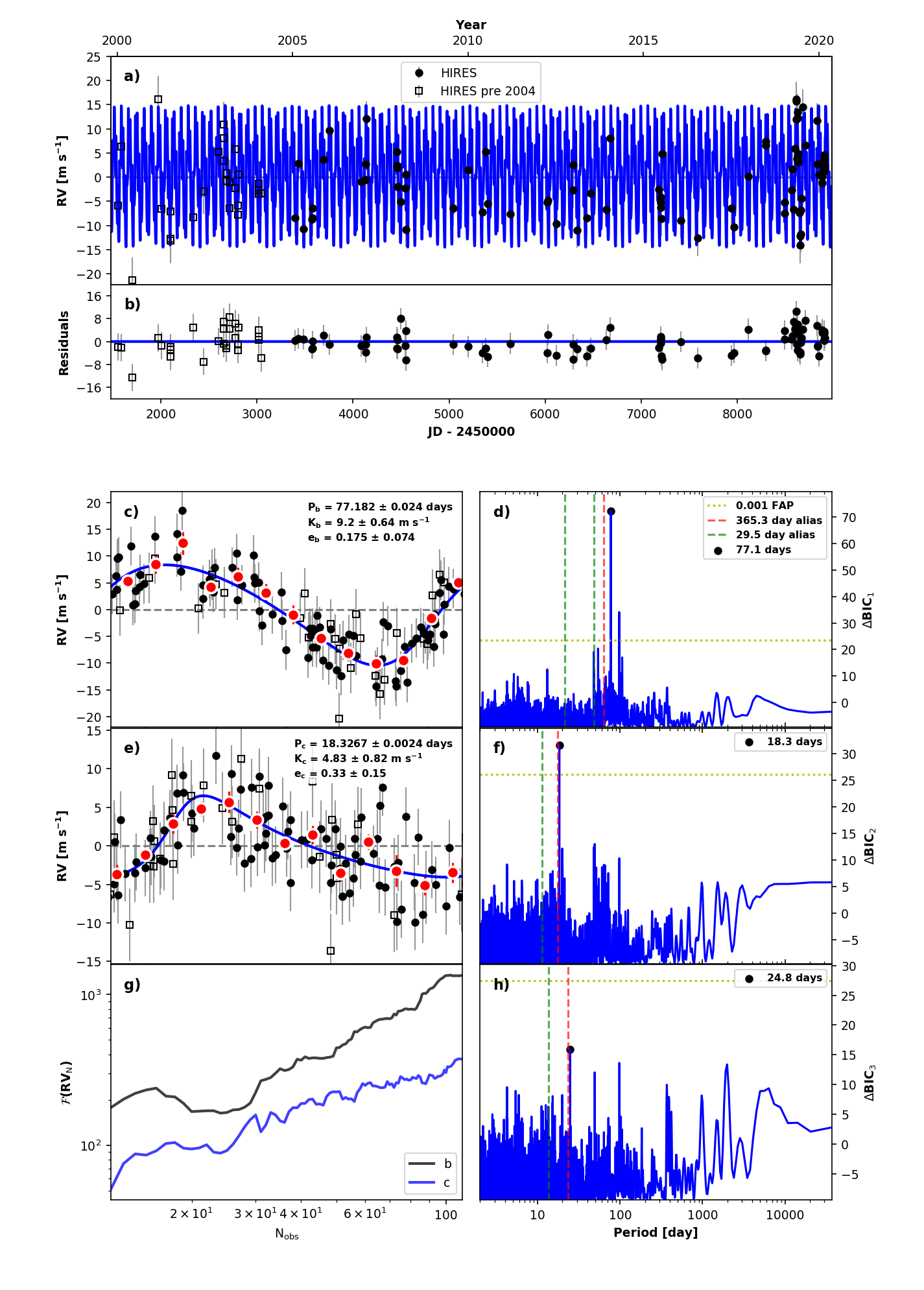}
\caption{\texttt{RVSearch} summary plot for HD 107148.  See Figure \ref{fig:hip109388_summary} for plot description.}
\label{fig:107148_summary}
\end{center}
\end{figure*}

HD 107148 is a G5 dwarf at a distance of 49.5 pc. Figure \ref{fig:107148_summary} shows the \texttt{RVSearch} results for this star. \cite{Butler06} reported a planet with a period of 44 days. They reported periodicity at 77 days, but determined that this was an alias of the 44 day signal. The 77 day signal is significantly stronger in our likelihood periodogram, as seen in Figure \ref{fig:107148_summary}, and better fits the data than a 44 day Keplerian by a significant \dBIC\. This constitutes strong evidence that the true period of this planet is 77 days. We report new orbital parameters for this planet in Table 3.

\begin{figure*}[ht!]
\begin{center}
\includegraphics[width=0.99\textwidth]{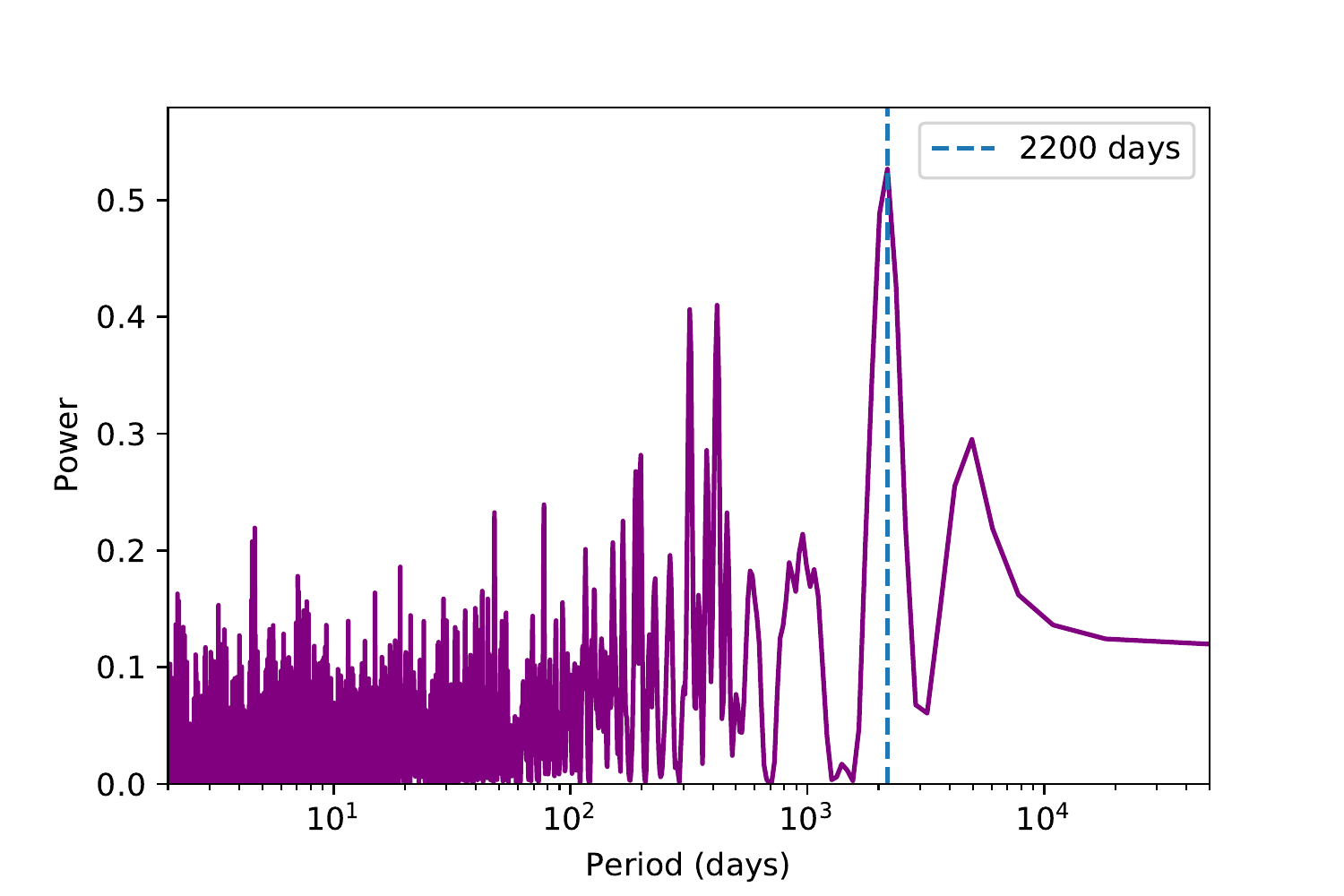}
\caption{Lomb-Scargle periodogram of HIRES S-values for HD 107148. Significant power at and beyond 4,300 days.}
\label{fig:107148_svals}
\end{center}
\end{figure*}

We also recovered a signal with a period of 18.3 days. There is significant periodicity in the S-value time series, a periodogram of which is shown in Figure \ref{fig:107148_svals}. However, it is concentrated around a period of 6 yr, and there is no significant power near 18.3 days. Furthermore, we find no correlation between the RVs and S-values. Thus, we report this signal as a confirmed planet, with \msini\ = $19.9 \pm 3.1$ \mearth\ and $a = 0.1406 \pm 0.0018$ AU. 

\clearpage

\subsection{HD 136925}

\begin{figure*}[ht!]
\begin{center}
\includegraphics[width=0.6\textwidth]{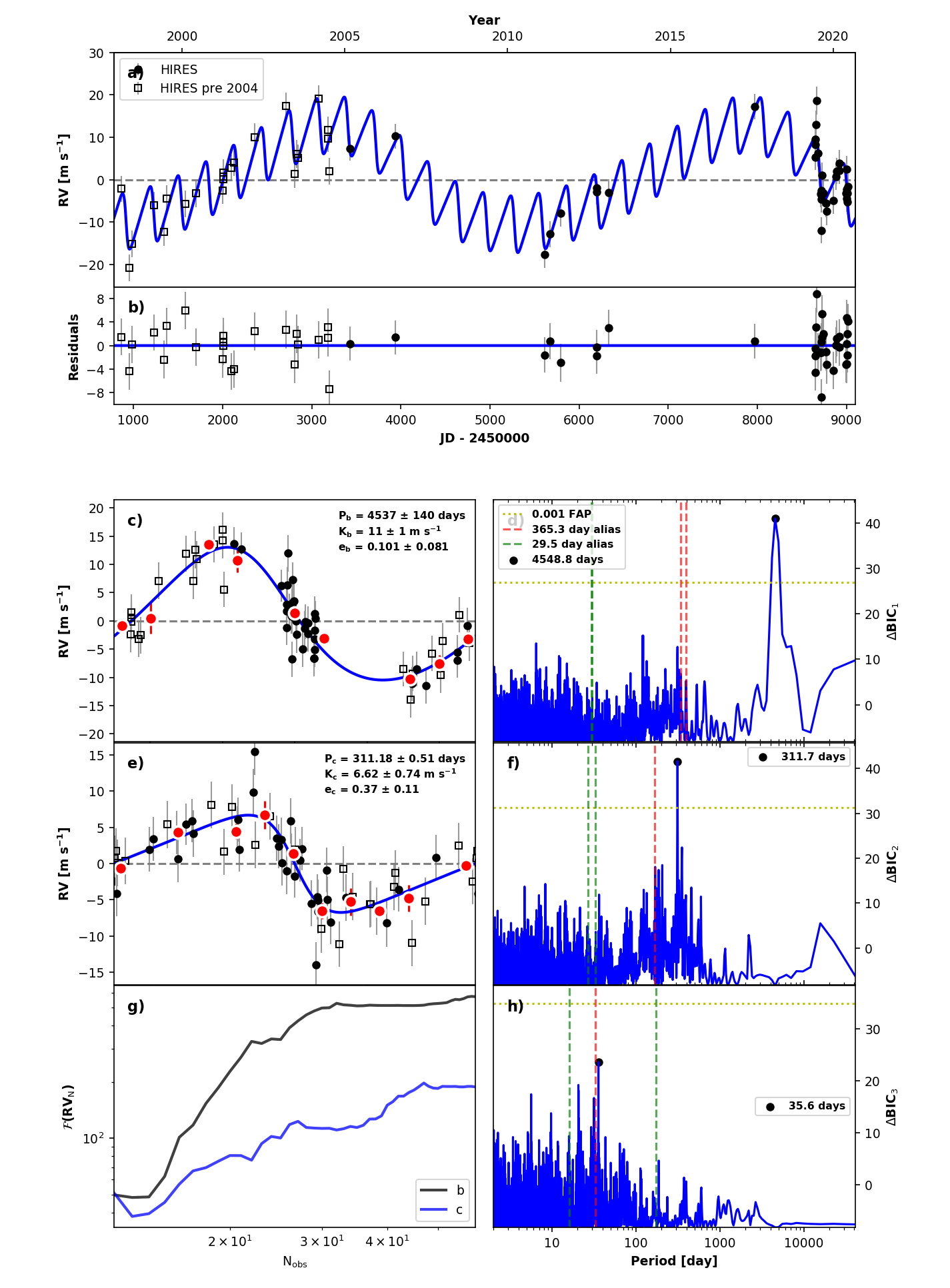}
\caption{\texttt{RVSearch} summary plot for HD 136925.  See Figure \ref{fig:hip109388_summary} for plot description.}
\label{fig:136925_summary}
\end{center}
\end{figure*}

HD 136925 is a G0 dwarf, found at a distance of 47.9 pc. \texttt{RVSearch} detected two periodic signals in this dataset, as seen in Figure \ref{fig:136925_summary}, at 311 days and 12.4 yr. This dataset is currently sparse, with two gaps of several years in the post-upgrade HIRES data, but there is clear long-period variation in the RVs. Keplerian modeling predicts \msini\ = 0.84 \mjup\ for the giant planet.

\begin{figure*}[ht!]
\begin{center}
\includegraphics[width=0.9\textwidth]{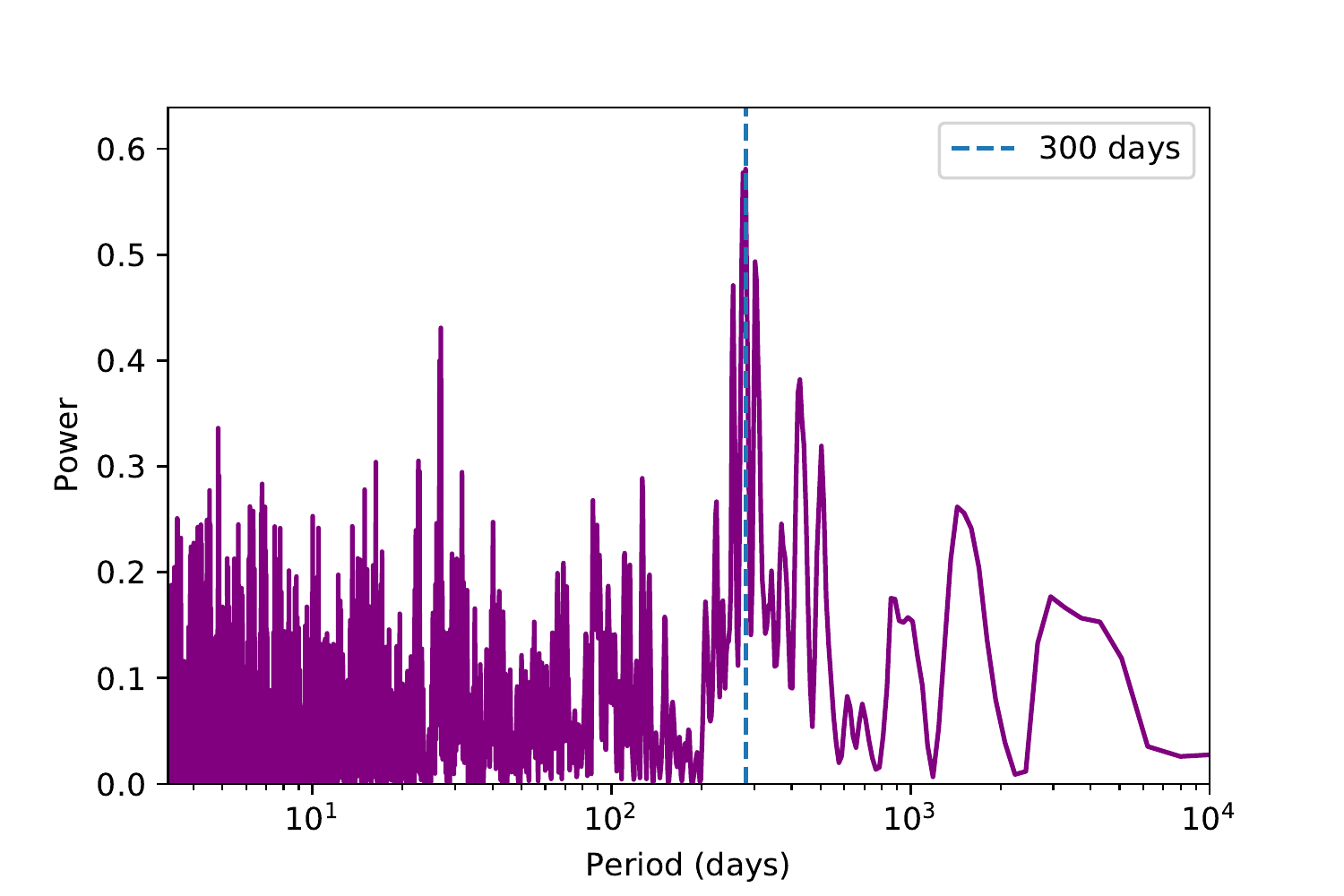}
\caption{RVSearch summary plot for HD 136925. Periodogram of HIRES S-values for HD 136925. Significant periodicity around 300 days, near the period of the inner signal.}
\label{fig:136925_activity_periodogram}
\end{center}
\end{figure*}

The S-value periodogram seen in Figure \ref{fig:136925_activity_periodogram} shows no significant power beyond 1000 days, suggesting that the long-period HD 136925 b is a real planet. There is broad power around 300 days, overlapping with the period of the inner signal. It is unclear whether this periodicity is caused by real stellar variability or is a product of sparse data. Table 1 reports current estimates of all planet parameters. We need more data in order to clarify our model, and determine whether the inner signal is caused by a planet or a product of stellar activity and sparse data. Therefore, we designated HD 136925 b as a planet, and the inner signal as a probable false positive, to be clarified with continued HIRES observing.

\clearpage

\subsection{HD 141004}

\begin{figure*}[ht!]
\begin{center}
\includegraphics[width=0.6\textwidth]{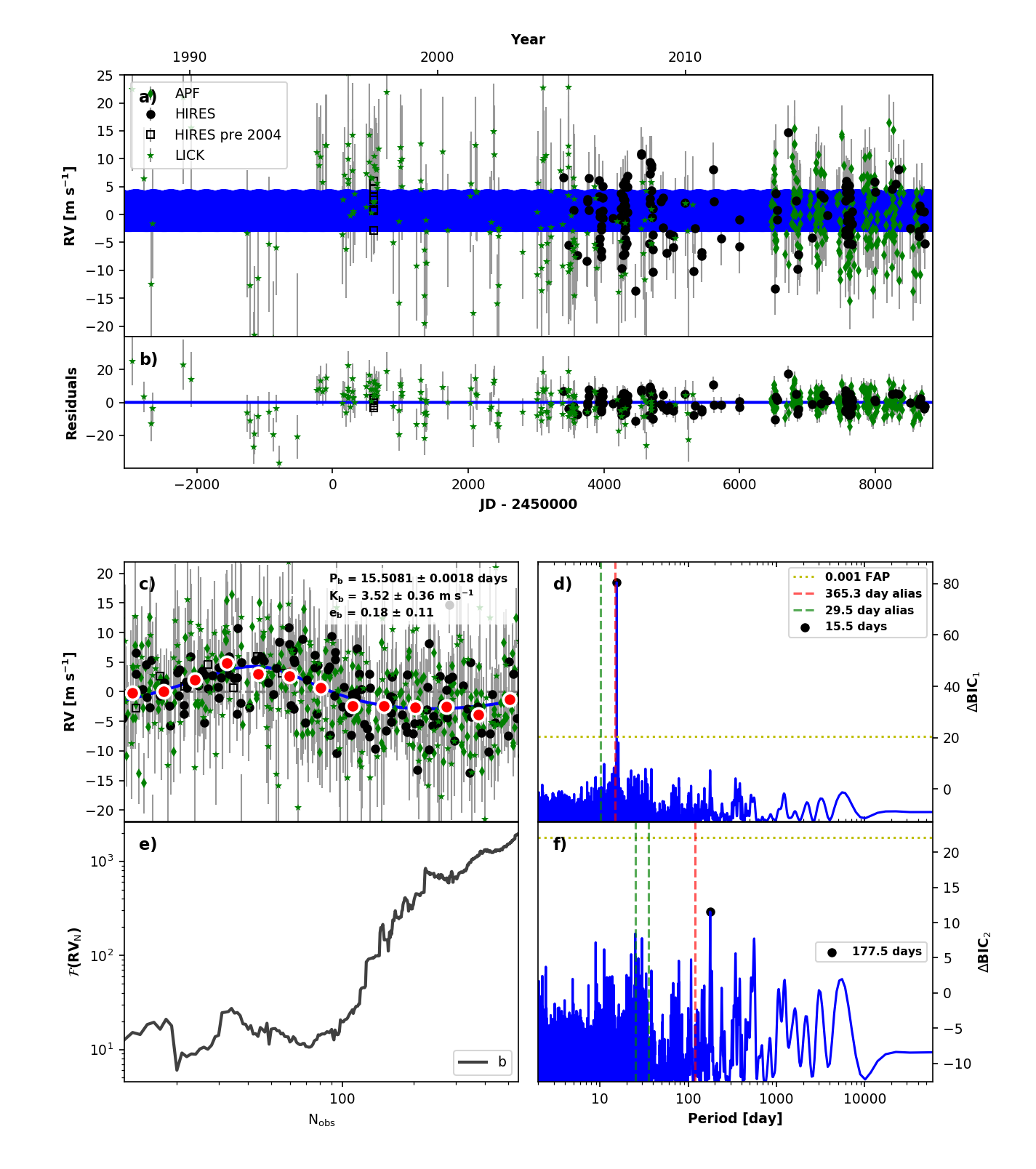}
\caption{\texttt{RVSearch} summary plot for HD 141004.   See Figure \ref{fig:hip109388_summary} for plot description.}
\label{fig:141004_summary}
\end{center}
\end{figure*}

HD 141004 is a G0 dwarf found at a distance of 11.8 pc. Figure \ref{fig:141004_summary} shows the RVSearch results for this star. Roy et al. (2021, in preparation) discovered a sub-Neptune at an orbital period of 15.5 days, with \msini\ = $13.9\ \pm 1.5$ \mearth, and will report on the analysis of this system in greater detail. Table 1 reports current estimates of all planet parameters.

\clearpage

\subsection{HD 145675}

\begin{figure*}[ht!]
\begin{center}
\includegraphics[width=0.6\textwidth]{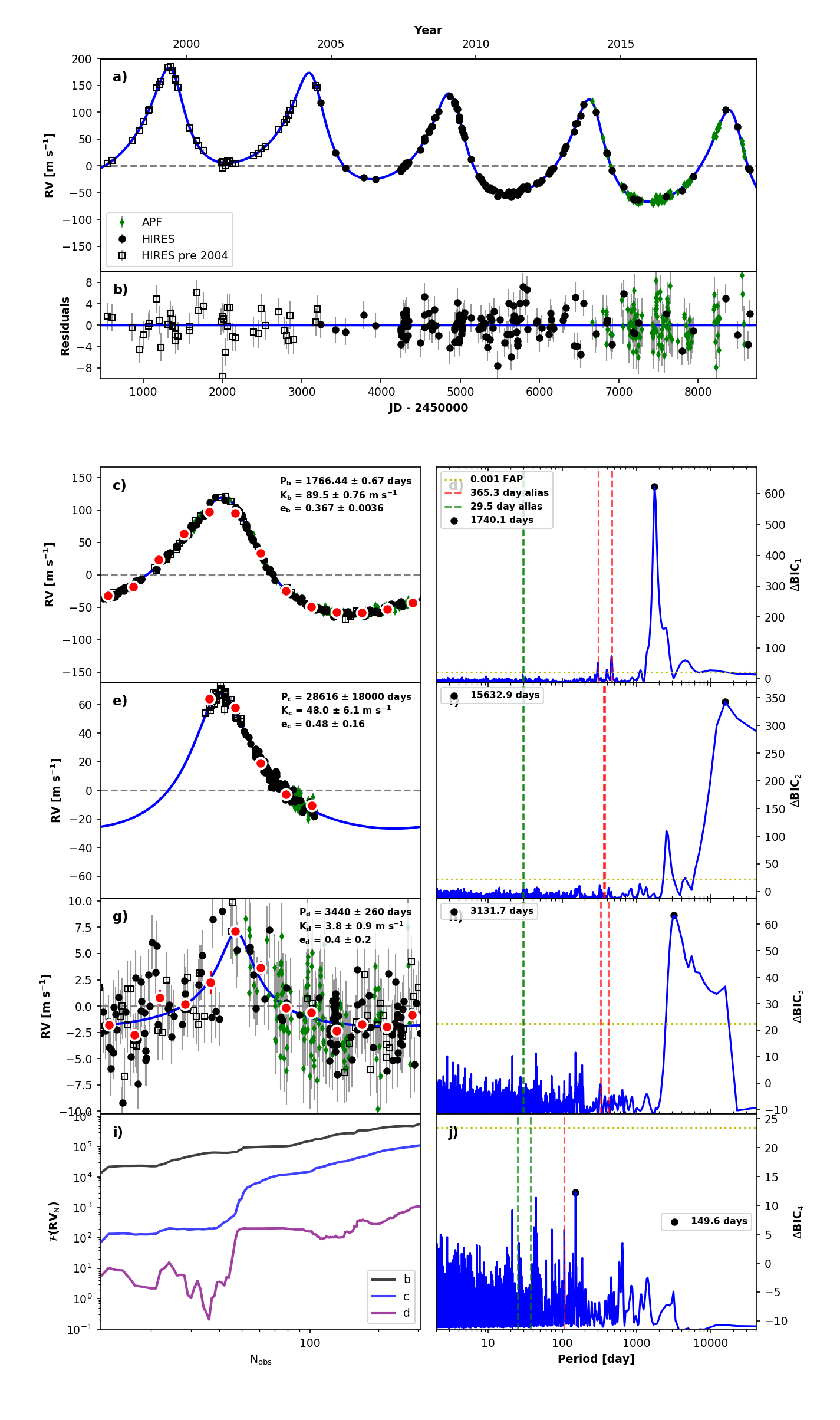} \\
\caption{\texttt{RVSearch} summary plot for HD 145675.  See Figure \ref{fig:hip109388_summary} for plot description.}
\label{fig:145675_summary}
\end{center}
\end{figure*}

HD 145675 (14 Her) is a K0 dwarf found at a distance of 17.9 pc. Figure \ref{fig:145675_summary} shows the \texttt{RVSearch} results for this star. This system has one known cold gas giant, with \msini = 5.10 \mjup\ and an orbital period of 4.84 yr, which was first reported in \cite{Butler03}. \cite{Wittenmyer07b} conducted further analysis with a longer observational baseline of twelve years, and noted a long-period trend. \cite{Wright07} used additional RV curvature constraints to show that this trend must correspond to a companion with $P > 12$ yr and \msini\ $>$ 5 \mjup. The observational baseline has since increased from 12 yr to 22, and regular observations with HIRES and APF allow us to place further constraints on this long-period companion. We find \msini\ = 5.8$\mathrm{\substack{+1.4 \\ -1.0}}$ \mjup, $P=68\mathrm{\substack{+64 \\ -25}}$ yr, semi-major axis $a = 16.4\mathrm{\substack{+9.3 \\ -4.3}}$ AU, and eccentricity $e=0.45\mathrm{\substack{+0.17 \\ -0.15}}$. Table 1 reports all planet parameters.

\begin{figure*}[ht!]
\begin{center}
\includegraphics[width=0.99\textwidth]{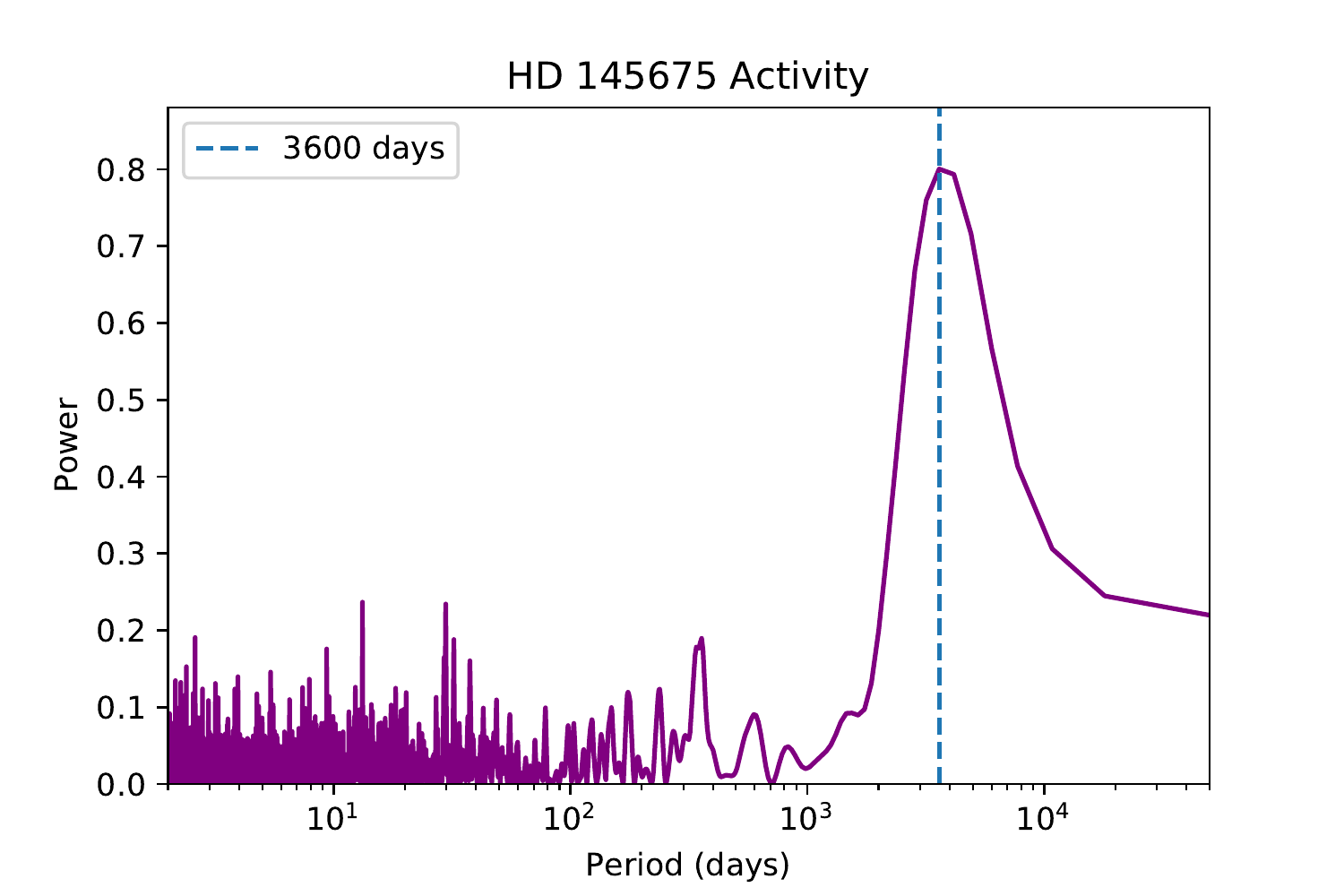}
\caption{Lomb-Scargle periodogram of HIRES S-values for HD 145675 showing significant power at 3,600 days.}
\label{fig:145675_svals}
\end{center}
\end{figure*}

\begin{figure*}[ht!]
\begin{center}
\includegraphics[width=0.49\textwidth]{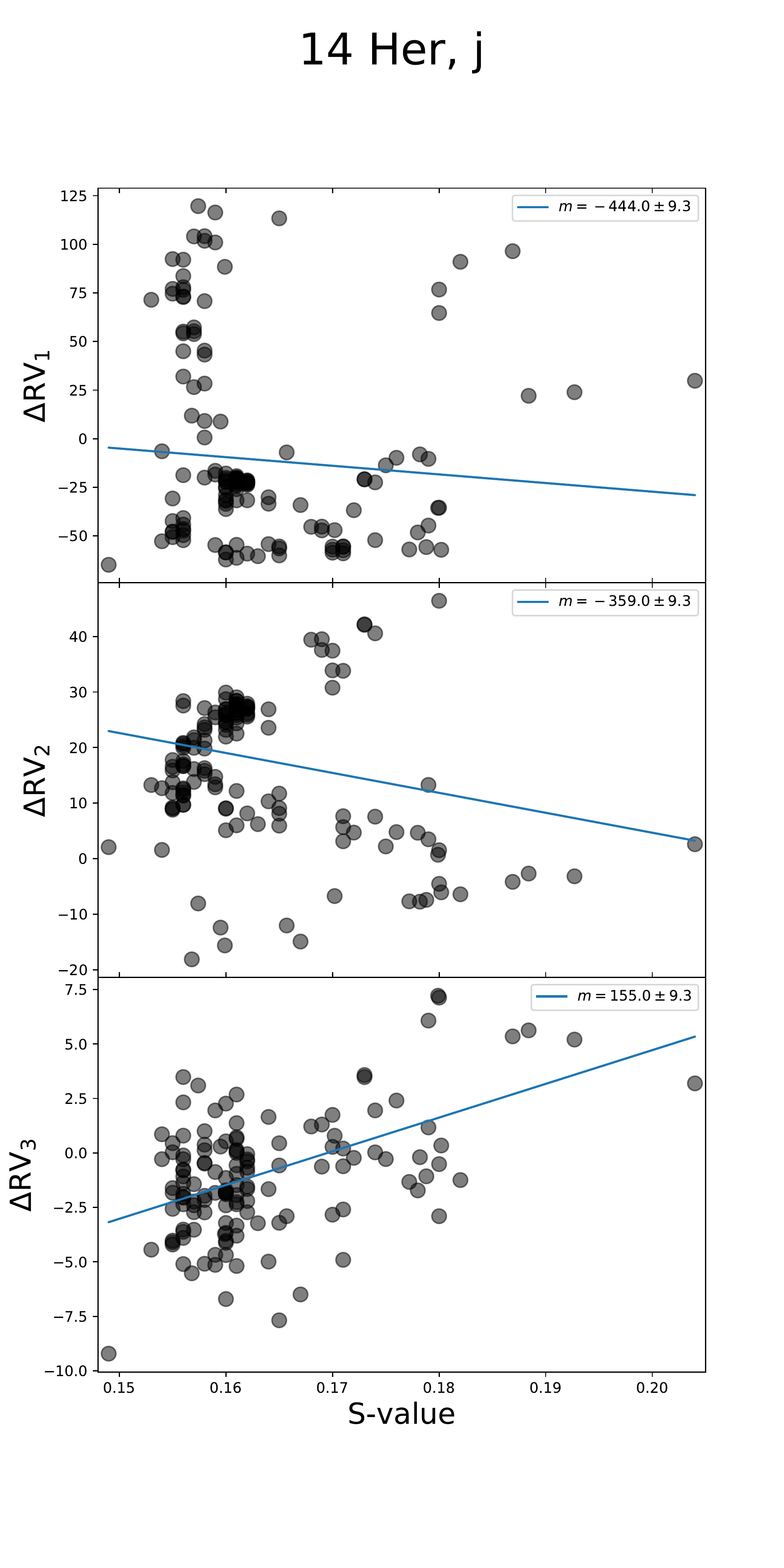}
\includegraphics[width=0.49\textwidth]{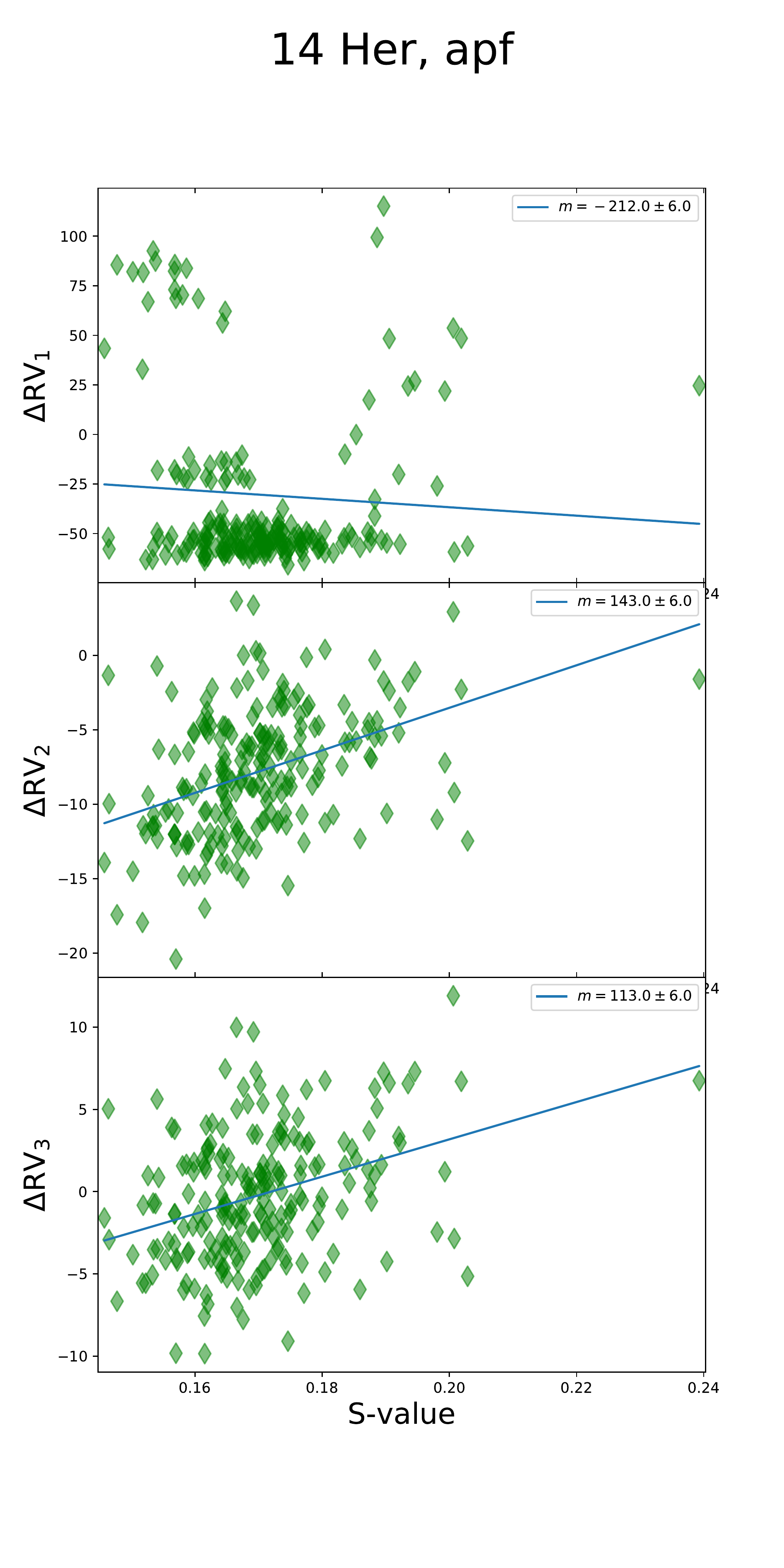} \\
\caption{Activity vetting plots for HD 145675. For all panels, the horizontal axis shows the S-value activity metric of each observation, while the vertical axis shows corresponding RV residuals for each individual Keplerian orbit. The left-hand panels show HIRES post-upgrade observations, while the right-hand panels show APF observations. Each row shows RVs with the model residuals of one Keplerian model, with the other Keplerian models subtracted from the data. The blue lines show linear correlations between these residuals and the corresponding S-values. In the HIRES and APF data, we measured $>3\sigma$ correlations for the third Keplerian signal. The APF and HIRES linear correlations are within $3\sigma$ of each other, implying that this signal is caused by stellar activity. We find correlations between the residuals and S-values for the second signal as well, but they are significantly different for HIRES and APF. Since the period of this signal is much greater than the APF baseline of this star, we discount this second correlation as caused by the limited baseline of the data with respect to the signal.}
\label{fig:145675_activity}
\end{center}
\end{figure*}

Figure \ref{fig:145675_svals} shows a Lomb--Scargle periodogram of the HIRES S-value time series. There is strong periodicity in the HIRES S-value time series, peaking around 10 yr, but no significant power near the supposed orbital period of the long-period candidate. These S-values strongly correlate with a third Keplerian signal picked up by our search, also with a period of 10 yr, as seen in the Figure \ref{fig:145675_activity}, therefore we designate this signal as stellar activity.

There is a potential complication owed to a stellar binary candidate. \cite{Roberts11} conducted a direct-imaging survey of known exoplanet hosts and reported a candidate stellar companion to 14 Her, with a differential magnitude of $10.9 \pm 1.0$, an angular separation of 4.3", and a minimum orbital separation of 78 AU. This is a single-epoch detection, and therefore could be only a visual binary. Additionally, \cite{Rodigas11} conducted a deep direct imaging study of 14 Her, to constrain the mass and orbital parameters of 14 Her c, which, at the time, presented only as a parabolic trend in RV data. They used the Clio-2 photometer on the MMT, which has a 9" x 30" field of view; the authors only looked at imaging data within 2", to filter out background stars. Although this deep imaging study did not mention any stellar companion, the candidate reported by \cite{Roberts11} falls outside of their considered imaging data, which corresponds to a minimum separation of 112.8 AU. \cite{Wittrock17} also found a null binary detection, using the Differential Speckle Survey Instrument (DSSI) at the Gemini-North Observatory. A 6 Jupiter mass object would not have been detected by the above surveys, as they were designed only to rule out stellar companions and therefore used shorter imaging exposures that would miss planetary-mass companions. 

Additionally, we used Gaia DR2 to search for bound stellar companions within 10", and found no such companions. We conclude that 14 Her does not have a bound stellar companion. Therefore, we designated 14 Her c as an eccentric, long-period planet. We will continue to monitor this star with Keck/HIRES and APF, to further constrain the orbit of this planet.

\clearpage

\subsection{HD 156668}

\begin{figure*}[ht!]
\begin{center}
\includegraphics[width=0.6\textwidth]{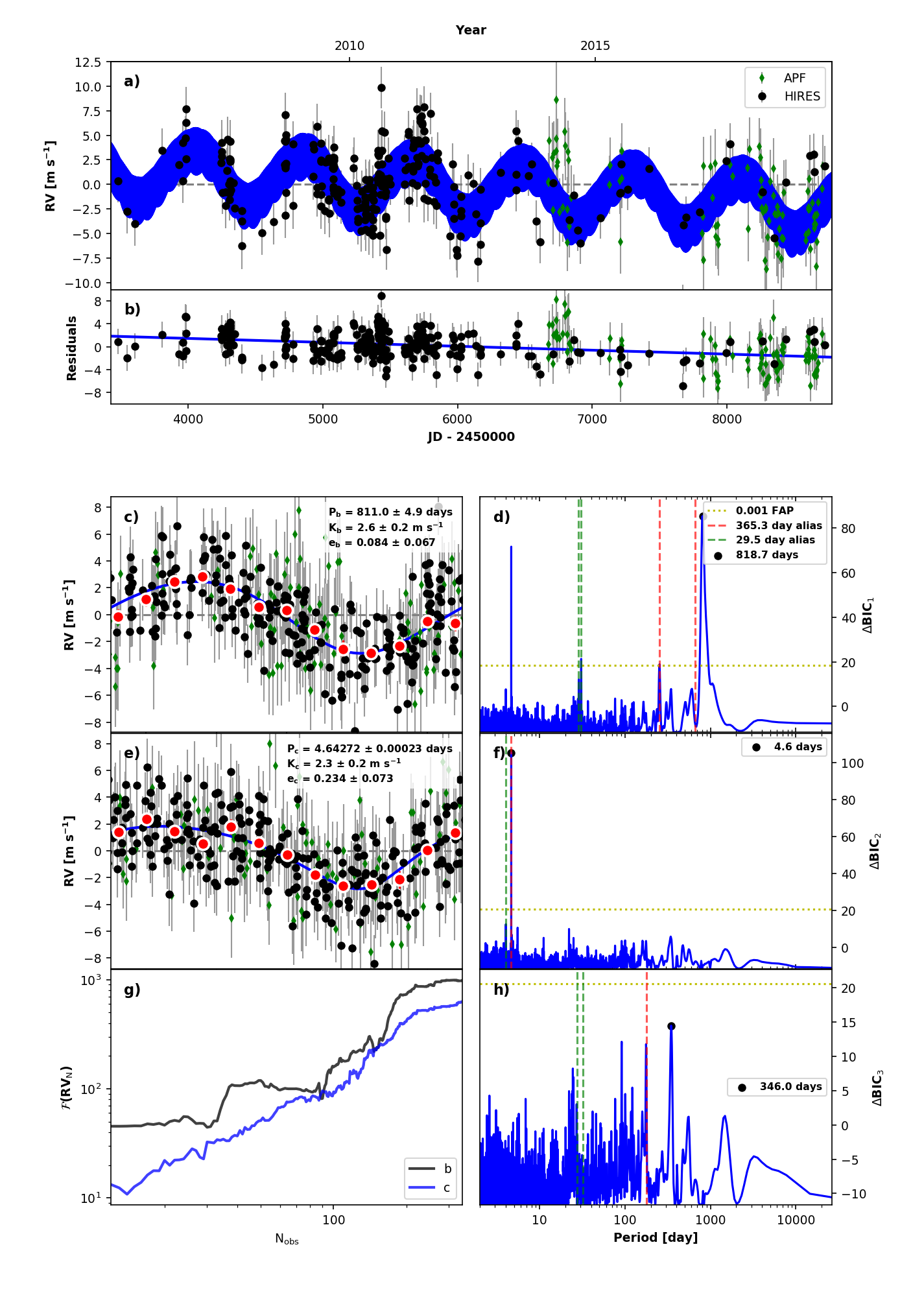}
\caption{\texttt{RVSearch} summary plot for HD 156668.  See Figure \ref{fig:hip109388_summary} for plot description.}
\label{fig:156668_summary}
\end{center}
\end{figure*}

HD 156668 is a K3 dwarf found at a distance of 24.4 pc. Figure \ref{fig:156668_summary} shows the \texttt{RVSearch} results for this star. This system has one known short-period super-Earth, with \msini = 4.15 \mearth and an orbital period of 4.64 days. This planet was first reported by \cite{Howard11}, who also noted a long-period ($P\approx 2.3$ yr) signal with insufficient RV observations or additional data for confirmation as a planet. The observational baseline has since increased from five years to fourteen, allowing us to confirm this long-period signal as a planet with \msini\ = 0.167 \mjup\ and an orbital period $P = 2.22$ yr. 

There is a strong periodicity in the HIRES S-value time series, peaking around 10 yr, but no significant power near the orbital period of the long-period candidate. If we do not model this activity, a one-year alias signal appears in the periodogram search (Fig.\ \ref{fig:156668_summary}). The data do not sufficiently constrain a Keplerian fit with a 10 yr period, but we find that a linear trend models the activity well enough to remove the one-year alias from the search. We opt to include this linear trend, which we treat as a nuisance parameter.


\clearpage

\subsection{HD 164922}

\begin{figure*}[ht!]
\begin{center}
\includegraphics[width=0.6\textwidth]{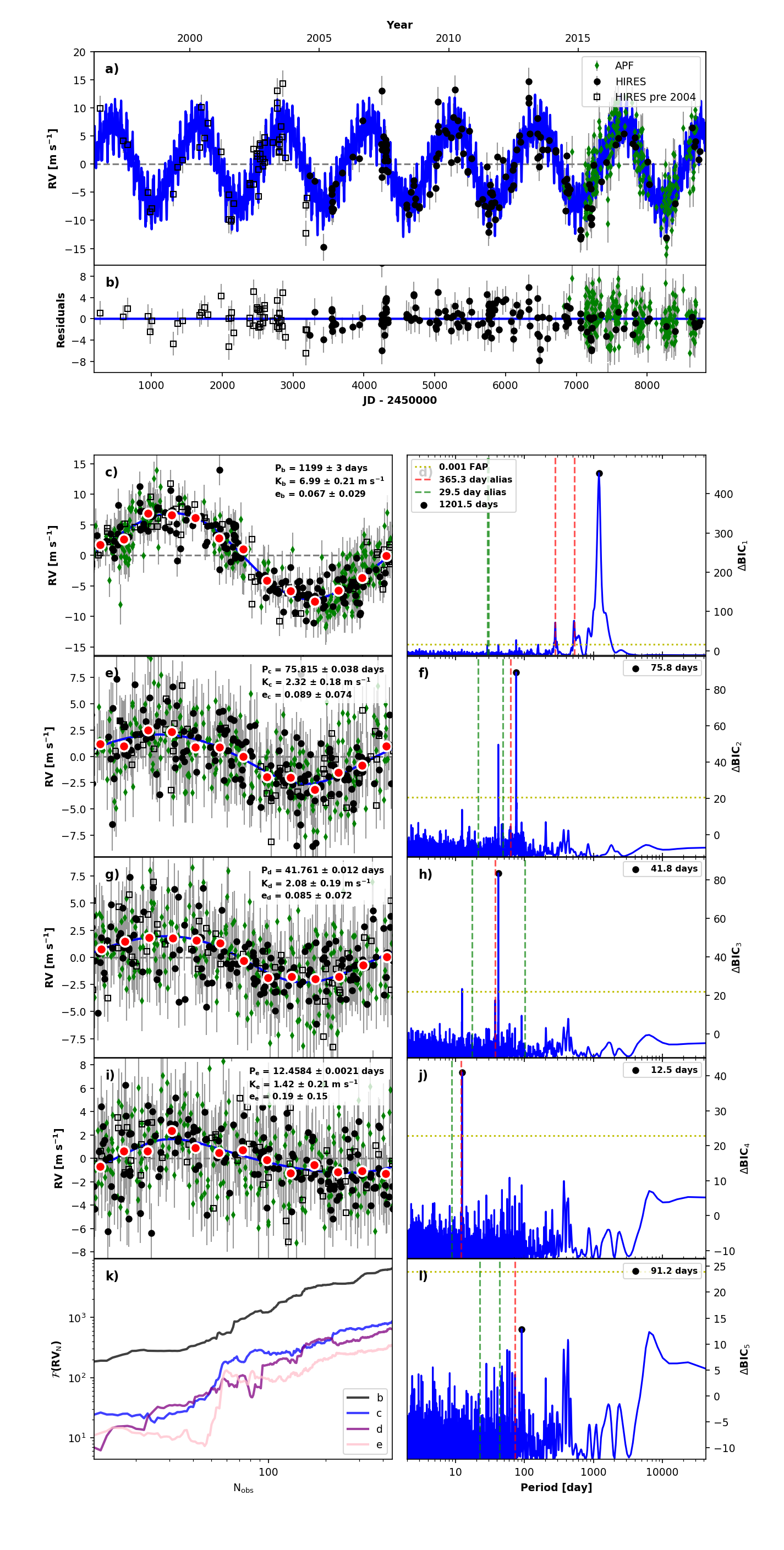}
\caption{\texttt{RVSearch} summary plot for HD 164922.  See Figure \ref{fig:hip109388_summary} for plot description.}
\label{fig:164922_summary}
\end{center}
\end{figure*}

HD 164922 is a G9 V dwarf located at a distance of 22.1 pc. Figure \ref{fig:164922_summary} shows the \texttt{RVSearch} results for this star. It hosts two known planets: a 0.3 \mjup\ planet with an orbital period of 1207 days \citep{Butler06}, and a super-Earth with \msini\ = 14.3 \mearth\ and an orbital period of 75.8 days. This super-Earth was reported by \cite{Fulton16}, who also reported residual power around 41.7 days but did not find it significant enough to merit candidate status. With approximately two more years of HIRES and APF data, we identified the 41.7 day signal as a strong planet candidate and confirmed the 12.5 day planet reported in \cite{Benatti20}. Both planets are of sub-Neptune mass and have eccentricity posteriors that are consistent with circular orbits. The 41.7 day planet has \msini\ = $10.7 \pm 1.0$ \mearth\, and a semi-major axis $a = 0.2294 \pm 0.0031$ AU. The 12.5 day planet has \msini\ = $4.63 \pm 0.70$ \mearth\, and a semi-major axis $a = 0.1024 \pm 0.0014$ AU. Table 1 reports all planet parameters.

\begin{figure*}[ht!]
\begin{center}
\includegraphics[width=0.8\textwidth]{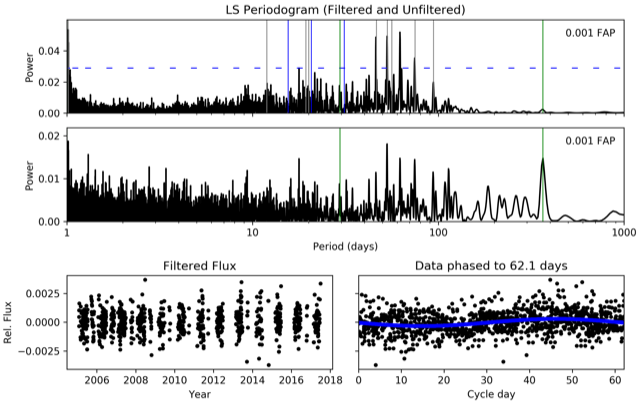}
\caption{Visualization of APT photometry analysis for HD 164922. The top panel shows a Lomb-Scargle periodogram of the photometry, with a moving-average filter to reduce alias issues. The middle panel shows an unfiltered periodogram.}
\label{fig:164922_APT}
\end{center}
\end{figure*}

\begin{figure*}[ht!]
\begin{center}
\includegraphics[width=0.8\textwidth]{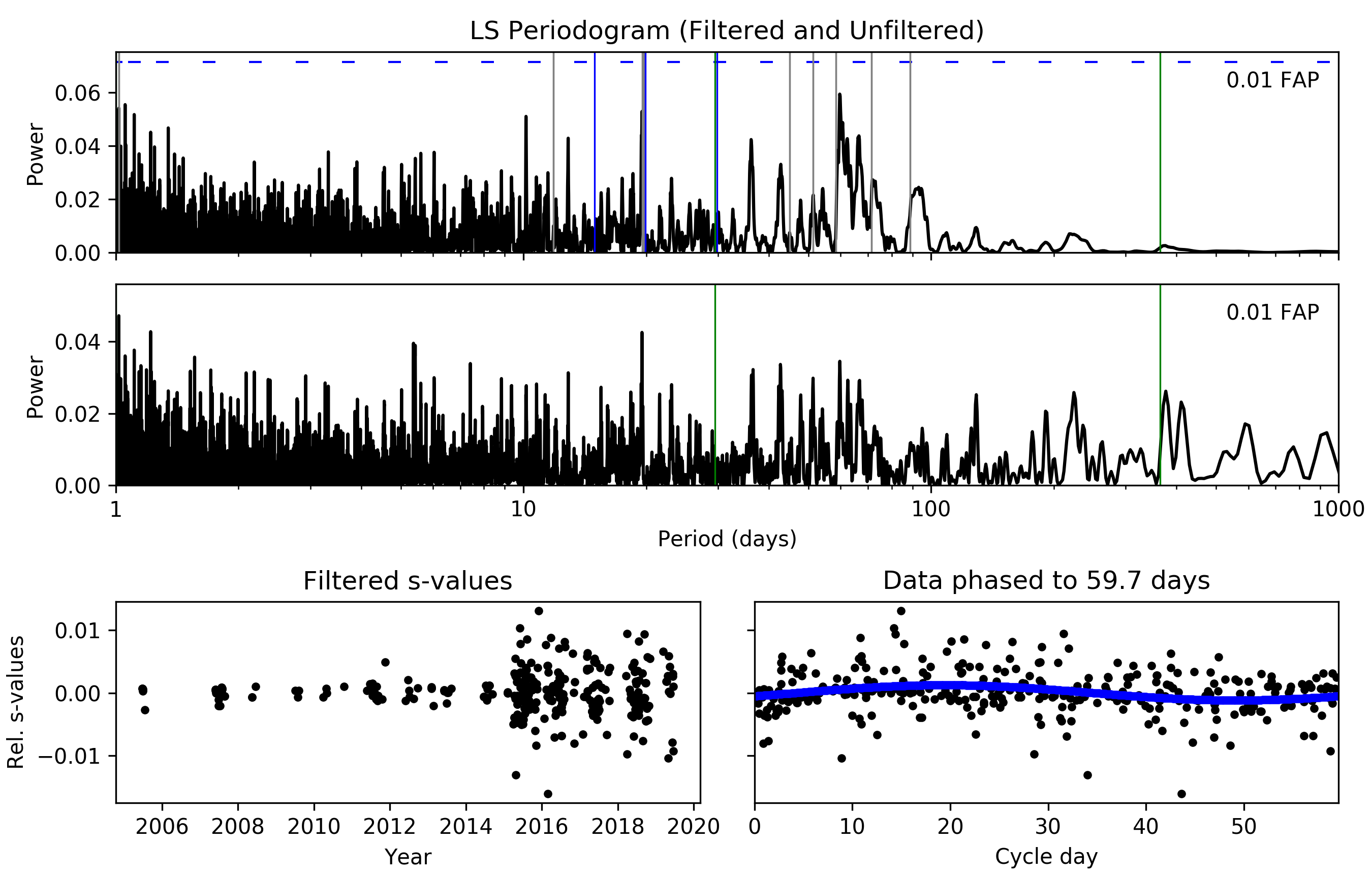}
\caption{Visualization of HIRES S-value analysis for HD 164922. The top panel shows a Lomb-Scargle periodogram of the S-values, with a moving-average filter to reduce alias issues. The middle panel shows an unfiltered periodogram.}
\label{fig:164922_sval}
\end{center}
\end{figure*}

To validate these candidates, we searched for periodicity in both S-value activity metrics and APT photometry. We found no evidence for stellar rotation in S-values, but estimated a stellar rotation period of 62.1 days from our APT photometry. Figure \ref{fig:164922_APT} shows periodograms and a phase-folded curve from this APT analysis, and Figure \ref{fig:164922_sval} shows equivalent analysis for HIRES S-values. The 1 yr alias of 62.1 days is 75.8 days, but the 75.8 day planet detection is high-amplitude and clean, without an additional peak near 62 days in any of the \texttt{RVSearch} periodograms. Therefore, within the limits of our activity metrics and vetting process, we ruled out stellar rotation as a cause of the 41.7 day signal. 

\cite{Benatti20} used multiple HARPS-N spectral activity indicators to estimate a stellar rotation period of 41.6 days, and they note that this rotation period is to be expected from empirical activity-rotation relationships. Therefore, they determined that the strong 42 day signal present in their HARPS RVs is caused by rotation. However, we find no evidence of significant 42 day periodicity in our analysis of spectral activity indicators or APT photometry, as seen in Figures \ref{fig:164922_APT} and \ref{fig:164922_sval}, and both datasets reflect significant periodicity near 60 days. Since our RV detection of this planet candidate is clean and does not conflict with our activity analysis, we chose to include this signal in our catalog as a planet candidate, to be confirmed or refuted by independent analysis.

\clearpage

\subsection{HD 168009}

\begin{figure*}[ht!]
\begin{center}
\includegraphics[width=0.6\textwidth]{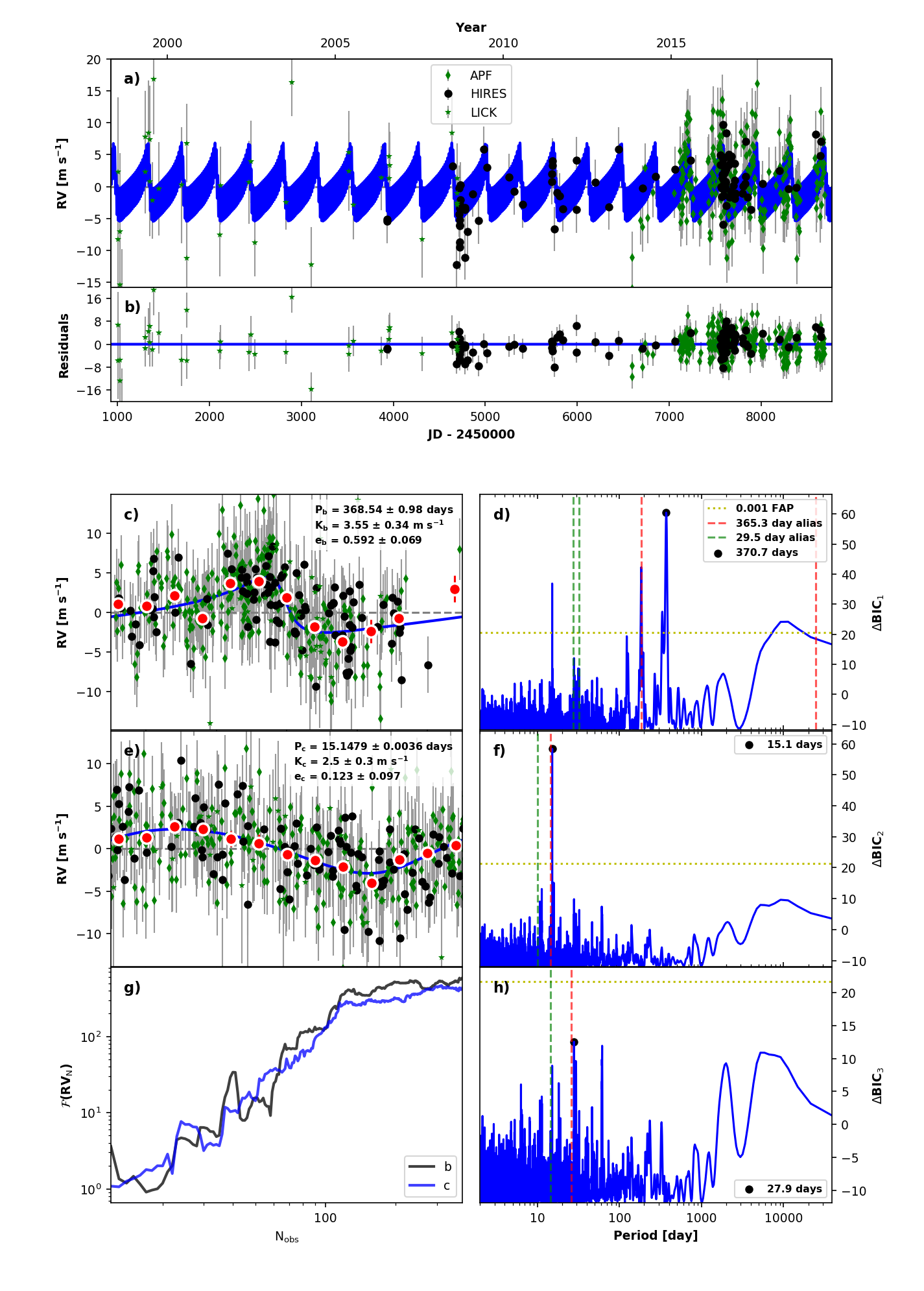}
\caption{\texttt{RVSearch} summary plot for HD 168009.  See Figure \ref{fig:hip109388_summary} for plot description.}
\label{fig:168009_summary}
\end{center}
\end{figure*}

HD 168009 is a G1 dwarf found at a distance of 23.3 pc. Figure \ref{fig:168009_summary} shows the RVSearch results for this star. Roy et al. (2021, in preparation) discovered a super-Earth candidate at an orbital period of 15.5 days, with \msini\ = $10.3\ \pm 1.1$ \mearth, and will report on the analysis of this candidate in greater detail. Table 1 reports current estimates of all planet parameters.

\texttt{RVSearch} also recovered a highly eccentric 1 yr signal, but this signal correlates with APF systematics. Therefore, we labeled it as a false positive. Roy et al. (2021, in preparation) will model these systematics in greater detail.

\clearpage

\subsection{HD 213472}

\begin{figure*}[ht!]
\begin{center}
\includegraphics[width=0.6\textwidth]{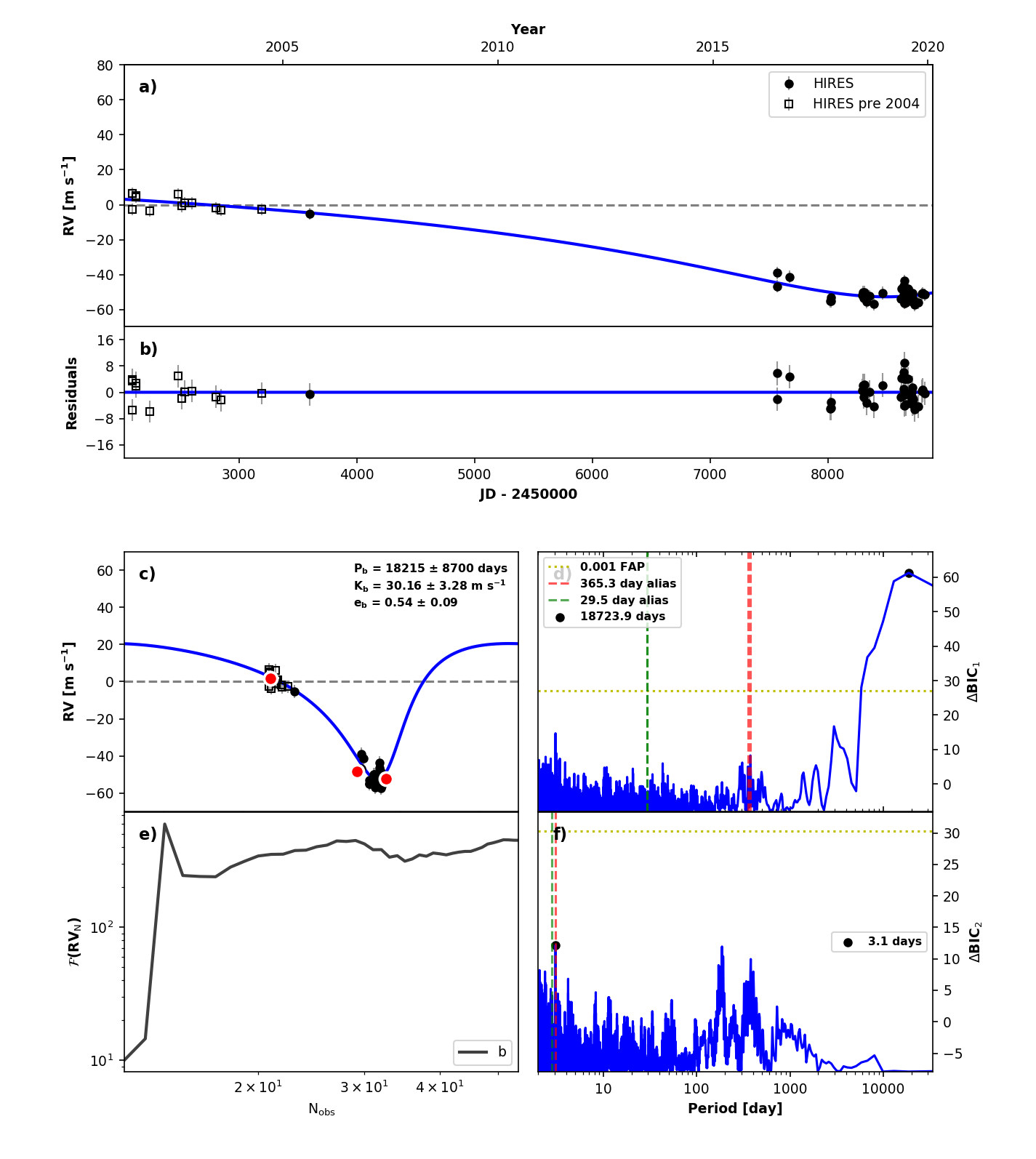}
\caption{\texttt{RVSearch} summary plot for HD 213472.  See Figure \ref{fig:hip109388_summary} for plot description.}
\label{fig:213472_summary}
\end{center}
\end{figure*}

\begin{figure*}[ht!]
\begin{center}
\includegraphics[width=\textwidth]{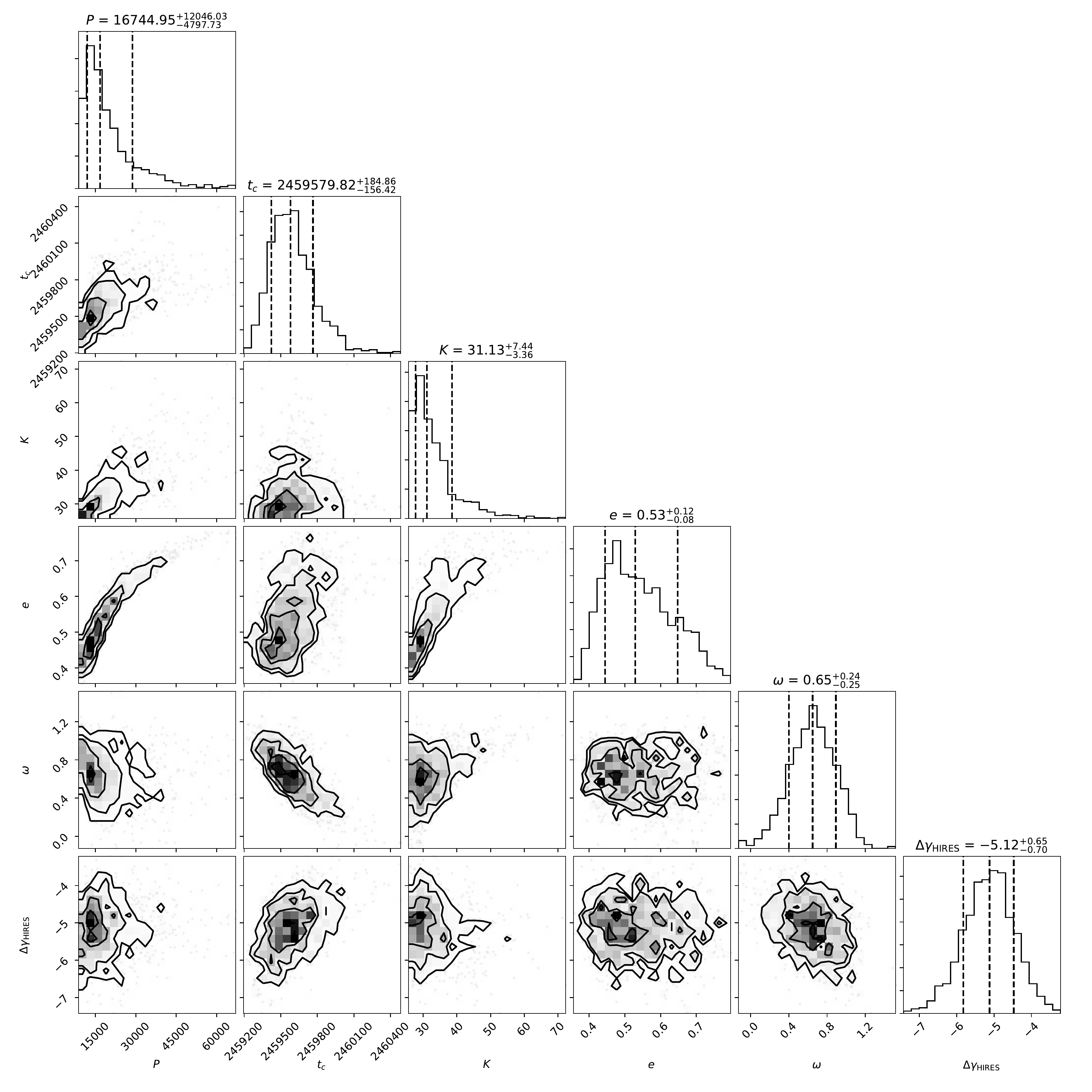}
\caption{Rejection sampling posterior for HD 213472 b orbital parameters. $\Delta \gamma$ is the relative linear offset between different instrumental datasets, in this case pre-upgrade and post-upgrade HIRES.}
\label{fig:213472_corner}
\end{center}
\end{figure*}

HD 213472 is a G5 dwarf located at a distance of 64.6 pc. Figure \ref{fig:213472_summary} shows the \texttt{RVSearch} results for this star. There is an approximately eleven-year gap in RV observations of this star. The first post-upgrade HIRES observation was measured in 2005, shortly after the last pre-upgrade observation, and the second post-upgrade observation was measured in 2016. The 40 \ms\ difference between these two observations prompted the CPS team to begin observing HD 213472 regularly. Together with observations since 2016, and the thirteen pre-upgrade HIRES measurements, the data are consistent with a long-period, eccentric, planetary companion. Our periodogram search detects such a long-period signal. Due to the sparseness of the data, traditional MCMC methods fail to return a well-sampled model posterior. We used the rejection sampling algorithm TheJoker \citep{Price-Whelan17} to estimate the posterior, and found it to be unimodal. This model is consistent with a very long-period gas giant, with \msini\ = $3.48\mathrm{\substack{+1.10 \\ -0.59}}$ \mjup\, orbital period $P = 46\mathrm{\substack{+33 \\ -13}}$ yr, semi-major axis $a = 13.0\mathrm{\substack{+5.7 \\ -2.6}}$ AU, and eccentricity of $e = 0.53\mathrm{\substack{+0.12 \\ -0.09}}$. Table 1 reports all planet parameters. Figure \ref{fig:213472_corner} shows the orbital parameter posteriors generated by TheJoker.

To investigate the possibility of a stellar or substellar companion, we compared this Keplerian model to a simple linear trend by computing the \dBIC\ between the two max-likelihood models. The Keplerian model is significantly preferred with \dBIC\ = 23.7. Additionally, we used Gaia to search for bound companions within 10'', and found no such companions. Therefore, we inferred that HD 213472 b is either a planet or low-mass substellar companion, and not a wide-orbit stellar companion. 

\begin{figure*}[ht!]
\begin{center}
\includegraphics[width=\textwidth]{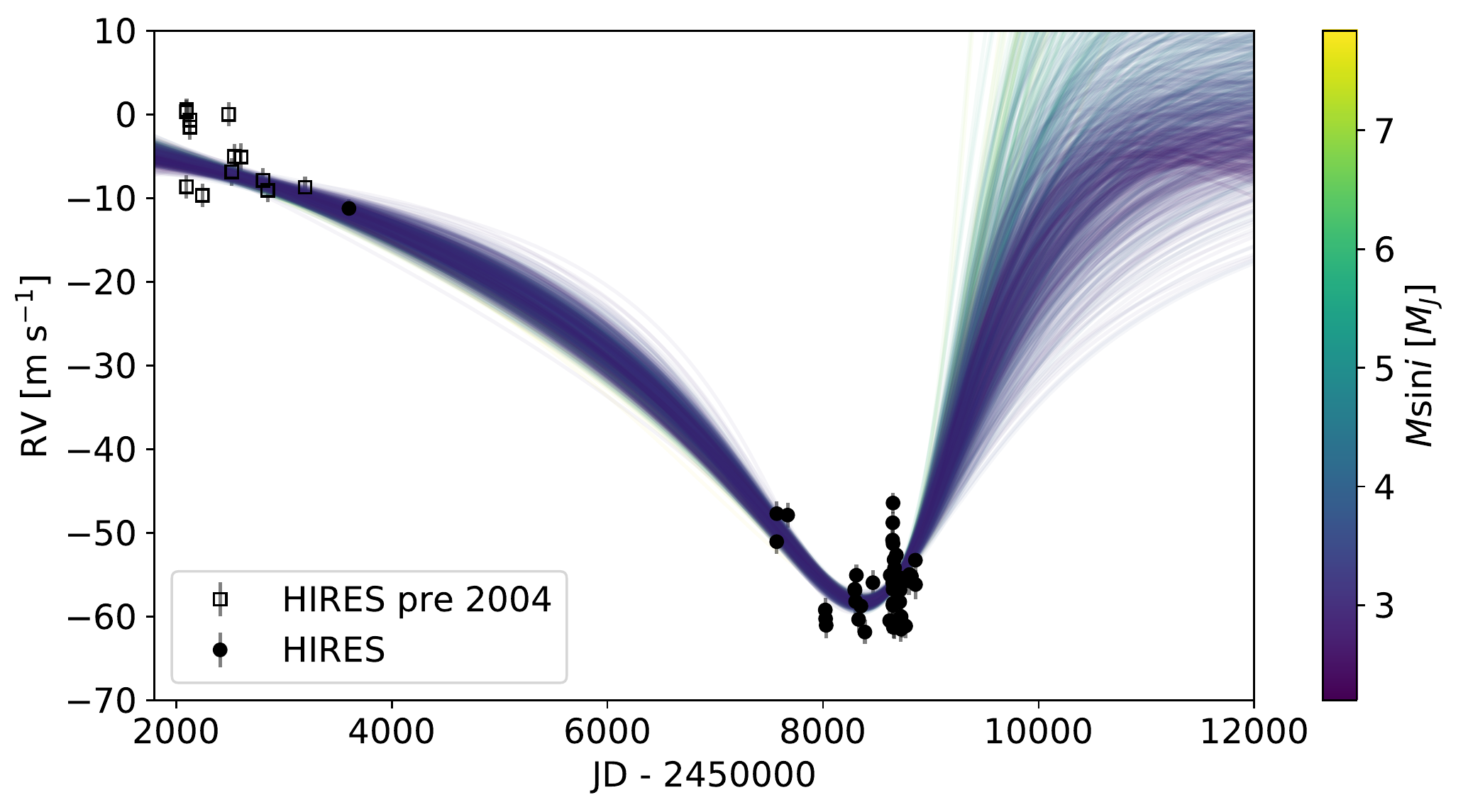}
\caption{Possible orbits for HD 213472 b. RV curves are drawn from the rejection sampling posterior generated with TheJoker. The color of each orbit drawn from the posterior scales with \msini\ .}
\label{fig:213472_orbits}
\end{center}
\end{figure*}

Figure \ref{fig:213472_orbits} shows a sample of possible orbits for HD 213472 b, drawn from our rejection sampling posteriors and projected over the next decade. More HIRES observations will further constrain this object's mass and orbital parameters.

\clearpage

\section{Stellar Catalog}

We record a subset of the stellar catalog and its associated stellar parameters in Table 2.  We make this table of CLS stars available in its entirety in machine-readable format.

\begin{longtable*}{llrrrrr}
\caption{Stellar Catalog} \\
\toprule 
\midrule 

CPS Name & Lit. Name & $T_{\textrm{eff}}$ [K] & [Fe/H] & log $g$ [log(cm s$^{-2}$)] & $R$ [\rsun] & $M$ [\msun] \\ 
\toprule 
\endfirsthead 
\caption[]{Stellar Catalog (Continued)} \\
\toprule 
\midrule 
CPS Name & Lit. Name & $T_{\textrm{eff}}$ [K] & [Fe/H] & log $g$ [log(cm s$^{-2}$)] & $R$ [\rsun] & $M$ [\msun] \\ 
\toprule 
\endhead 
HD 10002 & HD 10002 & $5320.0\pm 100.0$ & $0.251\pm 0.058$ & $4.449\pm 0.03$ & $0.949\pm 0.021$ & $0.928\pm 0.04$ \\ 
HD 10008 & HIP 7576 & $5390.0\pm 100.0$ & $0.024\pm 0.059$ & $4.553\pm 0.018$ & $0.832\pm 0.014$ & $0.897\pm 0.028$ \\ 
HD 100180 & HIP 56242 A & $5990.0\pm 100.0$ & $0.009\pm 0.06$ & $4.361\pm 0.032$ & $1.125\pm 0.026$ & $1.062\pm 0.049$ \\ 
HD 100623 & HIP 56452 & $5110.0\pm 100.0$ & $-0.321\pm 0.059$ & $4.576\pm 0.022$ & $0.734\pm 0.012$ & $0.748\pm 0.028$ \\ 
HD 101259 & HD 101259 & $4960.0\pm 100.0$ & $-0.634\pm 0.061$ & $3.178\pm 0.059$ & $4.72\pm 0.11$ & $1.23\pm 0.15$ \\ 
HD 10145 & HD 10145 & $5610.0\pm 100.0$ & $0.004\pm 0.058$ & $4.333\pm 0.027$ & $1.08\pm 0.025$ & $0.921\pm 0.036$ \\ 
\bottomrule 
\end{longtable*}

\section{Known Planets}

We record all planets recovered by \texttt{RVSearch} in Table 3, with \msini\ and key orbital parameter medians and uncertainties. We record all fitting parameter values in machine-readable format.

\begin{longtable*}{llrrrrr}
\caption{Planet Catalog} \\
\toprule
\midrule

CPS Name & Lit. Name & \msini\ [\mjup] & $a$ [AU] & $e$ & Ref. \\
\toprule
\endfirsthead
\caption[]{Planet Catalog (Continued)} \\
\toprule
\midrule
CPS Name & Lit. Name & \msini\ [\mjup] & $a$ [AU] & $e$ & Ref. \\
\toprule
\endhead
HD 104067 b & HD 104067 b & $0.202^{+0.017}_{-0.017}$ & $0.2673^{+0.0032}_{-0.0033}$ & $0.247^{+0.080}_{-0.082}$ & \cite{Vogt00} & \\
HD 10697 b & HD 10697 b & $6.39^{+0.13}_{-0.13}$ & $2.156^{+0.02}_{-0.02}$ & $0.0998^{+0.0082}_{-0.0082}$ & \cite{Butler06} & \\
HD 107148 b & HD 107148 b & $0.203^{+0.015}_{-0.015}$ & $0.3668^{+0.0047}_{-0.0048}$ & $0.174^{+0.071}_{-0.075}$ & \cite{Butler06} & \\
HD 107148 c & HD 107148 c & $0.0626^{+0.0097}_{-0.0098}$ & $0.1406^{+0.0018}_{-0.0018}$ & $0.34^{+0.13}_{-0.16}$ & This work & \\
HD 108874 b & HD 108874 b & $1.32^{+0.047}_{-0.047}$ & $1.053^{+0.016}_{-0.016}$ & $0.144^{+0.016}_{-0.016}$ & \cite{Butler03} & \\
HD 108874 c & HD 108874 c & $1.14^{+0.052}_{-0.050}$ & $2.83^{+0.044}_{-0.045}$ & $0.265^{+0.030}_{-0.031}$ & \cite{Vogt05} & \\
HD 114729 b & HD 114729 b & $0.892^{+0.053}_{-0.053}$ & $2.094^{+0.022}_{-0.022}$ & $0.098^{+0.048}_{-0.050}$ & \cite{Butler03} & \\
HD 114783 b & HD 114783 b & $1.033^{+0.034}_{-0.033}$ & $1.164^{+0.016}_{-0.016}$ & $0.126^{+0.016}_{-0.016}$ & \cite{Vogt02} & \\
HD 114783 c & HD 114783 c & $0.66^{+0.046}_{-0.047}$ & $4.97^{+0.12}_{-0.11}$ & $0.114^{+0.058}_{-0.058}$ & \cite{Bryan16} & \\
HD 115617 b & 61 Vir b & $0.0507^{+0.0035}_{-0.0036}$ & $0.215^{+0.0028}_{-0.0029}$ & $0.068^{+0.067}_{-0.047}$ & \cite{Vogt10} & \\
HD 115617 c & 61 Vir c & $0.0161^{+0.0017}_{-0.0016}$ & $0.04956^{+0.00066}_{-0.00067}$ & $0.099^{+0.091}_{-0.070}$ & \cite{Vogt10} & \\
HD 117176 b & 70 Vir b & $7.24^{+0.13}_{-0.13}$ & $0.4766^{+0.0043}_{-0.0044}$ & $0.3989^{+0.0011}_{-0.0012}$ & \cite{Butler96} & \\
HD 117207 b & HD 117207 b & $1.87^{+0.076}_{-0.075}$ & $3.744^{+0.059}_{-0.060}$ & $0.142^{+0.027}_{-0.026}$ & \cite{Marcy05} & \\
HD 11964a b & HD 11964a b & $0.631^{+0.027}_{-0.027}$ & $3.185^{+0.032}_{-0.032}$ & $0.101^{+0.044}_{-0.046}$ & \cite{Butler06} & \\
HD 11964a c & HD 11964a c & $0.0766^{+0.0063}_{-0.0061}$ & $0.2315^{+0.0022}_{-0.0022}$ & $0.106^{+0.088}_{-0.071}$ & \cite{Wright09} & \\
HD 120066 b & HR 5183 b & $3.15^{+0.18}_{-0.17}$ & $25.0^{+19.0}_{-8.3}$ & $0.886^{+0.049}_{-0.056}$ & \cite{Blunt19} & \\
HD 120136 b & tau Boo b & $4.3^{+0.075}_{-0.075}$ & $0.04869^{+0.00039}_{-0.00040}$ & $0.0074^{+0.0059}_{-0.0048}$ & \cite{Butler97} & \\
HD 12661 b & HD 12661 b & $2.283^{+0.062}_{-0.063}$ & $0.824^{+0.011}_{-0.011}$ & $0.3597^{+0.0046}_{-0.0044}$ & \cite{Fischer01} & \\
HD 12661 c & HD 12661 c & $1.855^{+0.054}_{-0.054}$ & $2.86^{+0.038}_{-0.039}$ & $0.025^{+0.012}_{-0.012}$ & \cite{Fischer03} & \\
HD 126614 b & HD 126614 b & $0.356^{+0.032}_{-0.030}$ & $2.291^{+0.026}_{-0.027}$ & $0.577^{+0.069}_{-0.065}$ & \cite{Howard10} & \\
HD 128311 b & HD 128311 b & $3.25^{+0.12}_{-0.12}$ & $1.742^{+0.020}_{-0.021}$ & $0.196^{+0.047}_{-0.057}$ & \cite{Butler03} & \\
HD 128311 c & HD 128311 c & $2.0^{+0.15}_{-0.15}$ & $1.088^{+0.012}_{-0.013}$ & $0.283^{+0.04}_{-0.04}$ & \cite{Vogt05} & \\
HD 130322 b & HD 130322 b & $1.149^{+0.037}_{-0.036}$ & $0.0929^{+0.0011}_{-0.0011}$ & $0.015^{+0.016}_{-0.010}$ & \cite{Udry00} & \\
HD 1326 b & HD 1326 b & $0.0171^{+0.0013}_{-0.0013}$ & $0.07321^{+0.00047}_{-0.00048}$ & $0.075^{+0.072}_{-0.052}$ & \cite{Howard14} & \\
HD 134987 b & HD 134987 b & $1.623^{+0.049}_{-0.049}$ & $0.817^{+0.012}_{-0.012}$ & $0.2281^{+0.0078}_{-0.0077}$ & \cite{Vogt00} & \\
HD 134987 c & HD 134987 c & $0.934^{+0.063}_{-0.060}$ & $6.62^{+0.16}_{-0.15}$ & $0.154^{+0.049}_{-0.050}$ & \cite{Jones10} & \\
HD 136925 b & HD 136925 b & $0.84^{+0.078}_{-0.074}$ & $5.13^{+0.12}_{-0.11}$ & $0.103^{+0.094}_{-0.070}$ & This work & \\
HD 13931 b & HD 13931 b & $1.911^{+0.077}_{-0.076}$ & $5.323^{+0.091}_{-0.091}$ & $0.02^{+0.021}_{-0.014}$ & \cite{Howard10} & \\
HD 141004 b & HIP 77257 b & $0.0428^{+0.0047}_{-0.0045}$ & $0.1238^{+0.002}_{-0.002}$ & $0.16^{+0.11}_{-0.10}$ & This work & \\
HD 141399 b & HD 141399 b & $1.329^{+0.046}_{-0.047}$ & $0.693^{+0.012}_{-0.012}$ & $0.0465^{+0.0068}_{-0.0068}$ & \cite{Vogt14} & \\
HD 141399 c & HD 141399 c & $1.263^{+0.047}_{-0.048}$ & $2.114^{+0.036}_{-0.037}$ & $0.044^{+0.013}_{-0.013}$ & \cite{Vogt14} & \\
HD 141399 d & HD 141399 d & $0.452^{+0.017}_{-0.017}$ & $0.4176^{+0.0070}_{-0.0073}$ & $0.053^{+0.015}_{-0.015}$ & \cite{Vogt14} & \\
HD 141399 e & HD 141399 e & $0.644^{+0.042}_{-0.040}$ & $4.5^{+0.11}_{-0.11}$ & $0.047^{+0.052}_{-0.033}$ & \cite{Vogt14} & \\
HD 143761 b & rho CrB b & $1.057^{+0.024}_{-0.024}$ & $0.2213^{+0.0025}_{-0.0025}$ & $0.0355^{+0.0035}_{-0.0037}$ & \cite{Noyes97} & \\
HD 143761 c & rho CrB c & $0.0885^{+0.0055}_{-0.0056}$ & $0.4157^{+0.0046}_{-0.0047}$ & $0.044^{+0.050}_{-0.031}$ & \cite{Fulton16} & \\
HD 145675 b & 14 Her b & $4.85^{+0.15}_{-0.14}$ & $2.83^{+0.040}_{-0.041}$ & $0.3674^{+0.0035}_{-0.0038}$ & \cite{Wittenmyer07} & \\
HD 145675 c & 14 Her c & $5.8^{+1.4}_{-1.0}$ & $16.4^{+9.3}_{-4.3}$ & $0.45^{+0.17}_{-0.15}$ & This work & \\
HD 145934 b & HD 145934 b & $2.04^{+0.24}_{-0.23}$ & $4.73^{+0.20}_{-0.21}$ & $0.057^{+0.057}_{-0.039}$ & \cite{Feng15} & \\
HD 1461 b & HD 1461 b & $0.0208^{+0.0019}_{-0.0018}$ & $0.06361^{+0.00096}_{-0.00099}$ & $0.062^{+0.060}_{-0.042}$ & \cite{Rivera10} & \\
HD 1461 c & HD 1461 c & $0.0222^{+0.0028}_{-0.0028}$ & $0.1121^{+0.0017}_{-0.0017}$ & $0.112^{+0.10}_{-0.077}$ & \cite{Mayor11} & \\
HD 147379A b & HD 147379A b & $0.096^{+0.012}_{-0.012}$ & $0.3315^{+0.0024}_{-0.0024}$ & $0.096^{+0.110}_{-0.068}$ & \cite{Reiners18} & \\
HD 154345 b & HD 154345 b & $0.905^{+0.071}_{-0.089}$ & $4.272^{+0.086}_{-0.080}$ & $0.038^{+0.036}_{-0.027}$ & \cite{Wright07} & \\
HD 156279 b & HD 156279 b & $9.5^{+0.31}_{-0.31}$ & $0.5039^{+0.0082}_{-0.0084}$ & $0.64779^{+0.00068}_{-0.00066}$ & \cite{Diaz12} & \\
HD 156279 c & HD 156279 c & $9.44^{+0.31}_{-0.32}$ & $5.46^{+0.091}_{-0.093}$ & $0.2597^{+0.0050}_{-0.0049}$ & \cite{Bryan16} & \\
HD 156668 b & HD 156668 b & $0.0991^{+0.0079}_{-0.0077}$ & $1.57^{+0.017}_{-0.017}$ & $0.089^{+0.084}_{-0.061}$ & \cite{Howard11} & \\
HD 156668 c & HD 156668 c & $0.0158^{+0.0013}_{-0.0013}$ & $0.05025^{+0.00050}_{-0.00051}$ & $0.235^{+0.072}_{-0.072}$ & This work & \\
HD 16141 b & HD 16141 b & $0.25^{+0.017}_{-0.017}$ & $0.3609^{+0.0072}_{-0.0075}$ & $0.195^{+0.051}_{-0.051}$ & \cite{Marcy00} & \\
HD 164922 b & HD 164922 b & $0.344^{+0.013}_{-0.013}$ & $2.149^{+0.025}_{-0.025}$ & $0.065^{+0.027}_{-0.029}$ & \cite{Butler06} & \\
HD 164922 c & HD 164922 c & $0.0451^{+0.0036}_{-0.0036}$ & $0.3411^{+0.0039}_{-0.0040}$ & $0.096^{+0.088}_{-0.066}$ & \cite{Fulton16} & \\
HD 164922 d & HD 164922 d & $0.0331^{+0.0031}_{-0.0031}$ & $0.2292^{+0.0026}_{-0.0027}$ & $0.086^{+0.083}_{-0.060}$ & This work & \\
HD 164922 e & HD 164922 e & $0.0149^{+0.0021}_{-0.0021}$ & $0.1023^{+0.0012}_{-0.0012}$ & $0.18^{+0.17}_{-0.12}$ & \cite{Benatti20} & \\
HD 167042 b & HD 167042 b & $1.59^{+0.13}_{-0.13}$ & $1.304^{+0.040}_{-0.043}$ & $0.092^{+0.061}_{-0.057}$ & \cite{Johnson08} & \\
HD 168009 b & HIP 89474 b & $0.03^{+0.0038}_{-0.0037}$ & $0.1192^{+0.0017}_{-0.0018}$ & $0.121^{+0.110}_{-0.082}$ & This work & \\
HD 168443 b & HD 168443 b & $7.92^{+0.16}_{-0.16}$ & $0.2977^{+0.0029}_{-0.0030}$ & $0.5313^{+0.00062}_{-0.00061}$ & \cite{Marcy99} & \\
HD 168443 c & HD 168443 c & $17.76^{+0.35}_{-0.35}$ & $2.881^{+0.028}_{-0.029}$ & $0.2112^{+0.0013}_{-0.0013}$ & \cite{Marcy01} & \\
HD 168746 b & HD 168746 b & $0.2294^{+0.0078}_{-0.0075}$ & $0.06504^{+0.00054}_{-0.00056}$ & $0.098^{+0.026}_{-0.026}$ & \cite{Pepe02} & \\
HD 169830 b & HD 169830 b & $2.957^{+0.069}_{-0.070}$ & $0.8131^{+0.0082}_{-0.0085}$ & $0.306^{+0.012}_{-0.013}$ & \cite{Naef01} & \\
HD 169830 c & HD 169830 c & $3.51^{+0.12}_{-0.12}$ & $3.283^{+0.035}_{-0.036}$ & $0.257^{+0.019}_{-0.019}$ & \cite{Mayor04} & \\
HD 170469 b & HD 170469 b & $0.555^{+0.075}_{-0.072}$ & $2.212^{+0.040}_{-0.044}$ & $0.15^{+0.16}_{-0.11}$ & \cite{Fischer07} & \\
HD 177830 b & HD 177830 b & $1.347^{+0.097}_{-0.098}$ & $1.179^{+0.040}_{-0.043}$ & $0.028^{+0.017}_{-0.017}$ & \cite{Vogt00} & \\
HD 177830 c & HD 177830 c & $0.104^{+0.016}_{-0.015}$ & $0.496^{+0.017}_{-0.018}$ & $0.54^{+0.12}_{-0.15}$ & \cite{Meschiari11} & \\
HD 178911B b & HD 178911B b & $7.07^{+0.22}_{-0.23}$ & $0.34^{+0.0053}_{-0.0055}$ & $0.1132^{+0.0025}_{-0.0025}$ & \cite{Zucker02} & \\
HD 179949 b & HD 179949 b & $0.966^{+0.031}_{-0.030}$ & $0.04439^{+0.00045}_{-0.00046}$ & $0.016^{+0.017}_{-0.011}$ & \cite{Tiney01} & \\
HD 181234 b & HD 181234 b & $8.9^{+1.90}_{-0.76}$ & $7.48^{+0.39}_{-0.23}$ & $0.793^{+0.086}_{-0.082}$ & \cite{Rickman19} & \\
HD 186427 b & 16 Cyg B b & $1.752^{+0.054}_{-0.054}$ & $1.676^{+0.025}_{-0.026}$ & $0.6832^{+0.0031}_{-0.0031}$ & \cite{Chochran97} & \\
HD 187123 b & HD 187123 b & $0.501^{+0.016}_{-0.016}$ & $0.04185^{+0.00065}_{-0.00067}$ & $0.004^{+0.0040}_{-0.0028}$ & \cite{Butler98} & \\
HD 187123 c & HD 187123 c & $1.713^{+0.058}_{-0.058}$ & $4.431^{+0.071}_{-0.072}$ & $0.227^{+0.016}_{-0.017}$ & \cite{Wright09} & \\
HD 188015 b & HD 188015 b & $1.455^{+0.060}_{-0.059}$ & $1.187^{+0.018}_{-0.019}$ & $0.17^{+0.025}_{-0.026}$ & \cite{Marcy05} & \\
HD 189733 b & HD 189733 b & $1.162^{+0.036}_{-0.035}$ & $0.03126^{+0.00036}_{-0.00037}$ & $0.027^{+0.021}_{-0.018}$ & \cite{Bouchy05} & \\
HD 190360 b & HD 190360 b & $1.492^{+0.043}_{-0.043}$ & $3.955^{+0.052}_{-0.053}$ & $0.3274^{+0.0081}_{-0.0087}$ & \cite{Naef03} & \\
HD 190360 c & HD 190360 c & $0.0674^{+0.0027}_{-0.0026}$ & $0.1294^{+0.0017}_{-0.0017}$ & $0.165^{+0.027}_{-0.028}$ & \cite{Vogt05} & \\
HD 192263 b & HD 192263 b & $0.658^{+0.03}_{-0.03}$ & $0.154^{+0.0019}_{-0.0020}$ & $0.04^{+0.037}_{-0.027}$ & \cite{Santos00} & \\
HD 192310 b & HD 192310 b & $0.0451^{+0.0064}_{-0.0060}$ & $0.3262^{+0.0035}_{-0.0037}$ & $0.14^{+0.120}_{-0.093}$ & \cite{Howard11} & \\
HD 195019 b & HD 195019 b & $3.655^{+0.091}_{-0.090}$ & $0.1376^{+0.0017}_{-0.0017}$ & $0.0198^{+0.0032}_{-0.0032}$ & \cite{Fischer99} & \\
HD 209458 b & HD 209458 b & $0.665^{+0.022}_{-0.022}$ & $0.04635^{+0.00068}_{-0.00070}$ & $0.0105^{+0.0110}_{-0.0074}$ & \cite{Henry00} & \\
HD 210277 b & HD 210277 b & $1.236^{+0.032}_{-0.033}$ & $1.123^{+0.014}_{-0.014}$ & $0.472^{+0.0055}_{-0.0056}$ & \cite{Marcy99} & \\
HD 213472 b & HD 213472 b & $3.48^{+1.10}_{-0.59}$ & $13.0^{+5.7}_{-2.6}$ & $0.53^{+0.120}_{-0.085}$ & This work & \\
HD 216520 b & HIP 112527 b & $0.0326^{+0.0036}_{-0.0038}$ & $0.1954^{+0.0025}_{-0.0025}$ & $0.19^{+0.13}_{-0.12}$ & Burt et al. 2020 & \\
HD 217014 b & 51 Peg b & $0.464^{+0.014}_{-0.014}$ & $0.05236^{+0.00076}_{-0.00079}$ & $0.0042^{+0.0046}_{-0.0030}$ & \cite{Mayor95} & \\
HD 217107 b & HD 217107 b & $1.385^{+0.039}_{-0.039}$ & $0.0739^{+0.0010}_{-0.0011}$ & $0.1279^{+0.0017}_{-0.0016}$ & \cite{Fischer99} & \\
HD 217107 c & HD 217107 c & $4.31^{+0.13}_{-0.13}$ & $5.944^{+0.083}_{-0.086}$ & $0.3928^{+0.0069}_{-0.0067}$ & \cite{Vogt05} & \\
HD 218566 b & HD 218566 b & $0.198^{+0.018}_{-0.018}$ & $0.6875^{+0.0082}_{-0.0085}$ & $0.268^{+0.10}_{-0.095}$ & \cite{Meschiari11} & \\
HD 219134 b & HD 219134 b & $0.308^{+0.014}_{-0.014}$ & $2.968^{+0.037}_{-0.037}$ & $0.025^{+0.027}_{-0.018}$ & \cite{Vogt15} & \\
HD 219134 c & HD 219134 c & $0.0516^{+0.0032}_{-0.0030}$ & $0.2346^{+0.0026}_{-0.0027}$ & $0.077^{+0.055}_{-0.050}$ & \cite{Vogt15} & \\
HD 219134 d & HD 219134 d & $0.013^{+0.0010}_{-0.0011}$ & $0.03838^{+0.00043}_{-0.00043}$ & $0.063^{+0.070}_{-0.045}$ & \cite{Vogt15} & \\
HD 219134 e & HD 219134 e & $0.0243^{+0.0023}_{-0.0022}$ & $0.1453^{+0.0016}_{-0.0016}$ & $0.072^{+0.078}_{-0.051}$ & \cite{Vogt15} & \\
HD 219134 f & HD 219134 f & $0.0112^{+0.0014}_{-0.0014}$ & $0.06466^{+0.00073}_{-0.00073}$ & $0.16^{+0.12}_{-0.11}$ & \cite{Vogt15} & \\
HD 22049 b & eps Eri b & $0.651^{+0.039}_{-0.039}$ & $3.5^{+0.039}_{-0.040}$ & $0.044^{+0.047}_{-0.031}$ & \cite{Hatzes00} & \\
HD 222582 b & HD 222582 b & $7.88^{+0.24}_{-0.24}$ & $1.335^{+0.02}_{-0.02}$ & $0.7615^{+0.0035}_{-0.0034}$ & \cite{Vogt00} & \\
HD 24040 b & HD 24040 b & $4.05^{+0.15}_{-0.15}$ & $4.708^{+0.076}_{-0.077}$ & $0.0117^{+0.0110}_{-0.0082}$ & \cite{Boisse12} & \\
HD 24040 c & HD 24040 c & $0.201^{+0.027}_{-0.027}$ & $1.3^{+0.021}_{-0.021}$ & $0.11^{+0.120}_{-0.079}$ & This work & \\
HD 26161 b & HD 26161 b & $13.5^{+8.5}_{-3.7}$ & $20.4^{+7.9}_{-4.9}$ & $0.82^{+0.061}_{-0.050}$ & This work & \\
HD 28185 b & HD 28185 b & $6.04^{+0.2}_{-0.2}$ & $1.045^{+0.016}_{-0.018}$ & $0.0629^{+0.0042}_{-0.0049}$ & \cite{Santos01} & \\
HD 285968 b & HD 285968 b & $0.0285^{+0.0043}_{-0.0043}$ & $0.06649^{+0.00042}_{-0.00043}$ & $0.16^{+0.14}_{-0.11}$ & \cite{Forveille09} & \\
HD 31253 b & HD 31253 b & $0.446^{+0.063}_{-0.063}$ & $1.296^{+0.024}_{-0.025}$ & $0.44^{+0.12}_{-0.17}$ & \cite{Meschiari11} & \\
HD 32963 b & HD 32963 b & $0.726^{+0.036}_{-0.035}$ & $3.416^{+0.058}_{-0.059}$ & $0.069^{+0.040}_{-0.039}$ & \cite{Rowan16} & \\
HD 33636 b & HIP 24205 b & $8.92^{+0.29}_{-0.30}$ & $3.238^{+0.051}_{-0.053}$ & $0.491^{+0.0050}_{-0.0049}$ & \cite{Vogt02} & \\
HD 34445 b & HD 34445 b & $0.658^{+0.04}_{-0.04}$ & $2.106^{+0.038}_{-0.040}$ & $0.103^{+0.059}_{-0.056}$ & \cite{Howard10} & \\
HD 3651 b & HD 3651 b & $0.2202^{+0.0077}_{-0.0077}$ & $0.2954^{+0.0043}_{-0.0044}$ & $0.614^{+0.015}_{-0.015}$ & \cite{Fischer03} & \\
HD 3765 b & HIP 3206 b & $0.173^{+0.014}_{-0.013}$ & $2.108^{+0.032}_{-0.033}$ & $0.298^{+0.078}_{-0.071}$ & This work & \\
HD 38529 b & HD 38529 b & $13.21^{+0.1}_{-0.1}$ & $3.736^{+0.01}_{-0.01}$ & $0.3545^{+0.0048}_{-0.0047}$ & \cite{Fischer01} & \\
HD 38529 c & HD 38529 c & $0.876^{+0.014}_{-0.014}$ & $0.1329^{+0.00035}_{-0.00035}$ & $0.26^{+0.014}_{-0.015}$ & \cite{Fischer03} & \\
HD 40979 b & HD 40979 b & $3.8^{+0.16}_{-0.16}$ & $0.8669^{+0.0074}_{-0.0076}$ & $0.251^{+0.028}_{-0.028}$ & \cite{Fischer03} & \\
HD 4203 b & HD 4203 b & $1.821^{+0.078}_{-0.077}$ & $1.177^{+0.021}_{-0.022}$ & $0.513^{+0.013}_{-0.014}$ & \cite{Vogt02} & \\
HD 4203 c & HD 4203 c & $2.68^{+0.99}_{-0.24}$ & $7.8^{+5.40}_{-0.78}$ & $0.19^{+0.290}_{-0.089}$ & \cite{Kane14} & \\
HD 4208 b & HD 4208 b & $0.771^{+0.036}_{-0.035}$ & $1.634^{+0.022}_{-0.022}$ & $0.059^{+0.044}_{-0.038}$ & \cite{Vogt02} & \\
HD 42618 b & HD 42618 b & $0.0478^{+0.0057}_{-0.0056}$ & $0.5337^{+0.0089}_{-0.0090}$ & $0.27^{+0.13}_{-0.12}$ & \cite{Fulton16} & \\
HD 45184 b & HD 45184 b & $0.0374^{+0.0040}_{-0.0038}$ & $0.0641^{+0.0010}_{-0.0011}$ & $0.18^{+0.1}_{-0.1}$ & \cite{Mayor11} & \\
HD 45184 c & HD 45184 c & $0.0345^{+0.0057}_{-0.0059}$ & $0.1095^{+0.0018}_{-0.0019}$ & $0.48^{+0.18}_{-0.24}$ & \cite{Udry19} & \\
HD 45350 b & HD 45350 b & $1.821^{+0.075}_{-0.070}$ & $1.958^{+0.029}_{-0.030}$ & $0.794^{+0.012}_{-0.012}$ & \cite{Marcy05} & \\
HD 46375 b & HD 46375 b & $0.2267^{+0.0087}_{-0.0087}$ & $0.03998^{+0.00059}_{-0.00060}$ & $0.063^{+0.024}_{-0.024}$ & \cite{Marcy00} & \\
HD 49674 b & HD 49674 b & $0.1149^{+0.0083}_{-0.0084}$ & $0.058^{+0.00082}_{-0.00084}$ & $0.06^{+0.060}_{-0.042}$ & \cite{Butler03} & \\
HD 50499 b & HD 50499 b & $1.346^{+0.084}_{-0.087}$ & $3.847^{+0.038}_{-0.040}$ & $0.348^{+0.046}_{-0.045}$ & \cite{Vogt05} & \\
HD 50499 c & HD 50499 c & $3.18^{+0.63}_{-0.46}$ & $10.1^{+2.00}_{-0.84}$ & $0.241^{+0.089}_{-0.075}$ & \cite{Rickman19} & \\
HD 50554 b & HD 50554 b & $4.35^{+0.19}_{-0.18}$ & $2.265^{+0.035}_{-0.036}$ & $0.459^{+0.019}_{-0.018}$ & \cite{Fischer02} & \\
HD 52265 b & HD 52265 b & $1.108^{+0.031}_{-0.031}$ & $0.506^{+0.0056}_{-0.0056}$ & $0.213^{+0.013}_{-0.014}$ & \cite{Butler00} & \\
HD 66428 b & HD 66428 b & $3.19^{+0.11}_{-0.11}$ & $3.455^{+0.049}_{-0.050}$ & $0.418^{+0.015}_{-0.014}$ & \cite{Butler06} & \\
HD 66428 c & HD 66428 c & $27^{+22}_{-17}$ & $23.0^{+19.0}_{-7.6}$ & $0.32^{+0.23}_{-0.16}$ & This work & \\
HD 68988 b & HD 68988 b & $1.915^{+0.053}_{-0.054}$ & $0.07021^{+0.00096}_{-0.0010}$ & $0.1581^{+0.0027}_{-0.0031}$ & \cite{Vogt02} & \\
HD 68988 c & HD 68988 c & $15.0^{+2.8}_{-1.5}$ & $13.2^{+5.3}_{-2.0}$ & $0.45^{+0.130}_{-0.081}$ & This work & \\
HD 69830 b & HD 69830 b & $0.0323^{+0.0021}_{-0.0020}$ & $0.0794^{+0.0012}_{-0.0013}$ & $0.112^{+0.054}_{-0.060}$ & \cite{Lovis06} & \\
HD 69830 c & HD 69830 c & $0.031^{+0.003}_{-0.003}$ & $0.1882^{+0.0029}_{-0.0030}$ & $0.114^{+0.096}_{-0.079}$ & \cite{Lovis06} & \\
HD 69830 d & HD 69830 d & $0.0444^{+0.0054}_{-0.0056}$ & $0.645^{+0.01}_{-0.01}$ & $0.104^{+0.10}_{-0.073}$ & \cite{Lovis06} & \\
HD 72659 b & HD 72659 b & $2.85^{+0.12}_{-0.12}$ & $4.652^{+0.067}_{-0.068}$ & $0.269^{+0.018}_{-0.026}$ & \cite{Butler03} & \\
HD 74156 b & HD 74156 b & $7.65^{+0.27}_{-0.26}$ & $3.726^{+0.064}_{-0.065}$ & $0.3691^{+0.0055}_{-0.0055}$ & \cite{Naef04} & \\
HD 74156 c & HD 74156 c & $1.745^{+0.061}_{-0.061}$ & $0.2844^{+0.0049}_{-0.0049}$ & $0.6536^{+0.0036}_{-0.0038}$ & \cite{Naef04} & \\
HD 75732 b & 55 Cnc b & $0.841^{+0.026}_{-0.026}$ & $0.1162^{+0.0018}_{-0.0018}$ & $0.0048^{+0.0036}_{-0.0031}$ & \cite{Butler97} & \\
HD 75732 c & 55 Cnc c & $0.171^{+0.0066}_{-0.0067}$ & $0.2432^{+0.0037}_{-0.0039}$ & $0.048^{+0.031}_{-0.029}$ & \cite{Marcy02} & \\
HD 75732 d & 55 Cnc d & $2.86^{+0.25}_{-0.25}$ & $5.54^{+0.1}_{-0.1}$ & $0.139^{+0.015}_{-0.016}$ & \cite{Marcy02} & \\
HD 75732 e & 55 Cnc e & $0.1475^{+0.0093}_{-0.0094}$ & $0.792^{+0.012}_{-0.013}$ & $0.142^{+0.060}_{-0.066}$ & \cite{McArthur04} & \\
HD 75732 f & 55 Cnc f & $0.0295^{+0.0014}_{-0.0013}$ & $0.01583^{+0.00024}_{-0.00025}$ & $0.036^{+0.034}_{-0.025}$ & \cite{Fischer08} & \\
HD 7924 b & HD 7924 b & $0.0259^{+0.0014}_{-0.0014}$ & $0.05596^{+0.00075}_{-0.00079}$ & $0.049^{+0.044}_{-0.035}$ & \cite{Fulton15} & \\
HD 7924 c & HD 7924 c & $0.0278^{+0.0019}_{-0.0019}$ & $0.1121^{+0.0015}_{-0.0016}$ & $0.1^{+0.069}_{-0.062}$ & \cite{Fulton15} & \\
HD 7924 d & HD 7924 d & $0.126^{+0.013}_{-0.012}$ & $3.51^{+0.10}_{-0.096}$ & $0.278^{+0.081}_{-0.082}$ & \cite{Fulton15} & \\
HD 80606 b & HD 80606 b & $4.16^{+0.13}_{-0.13}$ & $0.4602^{+0.0069}_{-0.0070}$ & $0.93043^{+0.00068}_{-0.00069}$ & \cite{Naef01} & \\
HD 82943 b & HD 82943 b & $1.652^{+0.085}_{-0.076}$ & $0.747^{+0.011}_{-0.011}$ & $0.421^{+0.014}_{-0.015}$ & \cite{Mayor04} & \\
HD 82943 c & HD 82943 c & $1.539^{+0.052}_{-0.052}$ & $1.189^{+0.017}_{-0.017}$ & $0.142^{+0.042}_{-0.046}$ & \cite{Mayor04} & \\
HD 8574 b & HD 8574 b & $1.765^{+0.075}_{-0.073}$ & $0.75^{+0.012}_{-0.012}$ & $0.306^{+0.022}_{-0.022}$ & \cite{Perrier03} & \\
HD 87883 b & HD 87883 b & $2.292^{+0.069}_{-0.069}$ & $4.073^{+0.056}_{-0.057}$ & $0.7121^{+0.0083}_{-0.0079}$ & \cite{Fischer09} & \\
HD 90156 b & HD 90156 b & $0.037^{+0.0062}_{-0.0060}$ & $0.2508^{+0.0037}_{-0.0037}$ & $0.16^{+0.18}_{-0.11}$ & \cite{Mordasini11} & \\
HD 92788 b & HD 92788 b & $3.52^{+0.1}_{-0.1}$ & $0.949^{+0.013}_{-0.013}$ & $0.3552^{+0.0070}_{-0.0072}$ & \cite{Fischer01} & \\
HD 92788 c & HD 92788 c & $2.81^{+0.18}_{-0.17}$ & $8.26^{+0.37}_{-0.28}$ & $0.355^{+0.057}_{-0.052}$ & \cite{Rickman19} & \\
HD 95128 b & 47 UMa b & $2.438^{+0.086}_{-0.085}$ & $2.059^{+0.031}_{-0.033}$ & $0.016^{+0.0076}_{-0.0080}$ & \cite{Butler96} & \\
HD 95128 c & 47 UMa c & $0.497^{+0.032}_{-0.030}$ & $3.403^{+0.055}_{-0.056}$ & $0.179^{+0.090}_{-0.092}$ & \cite{Fischer02} & \\
HD 95128 d & 47 UMa d & $1.51^{+0.22}_{-0.18}$ & $13.8^{+4.8}_{-2.1}$ & $0.38^{+0.16}_{-0.15}$ & \cite{Gregory10} & \\
HD 95735 b & GJ 411 b & $0.0568^{+0.0091}_{-0.0083}$ & $3.1^{+0.13}_{-0.11}$ & $0.14^{+0.160}_{-0.095}$ & \cite{Diaz19} & \\
HD 95735 c & GJ 411 c & $0.00882^{+0.00092}_{-0.00098}$ & $0.07892^{+0.00055}_{-0.00054}$ & $0.095^{+0.099}_{-0.066}$ & This work & \\
HD 97101 b & HD 97101 b & $0.184^{+0.010}_{-0.011}$ & $1.424^{+0.011}_{-0.011}$ & $0.185^{+0.069}_{-0.073}$ & Dedrick et al. submitted & \\
HD 97101 c & HD 97101 c & $0.0322^{+0.0043}_{-0.0038}$ & $0.2403^{+0.0017}_{-0.0017}$ & $0.28^{+0.13}_{-0.13}$ & Dedrick et al. submitted & \\
HD 97658 b & HD 97658 b & $0.0247^{+0.0018}_{-0.0017}$ & $0.0805^{+0.0010}_{-0.0011}$ & $0.063^{+0.061}_{-0.044}$ & \cite{Howard11} & \\
HD 9826 b & ups And b & $0.675^{+0.016}_{-0.016}$ & $0.05914^{+0.00062}_{-0.00063}$ & $0.0069^{+0.0072}_{-0.0046}$ & \cite{Butler97} & \\
HD 9826 c & ups And c & $1.965^{+0.049}_{-0.050}$ & $0.8265^{+0.0087}_{-0.0088}$ & $0.266^{+0.012}_{-0.012}$ & \cite{Butler99} & \\
HD 9826 d & ups And d & $4.1^{+0.1}_{-0.1}$ & $2.518^{+0.026}_{-0.027}$ & $0.294^{+0.011}_{-0.012}$ & \cite{Butler99} & \\
HD 99109 b & HD 99109 b & $0.474^{+0.035}_{-0.035}$ & $1.122^{+0.017}_{-0.017}$ & $0.06^{+0.064}_{-0.042}$ & \cite{Butler06} & \\
HD 99492 b & HD 99492 b & $0.0841^{+0.0061}_{-0.0062}$ & $0.1231^{+0.0014}_{-0.0015}$ & $0.085^{+0.070}_{-0.057}$ & \cite{Marcy05} & \\
GL 317 b & GL 317 b & $1.852^{+0.037}_{-0.037}$ & $1.1799^{+0.0076}_{-0.0076}$ & $0.098^{+0.016}_{-0.016}$ & \cite{Johnson07b} & \\
GL 317 c & GL 317 c & $1.673^{+0.078}_{-0.075}$ & $5.78^{+0.74}_{-0.38}$ & $0.248^{+0.110}_{-0.074}$ & \cite{Anglada-Escude12} & \\
GL 687 b & GL 687 b & $0.0553^{+0.0047}_{-0.0047}$ & $0.1658^{+0.0012}_{-0.0012}$ & $0.153^{+0.077}_{-0.080}$ & \cite{Burt14} & \\
HIP 109388 b & GJ 849 b & $0.891^{+0.035}_{-0.036}$ & $2.41^{+0.017}_{-0.017}$ & $0.05^{+0.029}_{-0.030}$ & \cite{Butler06_GJ849} & \\
HIP 109388 c & GJ 849 c & $1.079^{+0.053}_{-0.053}$ & $4.974^{+0.082}_{-0.074}$ & $0.099^{+0.041}_{-0.040}$ & \cite{Feng15} & \\
HIP 22627 b & GJ 179 b & $0.752^{+0.041}_{-0.041}$ & $2.522^{+0.034}_{-0.033}$ & $0.169^{+0.048}_{-0.050}$ & \cite{Howard10} & \\
HIP 57050 b & GJ 1148 b & $0.26^{+0.030}_{-0.029}$ & $0.9302^{+0.0088}_{-0.0086}$ & $0.27^{+0.14}_{-0.17}$ & \cite{Haghighipour10} & \\
HIP 57050 c & GJ 1148 c & $0.339^{+0.014}_{-0.014}$ & $0.1686^{+0.0014}_{-0.0014}$ & $0.366^{+0.027}_{-0.028}$ & \cite{Trifonov17} & \\
HIP 74995 b & GJ 581 b & $0.051^{+0.0020}_{-0.0019}$ & $0.04099^{+0.00043}_{-0.00044}$ & $0.02^{+0.022}_{-0.014}$ & \cite{Bonfils05} & \\
HIP 74995 c & GJ 581 c & $0.0159^{+0.0022}_{-0.0022}$ & $0.07358^{+0.00077}_{-0.00078}$ & $0.12^{+0.120}_{-0.084}$ & \cite{Mayor09} & \\
HIP 83043 b & GJ 649 b & $0.275^{+0.022}_{-0.021}$ & $1.1325^{+0.0087}_{-0.0087}$ & $0.17^{+0.1}_{-0.1}$ & \cite{Johnson10} & \\
HIP 57087 b & GJ 436 b & $0.0668^{+0.0022}_{-0.0022}$ & $0.02849^{+0.0002}_{-0.0002}$ & $0.145^{+0.027}_{-0.027}$ & \cite{Butler04} & \\
BD-103166 b & BD-103166 b & $0.449^{+0.017}_{-0.017}$ & $0.04501^{+0.00064}_{-0.00066}$ & $0.022^{+0.024}_{-0.015}$ & \cite{Butler00} & \\
HD 175541 b & HD 175541 b & $0.577^{+0.049}_{-0.047}$ & $0.99^{+0.030}_{-0.032}$ & $0.057^{+0.053}_{-0.039}$ & \cite{Johnson07} & \\
HD 37124 b & HD 37124 b & $0.65^{+0.016}_{-0.016}$ & $0.5253^{+0.0032}_{-0.0032}$ & $0.05^{+0.019}_{-0.022}$ & \cite{Butler03} & \\
HD 37124 c & HD 37124 c & $0.647^{+0.08}_{-0.09}$ & $1.687^{+0.011}_{-0.011}$ & $0.126^{+0.047}_{-0.045}$ & \cite{Butler03} & \\
HD 37124 d & HD 37124 d & $0.659^{+0.034}_{-0.036}$ & $2.67^{+0.022}_{-0.021}$ & $0.16^{+0.12}_{-0.11}$ & \cite{Vogt05} & \\
GL 876 b & GL 876 b & $2.108^{+0.036}_{-0.036}$ & $0.2177^{+0.0018}_{-0.0018}$ & $0.0019^{+0.0022}_{-0.0013}$ & \cite{Marcy98} & \\
GL 876 c & GL 876 c & $0.698^{+0.013}_{-0.013}$ & $0.1363^{+0.0011}_{-0.0011}$ & $0.0055^{+0.0064}_{-0.0040}$ & \cite{Marcy01} & \\
GL 876 d & GL 876 d & $0.0184^{+0.0016}_{-0.0015}$ & $0.02183^{+0.00018}_{-0.00019}$ & $0.273^{+0.093}_{-0.098}$ & \cite{Rivera05} & \\
HD 83443 b & HD 83443 b & $0.409^{+0.019}_{-0.019}$ & $0.04067^{+0.00060}_{-0.00062}$ & $0.074^{+0.031}_{-0.032}$ & \cite{Butler02} & \\
HD 183263 b & HD 183263 b & $3.705^{+0.098}_{-0.10}$ & $1.508^{+0.02}_{-0.02}$ & $0.3786^{+0.004}_{-0.004}$ & \cite{Marcy05} & \\
HD 183263 c & HD 183263 c & $7.97^{+0.22}_{-0.22}$ & $6.038^{+0.082}_{-0.085}$ & $0.0595^{+0.0064}_{-0.0063}$ & \cite{Wright09} & \\
\bottomrule
\end{longtable*}

\section{Resolved Binaries and Substellar Companions}

We record all stellar binaries and substellar companions recovered by \texttt{RVSearch} in Table 4.

\begin{longtable*}{llrrr}
\caption{Binary and Substellar Catalog} \\
\toprule
\midrule

CPS Name & Lit. Name & \msini\ [\mjup] & $a$ [AU] & $e$ \\
\toprule
\endfirsthead
\caption[]{Binary and Substellar Catalog (Continued)} \\
\toprule
\midrule
CPS Name & Lit. Name & \msini\ [\mjup] & $a$ [AU] & $e$ \\
\toprule
\endhead
HD 104304 & HIP 58576 & $110.0^{+5.3}_{-6.1}$ & $18.1^{+1.4}_{-1.6}$ & $0.37^{+0.031}_{-0.035}$ \\
HD 10790 & HD 10790 & $285^{+29}_{-20}$ & $1.544^{+0.02}_{-0.02}$ & $0.847^{+0.015}_{-0.011}$ \\
HD 111031 & HD 111031 & $66^{+47}_{-33}$ & $32^{+15}_{-10}$ & $0.49^{+0.14}_{-0.17}$ \\
HD 112914 & HIP 63406	& $223.1^{+4.6}_{-4.7}$ & $1.389^{+0.016}_{-0.016}$ & $0.32416^{+0.00036}_{-0.00037}$ \\
HD 120136 & tau Boo & $350^{+230}_{-160}$ & $46^{+46}_{-18}$ & $0.84^{+0.093}_{-0.140}$ \\
HD 126614 & HD 126614 & $27.6^{+6.4}_{-5.1}$ & $16.5^{+1.7}_{-1.5}$ & $0.043^{+0.047}_{-0.031}$ \\
HD 142229 & HD 142229 & $850^{+710}_{-470}$ & $28^{+16}_{-11}$ & $0.66^{+0.11}_{-0.13}$ \\
HD 142267 & HIP 77801 & $141.2^{+2.6}_{-2.7}$ & $0.4924^{+0.0048}_{-0.0050}$ & $0.54768^{+0.00018}_{-0.00018}$ \\
HD 144287 & HIP 78709 & $343^{+39}_{-29}$ & $5.043^{+0.052}_{-0.053}$ & $0.666^{+0.026}_{-0.022}$ \\
HD 157338 & HD 157338 & $154^{+54}_{-47}$ & $24.1^{+13.0}_{-6.1}$ & $0.689^{+0.088}_{-0.062}$ \\
HD 16160 & HIP 12114 & $67.3^{+1.5}_{-1.5}$ & $16.37^{+0.28}_{-0.28}$ & $0.6427^{+0.0038}_{-0.0039}$ \\
HD 161797 & HIP 86974 & $238^{+60}_{-44}$ & $22.2^{+4.4}_{-3.2}$ & $0.427^{+0.069}_{-0.058}$ \\
HD 167215 & HD 167215 & $201.1^{+4.2}_{-4.2}$ & $7.205^{+0.077}_{-0.076}$ & $0.8522^{+0.00059}_{-0.00066}$ \\
HD 167215 & HD 167215 & $24.0^{+19.0}_{-4.3}$ & $7.99^{+1.00}_{-0.51}$ & $0.3^{+0.27}_{-0.20}$ \\
HD 17382 & HIP 13081 & $197.9^{+4.5}_{-4.5}$ & $5.967^{+0.072}_{-0.073}$ & $0.6563^{+0.0058}_{-0.0061}$ \\
HD 18445 & HIP 13769 C & $33.8^{+7.0}_{-4.3}$ & $1.209^{+0.016}_{-0.017}$ & $0.67^{+0.120}_{-0.098}$ \\
HD 185414 & HIP 96395 & $109.4^{+3.6}_{-3.7}$ & $6.02^{+0.10}_{-0.11}$ & $0.69345^{+0.00058}_{-0.00059}$ \\
HD 190406 & HIP 98819 & $67.4^{+2.0}_{-2.0}$ & $15.54^{+0.32}_{-0.31}$ & $0.4618^{+0.0054}_{-0.0053}$ \\
HD 211681 & HD 211681 & $76.4^{+3.3}_{-3.2}$ & $7.79^{+0.21}_{-0.19}$ & $0.4413^{+0.0076}_{-0.0064}$ \\
HD 215578 & HD 215578 & $1405^{+50}_{-46}$ & $25.8^{+1.3}_{-1.1}$ & $0.477^{+0.018}_{-0.016}$ \\
HD 239960 & HIP 110893 A & $54^{+10}_{-13}$ & $15.1^{+7.1}_{-5.0}$ & $0.61^{+0.11}_{-0.14}$ \\
HD 28185 & HD 28185 & $40^{+43}_{-28}$ & $15.9^{+7.3}_{-5.1}$ & $0.26^{+0.120}_{-0.093}$ \\
HD 29461 & HD 29461 & $88.8^{+4.8}_{-3.4}$ & $4.907^{+0.062}_{-0.064}$ & $0.596^{+0.0140}_{-0.0088}$ \\
HD 30649 & HD 30649 & $213.5^{+2.8}_{-2.8}$ & $16.19^{+0.25}_{-0.24}$ & $0.5878^{+0.0054}_{-0.0051}$ \\
HD 31412 & HD 31412 & $359.9^{+8.9}_{-9.0}$ & $19.78^{+0.3}_{-0.3}$ & $0.97822^{+0.00051}_{-0.00042}$ \\
HD 3795 & HIP 3185 & $370^{+100}_{-86}$ & $18.7^{+3.3}_{-2.7}$ & $0.296^{+0.064}_{-0.057}$ \\
HD 5470 & HD 5470 & $229.8^{+6.0}_{-6.0}$ & $8.08^{+0.11}_{-0.11}$ & $0.35749^{+0.00110}_{-0.00088}$ \\
HD 6558 & HIP 5189 & $220^{+260}_{-120}$ & $24.1^{+9.8}_{-7.4}$ & $0.35^{+0.12}_{-0.11}$ \\
HD 4747 & HIP 3850 & $49.2^{+1.6}_{-1.6}$ & $9.85^{+0.17}_{-0.17}$ & $0.731^{+0.0015}_{-0.0015}$ \\
HD 50639 & HD 50639 & $940^{+440}_{-300}$ & $31.0^{+29.0}_{-9.4}$ & $0.69^{+0.12}_{-0.09}$ \\
HD 65430 & HIP 39064 & $95.5^{+1.7}_{-1.7}$ & $4.24^{+0.038}_{-0.039}$ & $0.37796^{+0.00068}_{-0.00066}$ \\
HD 68017 & HIP 40118 & $33.8^{+5.7}_{-5.8}$ & $21.2^{+4.8}_{-4.4}$ & $0.432^{+0.075}_{-0.090}$ \\
HD 6872 B & HD 6872 B & $950^{+150}_{-110}$ & $18.4^{+6.6}_{-3.4}$ & $0.718^{+0.060}_{-0.043}$ \\
HD 72659 & HD 72659 & $230^{+730}_{-190}$ & $49^{+24}_{-19}$ & $0.21^{+0.19}_{-0.14}$ \\
HD 8375 & HD 8375 & $146.4^{+5.8}_{-6.0}$ & $0.44^{+0.0087}_{-0.0092}$ & $0.01707^{+0.00057}_{-0.00061}$ \\
HD 8375 & HD 8375 & $500^{+2200}_{-450}$ & $27^{+18}_{-10}$ & $0.28^{+0.25}_{-0.20}$ \\
HD 87359 & HD 87359 & $200^{+170}_{-93}$ & $28.1^{+9.0}_{-8.7}$ & $0.28^{+0.11}_{-0.13}$ \\
HD 8765 & HD 8765 & $43.0^{+1.5}_{-1.5}$ & $3.356^{+0.050}_{-0.052}$ & $0.3936^{+0.0099}_{-0.0095}$ \\
HIP 42220 & HIP 42220 & $229.9^{+8.0}_{-7.7}$ & $7.06^{+0.61}_{-0.34}$ & $0.721^{+0.035}_{-0.022}$ \\
HIP 63510 & HIP 63510 & $74.1^{+3.1}_{-3.2}$ & $4.775^{+0.057}_{-0.054}$ & $0.309^{+0.028}_{-0.025}$ \\
HD 16287 & HIP 12158 & $126.1^{+2.4}_{-2.4}$ & $0.1129^{+0.0011}_{-0.0011}$ & $0.20693^{+0.00027}_{-0.00027}$ \\
HD 40647 & HD 40647 & $193.9^{+5.3}_{-5.2}$ & $7.01^{+0.16}_{-0.15}$ & $0.5085^{+0.0054}_{-0.0047}$ \\
HD 103829 & HD 103829 & $1010^{+240}_{-170}$ & $20.4^{+3.7}_{-2.4}$ & $0.468^{+0.058}_{-0.045}$ \\
HD 139457 & HD 139457 & $240^{+430}_{-140}$ & $37^{+29}_{-15}$ & $0.64^{+0.16}_{-0.20}$ \\
HD 167665 & HD 167665 & $48.4^{+1.4}_{-1.4}$ & $5.39^{+0.079}_{-0.080}$ & $0.3421^{+0.0040}_{-0.0044}$ \\
HD 200565 & HD 200565 & $177^{+57}_{-34}$ & $7.22^{+1.0}_{-0.7}$ & $0.6^{+0.16}_{-0.14}$ \\
HD 217165 & HD 217165 & $520^{+35}_{-25}$ & $24.1^{+3.9}_{-2.0}$ & $0.499^{+0.062}_{-0.039}$ \\
HIP 52942 A & HIP 52942 A & $201^{+42}_{-24}$ & $7.99^{+0.15}_{-0.16}$ & $0.901^{+0.021}_{-0.016}$ \\
\bottomrule
\end{longtable*}

\section{Long-Term Trends}

We record all linear and parabolic trends recovered by \texttt{RVSearch} in Table 5.

\begin{longtable*}{llrr}
\caption{Long Term Trends} \\
\toprule
\midrule

CPS Name & Lit. Name & $\dot \gamma$ [m s$^{-1}$ d$^{-1}$] & $\ddot \gamma$ [m s$^{-1}$ d$^{-2}$] \\
\toprule
\endfirsthead
\caption[]{Long Term Trends (Continued)} \\
\toprule
\midrule
CPS Name & Lit. Name & $\dot \gamma$ [m s$^{-1}$ d$^{-1}$] & $\ddot \gamma$ [m s$^{-1}$ d$^{-2}$] \\
\toprule
\endhead
HD 100623 & HIP 56452 & $0.00475\pm 0.00028$ & 0 \\
HD 110315 & HIP 61901 & $-0.05545\pm 0.00039$ & $1.85*10^{-6}\pm 1.1*10^{-7}$ \\
HD 110537 & HIP 62039 & $0.01936\pm 0.00051$ & 0 \\
HD 114174 & HIP 64150 & $0.16278\pm 0.00058$ & $-2.5*10^{-6}\pm 1.4*10^{-7}$ \\
HD 115404 A & HIP 64797 A & $0.0156\pm 0.0011$ & 0 \\
HD 131156 & HD 131156 & $0.066\pm 0.0019$ & $2.21*10^{-6}\pm 2.9*10^{-7}$ \\
HD 136925 & HD 136925 & $0.0026\pm 0.0012$ & $7.6*10^{-7}\pm 6.2*10^{-7}$ \\
HD 1388 & HIP 1444 & $0.05279\pm 0.00094$ & $-4.22*10^{-6}\pm 2.2*10^{-7}$ \\
HD 140538 A & HIP 77052 & $-0.00647\pm 0.00038$ & 0 \\
HD 145934 & HD 145934 & $-0.0589\pm 0.0011$ & $-2.05*10^{-6}\pm 3.3*10^{-7}$ \\
HD 145958 A & HD 145958 A & $-0.01751\pm 0.0005$ & 0 \\
HD 145958 B & HD 145958 B & $0.01627\pm 0.00038$ & $4.1*10^{-7}\pm 1.1*10^{-7}$ \\
HD 146362 B & HD 146362 B & $0.0139\pm 0.0011$ & 0 \\
HD 149806 & HIP 81375 & $0.00597\pm 0.00042$ & 0 \\
HD 153557 & HIP 83020 A & $0.0092\pm 0.0016$ & 0 \\
HD 156668 & HD 156668 & $-0.00074\pm 0.00014$ & 0 \\
HD 159062 & HD 159062 & $-0.0361\pm 0.0002$ & $1.14*10^{-6}\pm 1.5*10^{-7}$ \\
HD 163489 & HD 163489 & $-0.00596\pm 0.00081$ & 0 \\
HD 165401 & HIP 88622 & $0.14734\pm 0.00098$ & $2.63*10^{-6}\pm 6*10^{-7}$ \\
HD 168443 & HD 168443 & $-0.00795\pm 0.00029$ & 0 \\
HD 17230 & HD 17230 & $-0.00587\pm 0.00038$ & 0 \\
HD 173739 & HIP 12929 & $0.01702\pm 0.00054$ & 0 \\
HD 173740 & HIP 91772 & $-0.02088\pm 0.00054$ & 0 \\
HD 179957 & HD 179957 & $-0.00632\pm 0.00017$ & 0 \\
HD 179958 & HIP 94336 A & $0.00533\pm 0.00022$ & 0 \\
HD 180617 & HD 180617 & $-0.00122\pm 0.00036$ & 0 \\
HD 18143 & HIP 13642 A & $0.01043\pm 0.00044$ & 0 \\
HD 182488 & HIP 95319 & $-0.00539\pm 0.00035$ & $-9.48*10^{-7}\pm 9*10^{-8}$ \\
HD 186408 & HIP 96895 & $-0.00459\pm 0.0002$ & 0 \\
HD 187123 & HD 187123 & $-0.00129\pm 0.00022$ & 0 \\
HD 188512 & HIP 98036 & $0.00207\pm 0.00023$ & 0 \\
HD 190067 & HIP 98677 & $0.01134\pm 0.00027$ & 0 \\
HD 191408 & HIP 99461 & $0.00831\pm 0.00021$ & 0 \\
HD 19467 & HD 19467 & $-0.00392\pm 0.0006$ & 0 \\
HD 195019 & HD 195019 & $0.00385\pm 0.00057$ & 0 \\
HD 195564 & HIP 101345 & $-0.091\pm 0.019$ & $-5.1*10^{-6}\pm 1.9*10^{-6}$ \\
HD 196201 & HD 196201 & $0.23\pm 0.0029$ & $-1.03*10^{-5}\pm 7.6*10^{-7}$ \\
HD 200968 & HIP 104239 A & $-0.02717\pm 0.00061$ & $-8*10^{-7}\pm 2.3*10^{-7}$ \\
HD 201091 & HIP 104214 & $-0.00737\pm 0.00036$ & $-3.19*10^{-7}\pm 5.4*10^{-8}$ \\
HD 201092 & HIP 104217 & $0.01301\pm 0.00038$ & $-4.02*10^{-7}\pm 5.9*10^{-8}$ \\
HD 21019 A & HD 21019 A & $-0.00452\pm 0.00096$ & 0 \\
HD 213519 & HD 213519 & $-0.00768\pm 0.00068$ & 0 \\
HD 219834 B & HD 219834 B & $0.00531\pm 0.00036$ & 0 \\
HD 23439 & HD 23439 & $0.00208\pm 0.00017$ & 0 \\
HD 24040 & HD 24040 & $0.00581\pm 0.00044$ & $-6.6*10^{-7}\pm 1.2*10^{-7}$ \\
HD 24496 & HIP 18267 & $-0.01997\pm 0.00036$ & 0 \\
HD 24916 & HD 24916 & $-0.00386\pm 0.00067$ & 0 \\
HD 3074 A & HD 3074 A & $-0.00977\pm 0.00066$ & 0 \\
HD 32923 & HIP 23835 & $0.00513\pm 0.00046$ & 0 \\
HD 34445 & HD 34445 & $-0.00328\pm 0.00043$ & 0 \\
HD 34721 & HIP 24786 & $0.003\pm 0.00027$ & 0 \\
HD 38230 & HIP 27207 & $0.01111\pm 0.00021$ & 0 \\
HD 38 A & HD 38 A & $0.0039\pm 0.0012$ & 0 \\
HD 38 B & HD 38 B & $-0.0072\pm 0.0012$ & 0 \\
HD 39715 & HIP 27918 & $-0.0076\pm 0.001$ & $1.72*10^{-6}\pm 3.4*10^{-7}$ \\
HD 40397 & HIP 28267 & $-0.02912\pm 0.00029$ & $-8.19*10^{-7}\pm 8.6*10^{-8}$ \\
HD 45184 & HD 45184 & $-0.00111\pm 0.0002$ & 0 \\
HD 4614 & HIP 3821 A & $0.02037\pm 0.00048$ & $-3.96*10^{-7}\pm 5.9*10^{-8}$ \\
HD 4614 B & HIP 3821 B & $-0.02807\pm 0.00054$ & 0 \\
HD 53665 & HD 53665 & $0.0104\pm 0.001$ & 0 \\
HD 6101 & HIP 4849 A & $0.259\pm 0.022$ & 0 \\
HD 63754 & HD 63754 & $-0.00599\pm 0.00069$ & $-6.1*10^{-7}\pm 1.6*10^{-7}$ \\
HD 65277 & HIP 38931 & $-0.00573\pm 0.00019$ & 0 \\
HD 6734 & HD 6734 & $-0.04096\pm 0.00069$ & $-4.54*10^{-6}\pm 1.5*10^{-7}$ \\
HD 79210 & HIP 45343 & $0.01045\pm 0.00071$ & 0 \\
HD 79211 & HIP 120005 & $-0.00965\pm 0.00087$ & 0 \\
HD 8375 & HD 8375 & $0.66\pm 0.32$ & 0 \\
HD 8648 & HIP 6653 & $0.02094\pm 0.0004$ & 0 \\
HD 88986 & HD 88986 & $-0.00143\pm 0.00086$ & $1.78*10^{-6}\pm 1.9*10^{-7}$ \\
HD 91204 & HD 91204 & $-0.3961\pm 0.0025$ & 0 \\
HD 9540 A & HIP 7235 & $-0.00314\pm 0.00059$ & 0 \\
HD 98618 & HD 98618 & $-0.0076\pm 0.00059$ & 0 \\
GL 397 & GL 397 & $0.00574\pm 0.00089$ & 0 \\
HIP 57050 & HIP 57050 & $0.0002\pm 0.0014$ & 0 \\
HIP 71898 & HIP 71898 & $0.0213\pm 0.0012$ & 0 \\
HD 71881 & HD 71881 & $-0.0314\pm 0.0012$ & 0 \\
HD 88218 & HD 88218 & $-0.1164\pm 0.0022$ & $-4.59*10^{-6}\pm 4.3*10^{-7}$ \\
HD 89391 & HD 89391 & $0.0116\pm 0.001$ & 0 \\
HD 105618 & HD 105618 & $0.0064\pm 0.0015$ & 0 \\
HD 111484 A & HD 111484 A & $-0.0862\pm 0.0054$ & 0 \\
GL 528 B & GL 528 B & $0.0812\pm 0.003$ & 0 \\
HD 120476 A & HD 120476 A & $-0.0708\pm 0.0011$ & $-3.18*10^{-6}\pm 3.8*10^{-7}$ \\
HD 129814 & HD 129814 & $-0.0431\pm 0.0011$ & $2.73*10^{-5}\pm 3.5*10^{-7}$ \\
HD 131509 & HD 131509 & $-0.01919\pm 0.00057$ & 0 \\
HD 147231 & HD 147231 & $-0.0754\pm 0.001$ & 0 \\
HD 151995 & HIP 82389 & $-0.00643\pm 0.00086$ & 0 \\
HD 156826 & HD 156826 & $-0.0242\pm 0.00087$ & 0 \\
HD 180684 & HD 180684 & $0.1709\pm 0.0019$ & $-6.73*10^{-6}\pm 5.8*10^{-7}$ \\
HD 201203 & HD 201203 & $-0.1447\pm 0.0096$ & $-3.14*10^{-5}\pm 4.6*10^{-6}$ \\
HD 183263 & HD 183263 & $-0.00659\pm 0.00044$ & 0 \\
\bottomrule
\end{longtable*}

\section{Data}

We include a sample table of RVs in Table 6.

\begin{longtable*}{lrrrrr}
\caption{Sample of RV Data} \\
\toprule 
\midrule 

Name & Inst. & BJD - 2455000 & $RV$ [m/s] & $\sigma_{RV}$ [m/s] & S-Value \\ 
\toprule 
\endfirsthead 
\caption[]{Sample of RV Data (Continued)} \\
\toprule 
\midrule 
Name & Inst. & BJD - 2455000 & $RV$ [m/s] & $\sigma_{RV}$ [m/s] & S-Value \\ 
\toprule 
\endhead 
HD 156668 & HIRES pre-2004 & -2167.0766 & -0.45 & 1.62 & nan \\ 
HD 156668 & HIRES pre-2004 & -1925.896 & -7.47 & 1.68 & nan \\ 
HD 156668 & HIRES pre-2004 & -1803.1301 & -3.06 & 1.72 & nan \\ 
HD 156668 & HIRES post-2004  & -1521.0223 & -0.01 & 1.02 & 0.201 \\ 
HD 156668 & HIRES post-2004  & -1452.0904 & -3.86 & 0.98 & 0.214 \\ 
HD 156668 & HIRES post-2004  & -1395.1611 & -4.85 & 0.98 & 0.212 \\ 
HD 156668 & HIRES post-2004  & -1192.8559 & 3.43 & 1.09 & 0.222 \\ 
HD 156668 & HIRES post-2004  & -1067.0813 & 1.23 & 0.99 & 0.229 \\ 
HD 156668 & HIRES post-2004  & -1039.086 & 3.95 & 0.92 & 0.22 \\ 
HD 156668 & HIRES post-2004  & -1038.1904 & 0.36 & 0.95 & 0.219 \\ 
HD 156668 & HIRES post-2004  & -1018.2294 & 3.12 & 0.88 & 0.231 \\ 
HD 156668 & HIRES post-2004  & -1017.1234 & 1.06 & 0.93 & 0.233 \\ 
HD 156668 & HIRES post-2004  & -1016.1809 & 6.45 & 0.87 & 0.23 \\ 
\bottomrule 
\end{longtable*}

\section{False Positives}

We record all RVSearch-detected false positives in Table 7. The 'cause' column denotes why a signal was labeled as a false positive. 'A' refers to a long-period magnetic activity cycle, 'R' refers to stellar rotation, and 'N' refers to an annual and/or instrumental systematic. Long-period instrumental systematics are occasionally caused by offsets between dewars in the Lick data. Several of these false positives correspond to reported planets in the literature, or to stars that have been discussed extensively in the literature. We elaborate on each of these cases in the subsections below.

\begin{longtable*}{llrrr}
\caption{False Positives} \\
\toprule
\midrule

CPS Name & Lit. Name & $P$ [d] & $K$ [m s$^{-1}$] & Cause \\
\endfirsthead
\caption[]{False Positives (Continued)} \\
\toprule
\midrule
CPS Name & Lit. Name & $P$ [d] & $K$ [m s$^{-1}$] & Cause \\
\toprule
\endhead
HD 103932 & HIP 58345 & $3660.0^{+190.0}_{-180.0}$ & $5.5^{+1.30}_{-0.98}$ & A \\
HD 10476 & HIP 7981 & $360.74^{+0.39}_{-0.64}$ & $2.84^{+0.59}_{-0.58}$ & N \\
HD 115617 & 61 Vir & $122.67^{+0.12}_{-0.14}$ & $1.94^{+0.30}_{-0.27}$ & N \\
HD 120467 & HIP 67487 & $66.464^{+0.056}_{-0.066}$ & $5.1^{+0.96}_{-0.82}$ & R \\
HD 122064 & HIP 68184 & $3100.0^{+310.0}_{-180.0}$ & $1.75^{+0.87}_{-0.39}$ & A \\
HD 136352 & HD 136352 & $244.5^{+1.5}_{-1.8}$ & $11.0^{+69.0}_{-3.3}$ & N \\
HD 136713 & HIP 75253 & $2710.0^{+140.0}_{-130.0}$ & $5.0^{+1.30}_{-0.88}$ & A \\
HD 136925 & HD 136925 & $311.2^{+0.55}_{-0.48}$ & $6.66^{+0.76}_{-0.73}$ & A \\
HD 139323 & HIP 76375 & $3310.0^{+110.0}_{-89.0}$ & $4.0^{+1.10}_{-0.97}$ & A \\
HD 140538 A & HD 140538 A & $1417.0^{+23.0}_{-25.0}$ & $4.73^{+0.70}_{-0.72}$ & A \\
HD 14412 & HIP 10798 & $364.0^{+1.5}_{-2.2}$ & $3.0^{+22.0}_{-1.1}$ & N \\
HD 144579 & HIP 78775 & $91.8^{+0.26}_{-0.34}$ & $2.17^{+0.58}_{-0.41}$ & N \\
HD 145675 & 14 Her & $3400.0^{+140.0}_{-240.0}$ & $3.52^{+1.10}_{-0.71}$ & A \\
HD 1461 & HD 1461 & $72.98^{+0.12}_{-0.10}$ & $1.33^{+0.25}_{-0.22}$ & N \\
HD 1461 & HD 1461 & $378.2^{+3.0}_{-2.2}$ & $3.8^{+1.3}_{-1.0}$ & A \\
HD 1461 & HD 1461 & $4060.0^{+280.0}_{-210.0}$ & $2.12^{+0.24}_{-0.24}$ & N \\
HD 146233 & HIP 79672 & $2426.0^{+60.0}_{-42.0}$ & $5.55^{+0.70}_{-0.67}$ & A \\
HD 154345 & HD 154345 & $2763.0^{+65.0}_{-69.0}$ & $5.5^{+1.10}_{-0.91}$ & A \\
HD 158633 & HIP 85235 & $367.5^{+1.2}_{-1.2}$ & $3.24^{+0.27}_{-0.26}$ & N \\
HD 161797 & HIP 86974 & $52.386^{+0.049}_{-0.049}$ & $1.69^{+0.32}_{-0.28}$ & R \\
HD 168009 & HIP 89474 & $368.57^{+1.00}_{-0.95}$ & $3.53^{+0.35}_{-0.32}$ & N \\
HD 168723 & HIP 89962 & $795.2^{+2.2}_{-11.0}$ & $8.1^{+2.3}_{-1.8}$ & N \\
HD 185144 & HIP 96100 & $2257.0^{+31.0}_{-30.0}$ & $1.97^{+0.13}_{-0.13}$ & A \\
HD 185144 & HIP 96100 & $347.11^{+0.92}_{-0.82}$ & $1.04^{+0.19}_{-0.18}$ & N \\
HD 18803 & HIP 14150 & $1960.0^{+36.0}_{-28.0}$ & $5.73^{+0.70}_{-0.66}$ & A \\
HD 190360 & HD 190360 & $90.34^{+0.11}_{-0.13}$ & $1.49^{+0.17}_{-0.18}$ & N \\
HD 190406 & HIP 98819 & $995.8^{+4.3}_{-4.3}$ & $8.22^{+0.73}_{-0.68}$ & A \\
HD 190406 & HIP 98819 & $4092.0^{+100.0}_{-73.0}$ & $9.1^{+1.2}_{-1.1}$ & A \\
HD 192310 & HD 192310 & $1630.0^{+51.0}_{-53.0}$ & $1.95^{+0.49}_{-0.36}$ & A \\
HD 193202 & HIP 99427 & $574.6^{+1.9}_{-6.7}$ & $3.8^{+0.48}_{-0.47}$ & N \\
HD 19373 & HIP 14632 & $367.5^{+1.90}_{-0.87}$ & $3.0^{+0.49}_{-0.45}$ & N \\
HD 195564 & HIP 101345 & $26000.0^{+16000.0}_{-11000.0}$ & $160.0^{+110.0}_{-120.0}$ & N \\
HD 197076 & HIP 102040 & $1620.0^{+29.0}_{-34.0}$ & $4.45^{+0.44}_{-0.42}$ & A \\
HD 197076 & HIP 102040 & $23.6803^{+0.0072}_{-0.0066}$ & $3.16^{+0.55}_{-0.49}$ & R \\
HD 199960 & HIP 103862 & $2357.0^{+65.0}_{-66.0}$ & $2.98^{+0.72}_{-0.48}$ & A \\
HD 201091 & HIP 104214 & $2571.0^{+55.0}_{-58.0}$ & $2.37^{+0.33}_{-0.32}$ & A \\
HD 201092 & HIP 104217 & $49.038^{+0.036}_{-0.032}$ & $1.73^{+0.26}_{-0.26}$ & R \\
HD 20165 & HIP 15099 & $2759.0^{+97.0}_{-91.0}$ & $4.56^{+0.57}_{-0.55}$ & A \\
HD 211080 & HIP 109836 & $677.6^{+7.2}_{-5.5}$ & $11.8^{+2.1}_{-1.8}$ & A \\
HD 213042 & HIP 110996 & $2530.0^{+210.0}_{-90.0}$ & $6.2^{+2.0}_{-1.4}$ & A \\
HD 214683 & HIP 111888 & $16.59^{+0.024}_{-0.540}$ & $33.0^{+120.0}_{-17.0}$ & R \\
HD 216520 & HIP 112527 & $181.63^{+0.64}_{-0.47}$ & $1.87^{+0.29}_{-0.29}$ & N \\
HD 216520 & HIP 112527 & $35.466^{+0.015}_{-0.016}$ & $2.42^{+0.27}_{-0.28}$ & R \\
HD 216520 & HIP 112527 & $5500.0^{+5100.0}_{-310.0}$ & $3.04^{+0.69}_{-0.46}$ & A \\
HD 217014 & 51 Peg & $100000.0^{+170000.0}_{-63000.0}$ & $10.0^{+280.0}_{-1.5}$ & N \\
HD 218868 & HIP 114456 & $1824.0^{+43.0}_{-56.0}$ & $7.3^{+1.6}_{-1.0}$ & A \\
HD 219134 & HD 219134 & $192.06^{+0.49}_{-0.40}$ & $2.0^{+0.21}_{-0.20}$ & N \\
HD 219134 & HD 219134 & $10230.0^{+830.0}_{-580.0}$ & $8.8^{+2.0}_{-1.7}$ & N \\
HD 219134 & HD 219134 & $364.3^{+1.9}_{-2.3}$ & $1.66^{+0.35}_{-0.31}$ & N \\
HD 22049 & eps Eri & $773.4^{+4.7}_{-4.8}$ & $4.1^{+0.70}_{-0.68}$ & N \\
HD 23439 & HD 23439 & $45.683^{+0.023}_{-0.022}$ & $2.63^{+0.40}_{-0.38}$ & R \\
HD 24496 & HIP 18267 & $182.13^{+0.87}_{-0.57}$ & $3.7^{+1.10}_{-0.68}$ & N \\
HD 26151 & HD 26151 & $113.57^{+0.13}_{-0.15}$ & $5.65^{+0.46}_{-0.46}$ & N \\
HD 26151 & HD 26151 & $32.879^{+0.015}_{-0.017}$ & $3.83^{+0.47}_{-0.49}$ & R \\
HD 26965 & HD 26965 & $42.305^{+0.015}_{-0.019}$ & $1.82^{+0.43}_{-0.31}$ & R \\
HD 26965 & HD 26965 & $3560.0^{+200.0}_{-580.0}$ & $1.89^{+0.37}_{-0.32}$ & A \\
HD 32147 & HD 32147 & $3444.0^{+91.0}_{-81.0}$ & $1.71^{+0.19}_{-0.18}$ & A \\
HD 32147 & HD 32147 & $51.997^{+0.078}_{-0.039}$ & $0.99^{+0.21}_{-0.19}$ & N \\
HD 34445 & HD 34445 & $214.74^{+0.49}_{-0.52}$ & $5.39^{+0.71}_{-0.69}$ & N \\
HD 34445 & HD 34445 & $117.69^{+0.16}_{-0.19}$ & $4.63^{+0.61}_{-0.58}$ & N \\
HD 34445 & HD 34445 & $49.231^{+0.048}_{-0.038}$ & $3.68^{+0.61}_{-0.53}$ & R \\
HD 36003 & HIP 23311 & $2790.0^{+140.0}_{-110.0}$ & $3.27^{+0.49}_{-0.46}$ & A \\
HD 3651 & HD 3651 & $5140.0^{+220.0}_{-290.0}$ & $2.45^{+0.97}_{-0.57}$ & A \\
HD 38858 & HD 38858 & $3113.0^{+82.0}_{-79.0}$ & $4.43^{+0.73}_{-0.64}$ & A \\
HD 42618 & HD 42618 & $4040.0^{+390.0}_{-260.0}$ & $2.72^{+0.27}_{-0.26}$ & A \\
HD 42618 & HD 42618 & $388.0^{+1.8}_{-2.2}$ & $1.64^{+0.28}_{-0.24}$ & N \\
HD 45184 & HD 45184 & $2479.0^{+110.0}_{-85.0}$ & $2.91^{+0.60}_{-0.56}$ & A \\
HD 4614 & HIP 3821 A & $91.011^{+0.093}_{-0.086}$ & $2.29^{+0.45}_{-0.39}$ & N \\
HD 4628 & HIP 3765 & $2468.0^{+16.0}_{-67.0}$ & $8.0^{+26.0}_{-5.2}$ & A \\
HD 48682 & HIP 32480 & $923.0^{+15.0}_{-28.0}$ & $4.1^{+0.84}_{-0.73}$ & A \\
HD 52265 & HD 52265 & $1383.0^{+25.0}_{-15.0}$ & $6.5^{+2.3}_{-1.2}$ & N \\
HD 52711 & HIP 34017 & $43000.0^{+140000.0}_{-27000.0}$ & $5.55^{+1.10}_{-0.74}$ & A \\
HD 55575 & HIP 35136 & $52.178^{+0.048}_{-0.045}$ & $2.6^{+0.37}_{-0.35}$ & R \\
HD 68988 & HD 68988 & $1922.0^{+49.0}_{-57.0}$ & $5.63^{+0.79}_{-0.74}$ & A \\
HD 69830 & HD 69830 & $200.61^{+0.68}_{-0.69}$ & $1.68^{+0.20}_{-0.21}$ & N \\
HD 69830 & HD 69830 & $381.8^{+4.9}_{-6.0}$ & $1.72^{+0.31}_{-0.25}$ & N \\
HD 75732 & 55 Cnc & $6110.0^{+280.0}_{-190.0}$ & $16.3^{+2.5}_{-2.9}$ & A \\
HD 75732 & 55 Cnc & $1966.0^{+23.0}_{-22.0}$ & $5.17^{+0.41}_{-0.36}$ & N \\
HD 7924 & HD 7924 & $24.4459^{+0.0090}_{-0.0091}$ & $1.59^{+0.18}_{-0.17}$ & A \\
HD 7924 & HD 7924 & $383.1^{+2.0}_{-1.8}$ & $1.69^{+0.27}_{-0.22}$ & N \\
HD 82943 & HD 82943 & $1078.0^{+15.0}_{-14.0}$ & $4.6^{+1.10}_{-0.94}$ & N \\
HD 9407 & HD 9407 & $121.91^{+0.140}_{-0.095}$ & $1.4^{+0.31}_{-0.25}$ & N \\
HD 9407 & HD 9407 & $178.62^{+0.46}_{-0.47}$ & $0.98^{+0.18}_{-0.17}$ & N \\
HD 95128 & 47 UMa & $18700.0^{+11000.0}_{-4000.0}$ & $12.4^{+1.4}_{-1.1}$ & N \\
HD 95128 & 47 UMa & $387.7^{+2.3}_{-2.0}$ & $2.34^{+0.37}_{-0.31}$ & N \\
HD 95735 & GJ 411 & $214.59^{+0.99}_{-0.64}$ & $1.45^{+0.42}_{-0.27}$ & N \\
HD 97101 & HD 97101 & $39.639^{+0.017}_{-0.019}$ & $3.39^{+0.29}_{-0.30}$ & R \\
HD 97101 & HD 97101 & $3080.0^{+170.0}_{-130.0}$ & $3.76^{+0.84}_{-0.64}$ & A \\
HD 97101 & HD 97101 & $19.6244^{+0.0064}_{-0.0064}$ & $1.71^{+0.30}_{-0.29}$ & R \\
HD 97658 & HD 97658 & $3660.0^{+170.0}_{-160.0}$ & $2.82^{+0.54}_{-0.54}$ & A \\
HD 9826 & ups And & $6290.0^{+140.0}_{-130.0}$ & $9.5^{+1.2}_{-1.2}$ & A \\
HD 9826 & ups And & $174.55^{+0.39}_{-0.34}$ & $5.07^{+0.93}_{-0.95}$ & N \\
HD 99491 & HIP 55846 & $2183.0^{+40.0}_{-37.0}$ & $6.49^{+0.62}_{-0.62}$ & A \\
HD 99492 & HD 99492 & $3680.0^{+190.0}_{-180.0}$ & $3.48^{+0.56}_{-0.54}$ & A \\
HD 9986 & HIP 7585 & $3103.0^{+51.0}_{-48.0}$ & $13.19^{+0.61}_{-0.62}$ & A \\
GL 388 & GL 388 & $2.068246^{+5.1e-05}_{-5.3e-05}$ & $27.6^{+5.6}_{-3.7}$ & R \\
GL 412 A & GL 412 A & $36.887^{+0.014}_{-0.017}$ & $3.14^{+0.81}_{-0.68}$ & R \\
GL 687 & GL 687 & $28.749^{+0.013}_{-0.012}$ & $2.92^{+0.51}_{-0.47}$ & N \\
HIP 42220 & HIP 42220 & $3600.0^{+2100.0}_{-770.0}$ & $41.0^{+50.0}_{-17.0}$ & N \\
HIP 74995 & HIP 74995 & $30.499^{+0.028}_{-0.021}$ & $2.71^{+0.51}_{-0.46}$ & R \\
GL 876 & GL 876 & $15.04336^{+0.00069}_{-0.00069}$ & $20.31^{+0.59}_{-0.59}$ & N \\
GL 876 & GL 876 & $10.014^{+0.0010}_{-0.0011}$ & $4.77^{+0.57}_{-0.59}$ & N \\
\bottomrule
\end{longtable*}

\subsection{HD 115617}

\begin{figure*}[ht!]
\begin{center}
\includegraphics[width=0.6\textwidth]{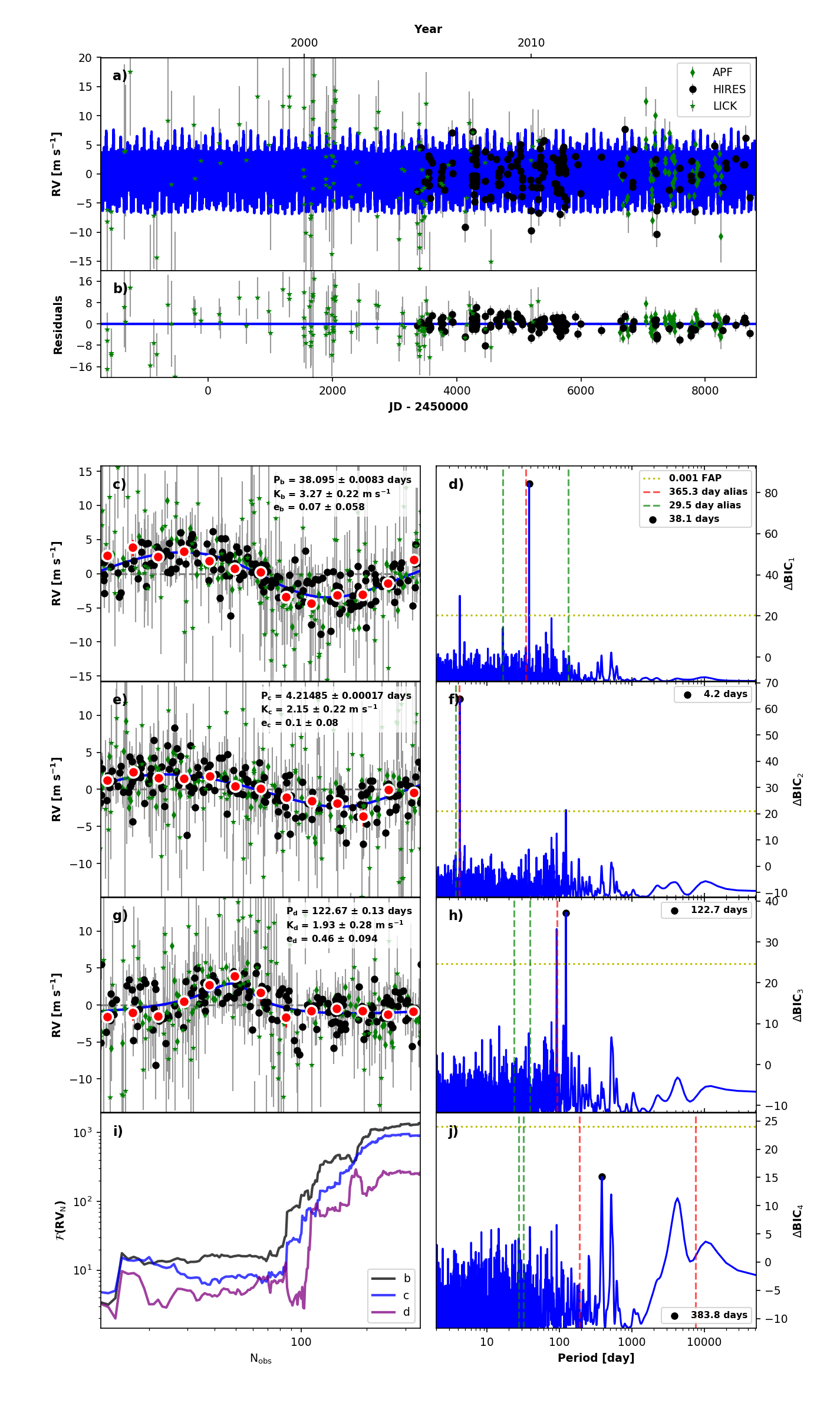}
\caption{\texttt{RVSearch} summary plot for HD 115617. See Figure  \ref{fig:hip109388_summary} for plot description. Note the nearly equivalent-height peaks at 1/3 and 1/4 year in panel h, corresponding to the 124 day reported planet. Panel j shows that there is residual power at 1 year after subtracting the 122 day signal, suggesting the presence of yearly systematic noise in the data.}
\label{fig:115617_summary}
\end{center}
\end{figure*}

\begin{figure*}[ht!]
\begin{center}
\includegraphics[width=0.9\textwidth]{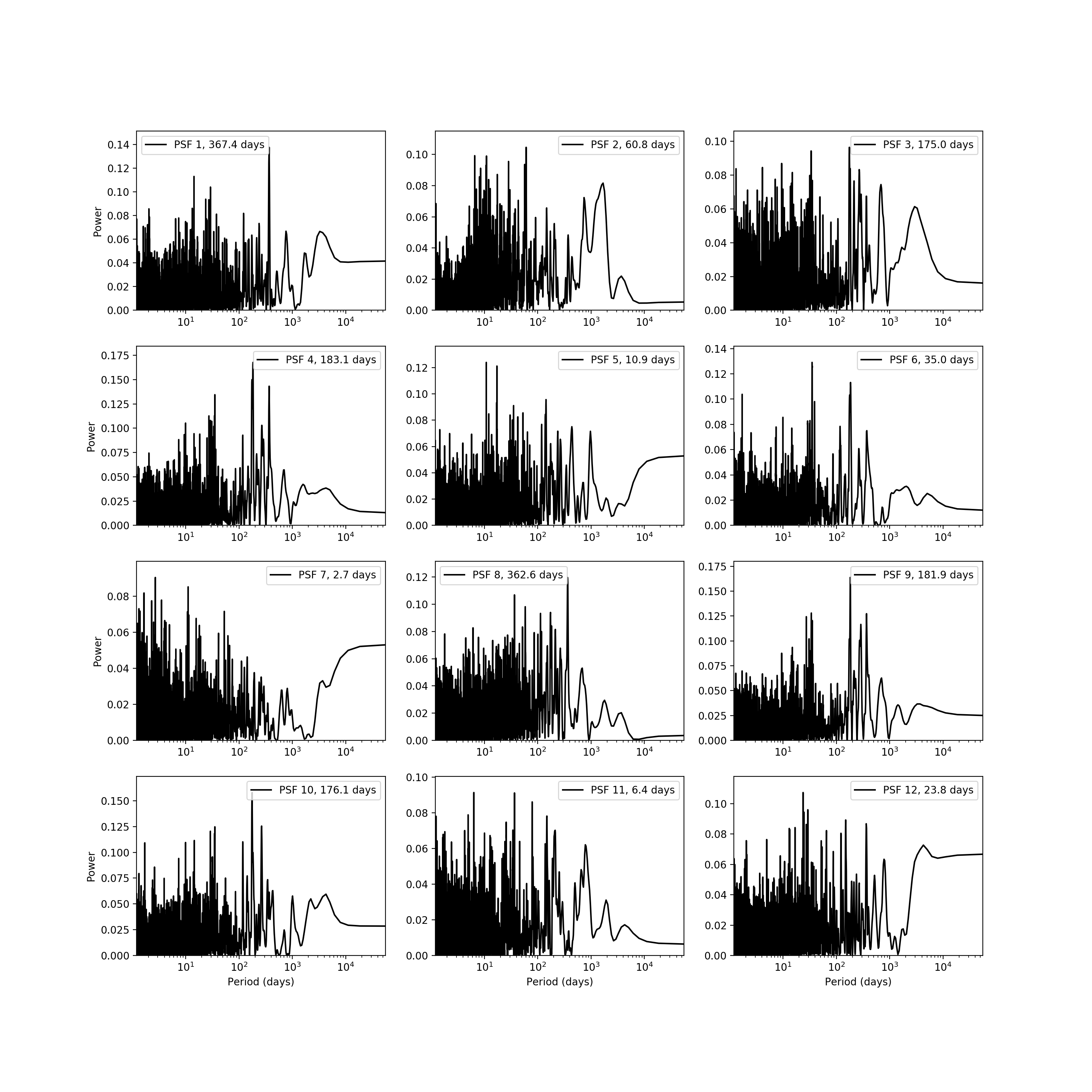}
\caption{PSF Lomb--Scargle periodograms for HD 115617. Each panel corresponds to a Doppler code PSF fitting parameter.}
\label{fig:115617_periodogram}
\end{center}
\end{figure*}

\begin{figure*}[ht!]
\begin{center}
\includegraphics[width=0.9\textwidth]{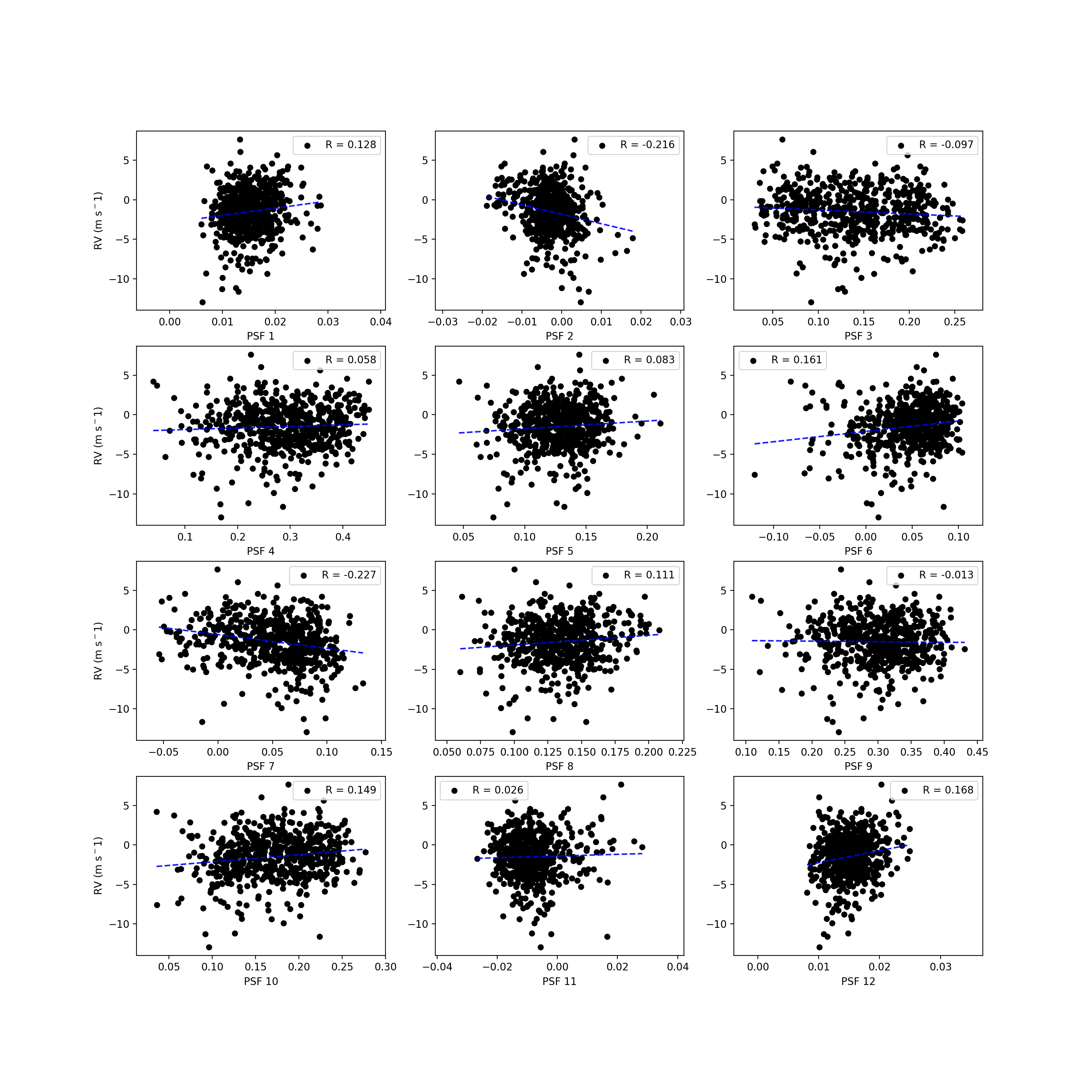}
\caption{PSF correlation plots for the candidate HD 115617 d. Each panel corresponds to a Doppler code PSF fitting parameter, with PSF value on the x-axis and RV without the signatures of the inner two planets on the y-axis. Dashed blue lines are least-squares linear fits. R is the Pearson correlation value; multiple PSF parameters have $|R|$ $>$ 0.15.}
\label{fig:115617_correlate}
\end{center}
\end{figure*}

\cite{Vogt10} reported three planets orbiting this star, with periods of 4.2, 38, and 124 days. \texttt{RVSearch} recovered signals at all three periods. However, the 124 day signal (1/3rd of a year) has a strong harmonic at 1/4th of a year, and there is significant residual power at roughly one year, as seen in panels h and j of Figure \ref{fig:115617_summary}. We investigated this candidate by computing periodograms for the 12 HIRES PSF parameters computed for each RV measurement, and found periodicity at 1 yr and harmonics of 1 yr for several parameters, as seen in Figure \ref{fig:115617_periodogram}. Additionally, several of these PSF parameters correlate strongly with the corresponding RVs, after subtracting the RV models of the two inner planets, as seen in Figure \ref{fig:115617_correlate}. Therefore, we designated the 124 day signal as a yearly systematic.

\subsection{HD 154345}

\begin{figure*}[ht!]
\begin{center}
\includegraphics[width=0.9\textwidth]{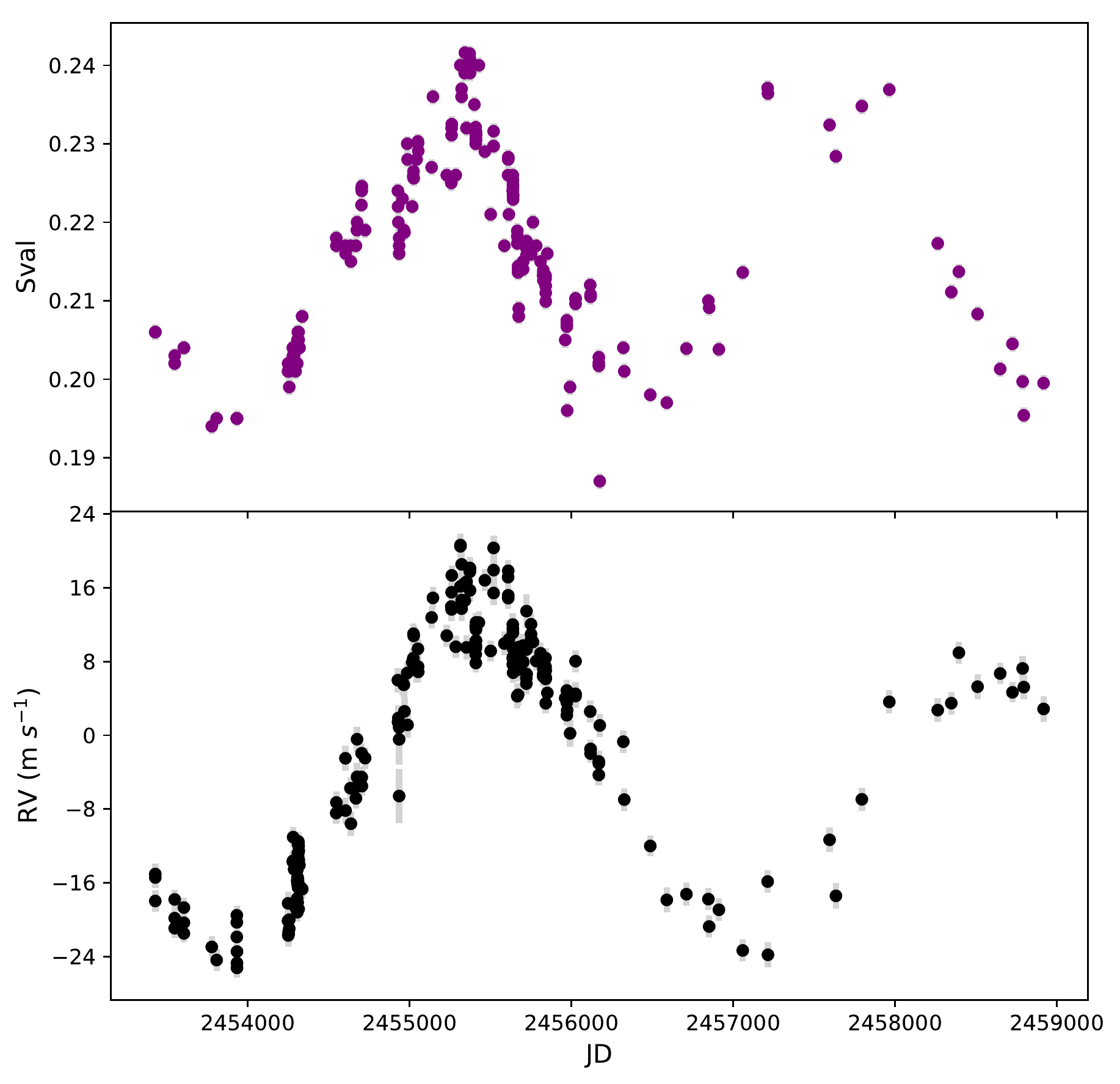}
\caption{HIRES post-upgrade RV and S-value activity timeseries for HD 154345. Note that the two datasets share minima and appear to be in phase when post-upgrade observations began, but have drifted completely out of phase over the following 23 years.}
\label{fig:154345_timeseries}
\end{center}
\end{figure*}

\begin{figure*}[ht!]
\begin{center}
\includegraphics[width=0.6\textwidth]{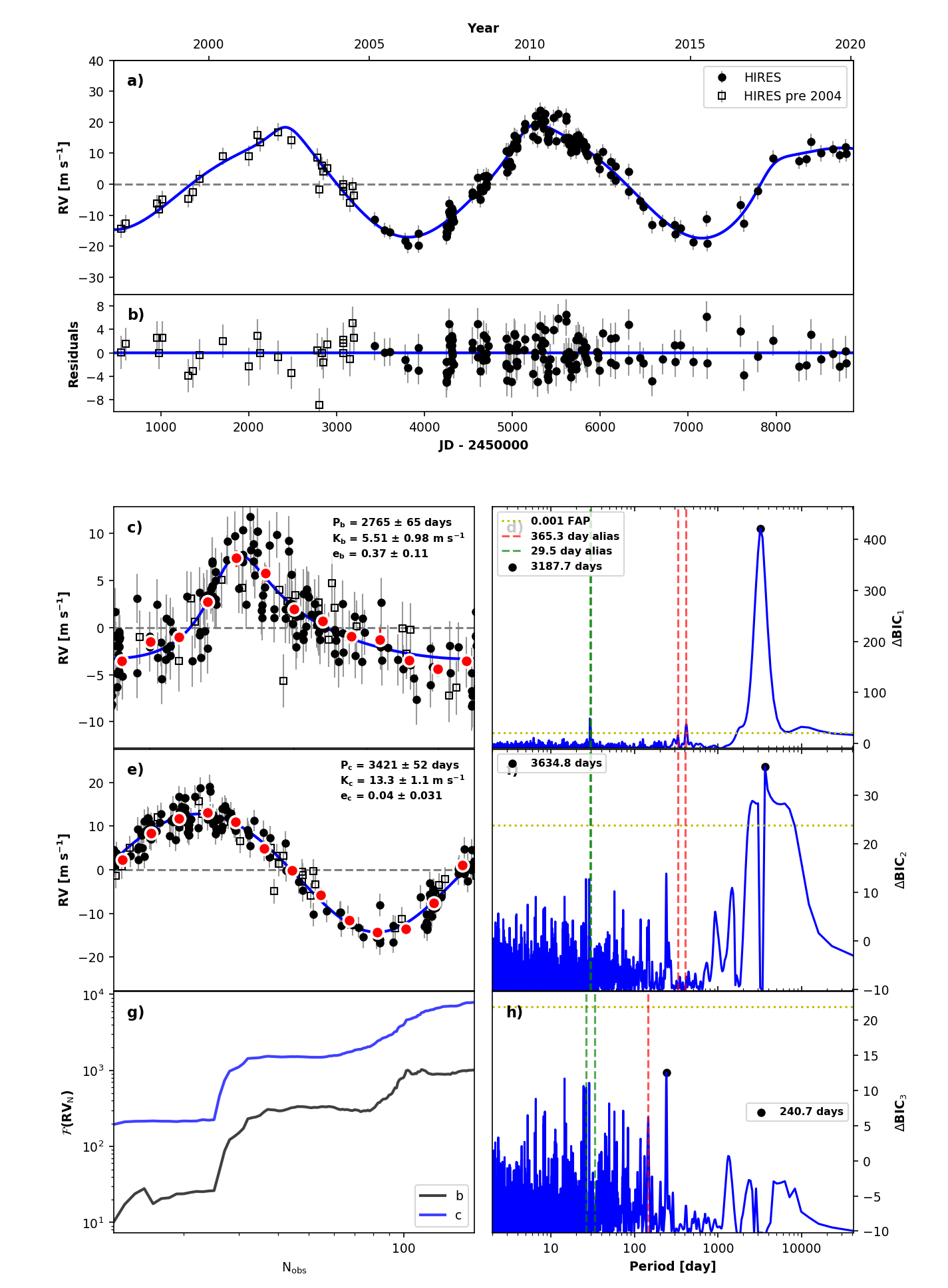}
\caption{\texttt{RVSearch} summary plot for HD 154345; see Figure \ref{fig:hip109388_summary} for description. \texttt{RVSearch} first recovered a strong signal at ~9 years, but then recovered additional power at a similar period due to stellar activity. The final orbit fit switched the two models, so that panels e) and d) show the planetary signal, while panels c) and f) show the stellar activity signal.}
\label{fig:154345_summary}
\end{center}
\end{figure*}

Here, we confirm the planetary status of the planet claim for HD 154345. \cite{Wright08} announced the detection of a true Jupiter analog, with \msini\ = 0.95 \mjup\ and an orbital period of 9.2 yr, corresponding to an orbital separation of 4.2 AU. This  paper also presented strong evidence for a stellar magnetic activity cycle with a periodic timescale of roughly nine years. As the CPS group continued to observe HD 154345 over the next few years, the planet candidate's RV signature and the corresponding S-values appeared to be strongly in phase, and \cite{Wright16} noted that the candidate may be a false positive. However, in the twelve years since HD 154345 b was initially reported, HIRES RV measurements and activity metrics have drifted from being completely in phase to being completely out of phase, as seen in Figure \ref{fig:154345_timeseries}, and therefore are not linearly correlated. This strongly implies that this Jupiter analog candidate cannot be attributed to stellar activity, and that this candidate should be cemented as a confirmed planet. \texttt{RVSearch} detects two signals in our HD 154345 dataset, both close to 9 yr, as seen in Figure \ref{fig:154345_summary}. We attribute the circular orbit with a greater RV amplitude to HD 154345 b, and the weak, eccentric signal to stellar activity.

\subsection{HD 26965}

\begin{figure*}[ht!]
\begin{center}
\includegraphics[width=\textwidth]{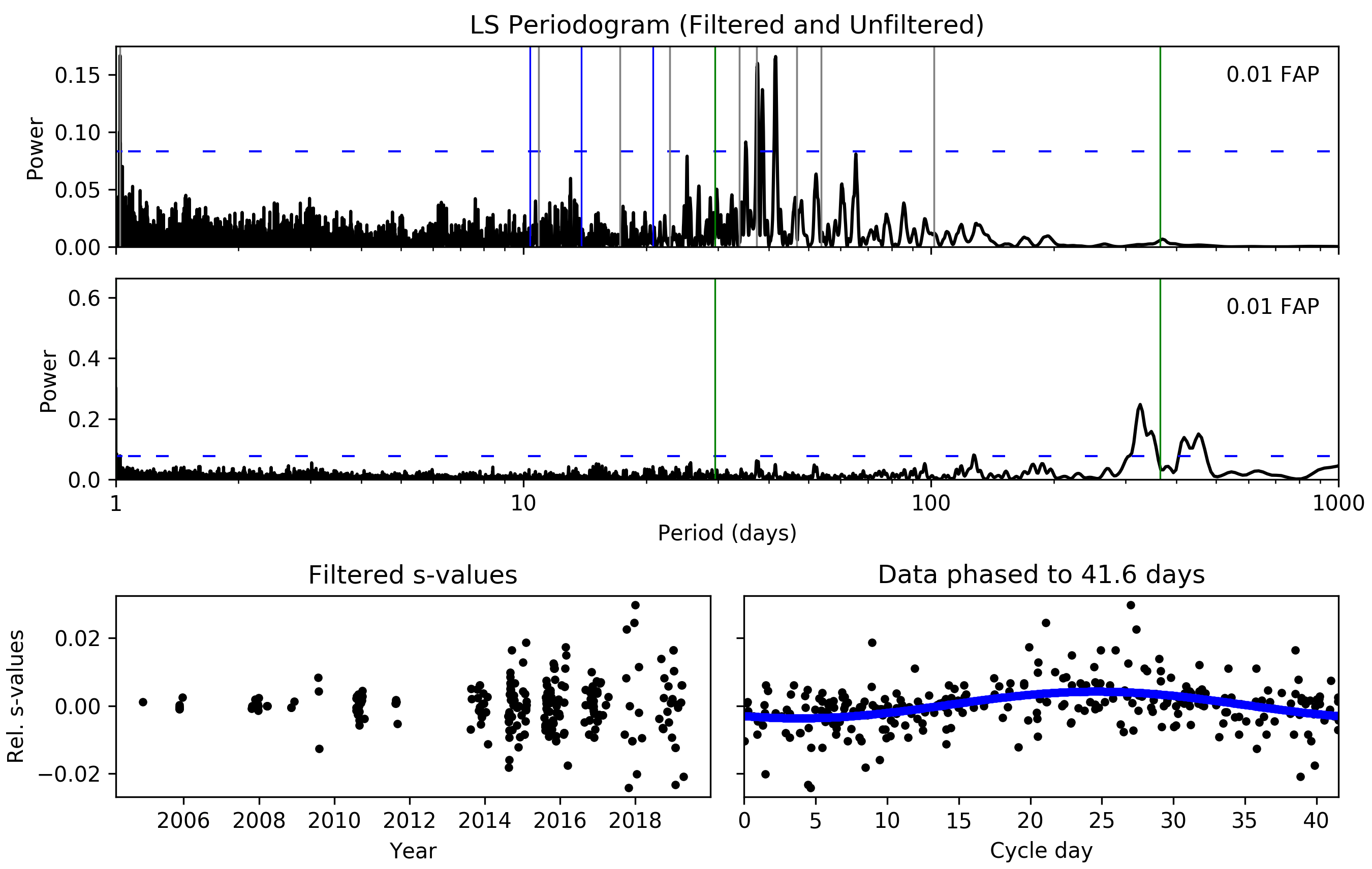}
\caption{Stellar rotation analysis of HIRES S-values for HD 26965. The top panel shows a Lomb--Scargle periodogram of the S-values after we applied a high-pass filter to them, to remove the impact of the long-period magnetic activity cycle. The middle panel shows a periodogram of the raw S-values. The top panel shows significant periodicity near 40 days, with a maximum at 41.6 days. The bottom-left panel shows the filtered S-values, while the bottom-right panel shows the filtered S-values phased to 41.6 days; there appears to be a coherent signal at this period, implying stellar rotation with this period.}
\label{fig:26965_rotation}
\end{center}
\end{figure*}

\begin{figure*}[ht!]
\begin{center}
\includegraphics[width=\textwidth]{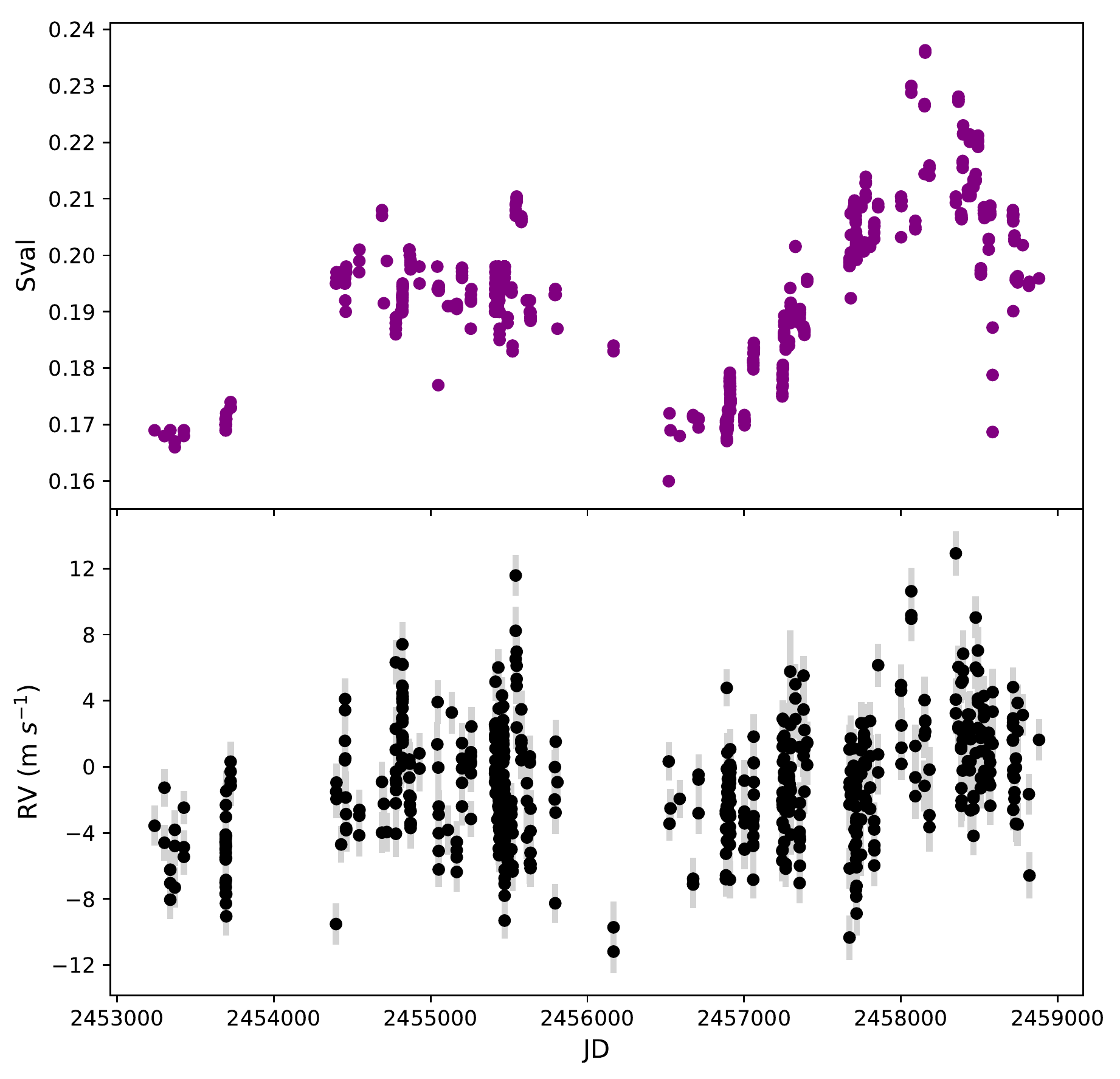}
\caption{HIRES post-upgrade S-values and RVs for HD 26965. The two datasets both have long-period power and are in phase with each other.}
\label{fig:26965_magnetic}
\end{center}
\end{figure*}

\cite{Ma18} reported a 42.4 day super-Earth orbiting the nearby star HD 26965, using datasets taken by multiple spectrographs, including HIRES. We detected significant periodicity at 42 days in the HIRES S-value measurements as seen in Figure \ref{fig:26965_rotation}, and determined that 42 days is the likely stellar rotation period of HD 26965. There is also evidence of a long-period magnetic activity cycle, as seen in the juxtaposition of S-values and RVs in Figure \ref{fig:26965_magnetic}.

\subsection{HD 34445}

\begin{figure*}[ht!]
\begin{center}
\includegraphics[width=0.6\textwidth]{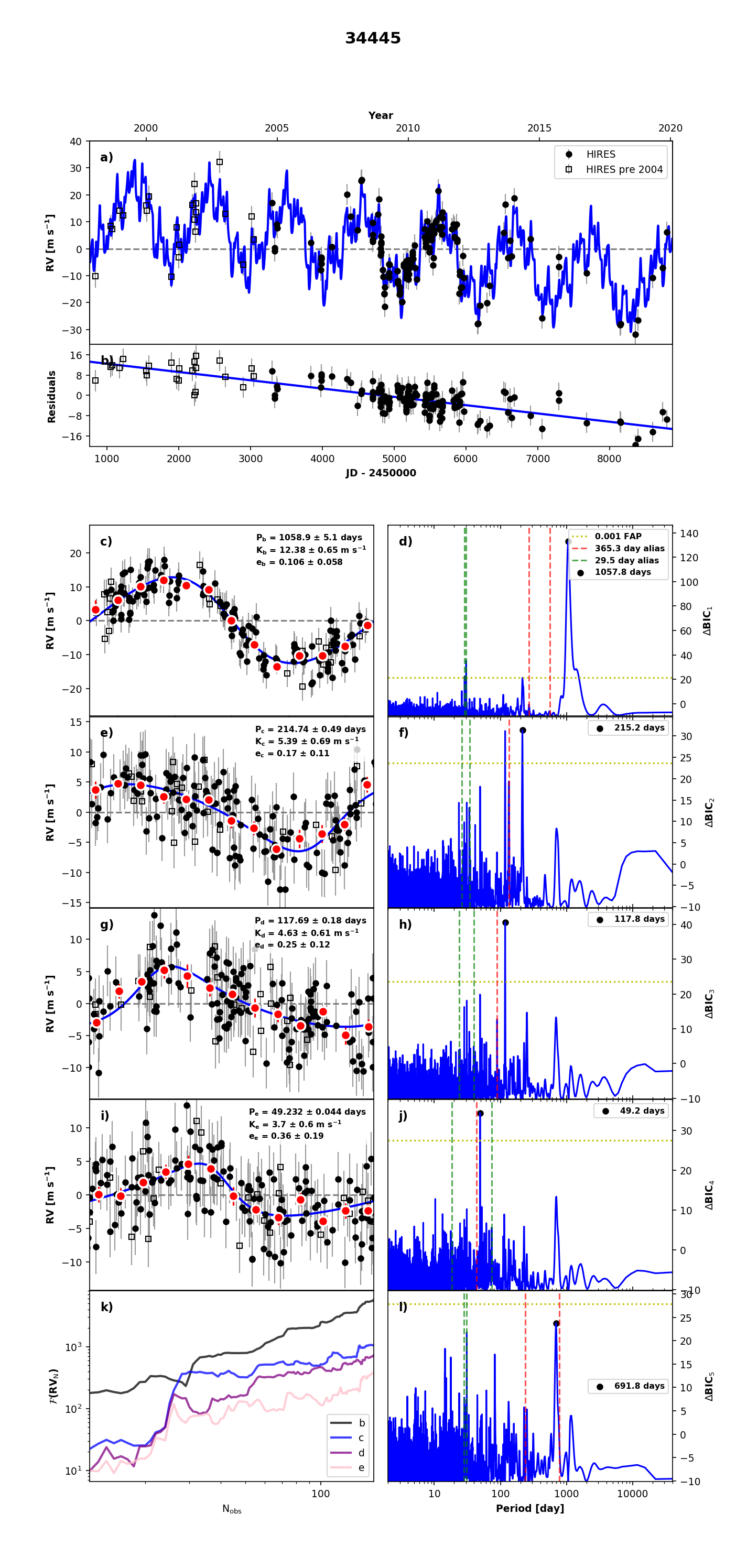}
\caption{\texttt{RVSearch} summary plot for HD 34445; see Figure \ref{fig:hip109388_summary} for description. \texttt{RVSearch} first recovered the known giant planet, }
\label{fig:34445_summary}
\end{center}
\end{figure*}

\begin{figure*}[ht!]
\begin{center}
\includegraphics[width=0.9\textwidth]{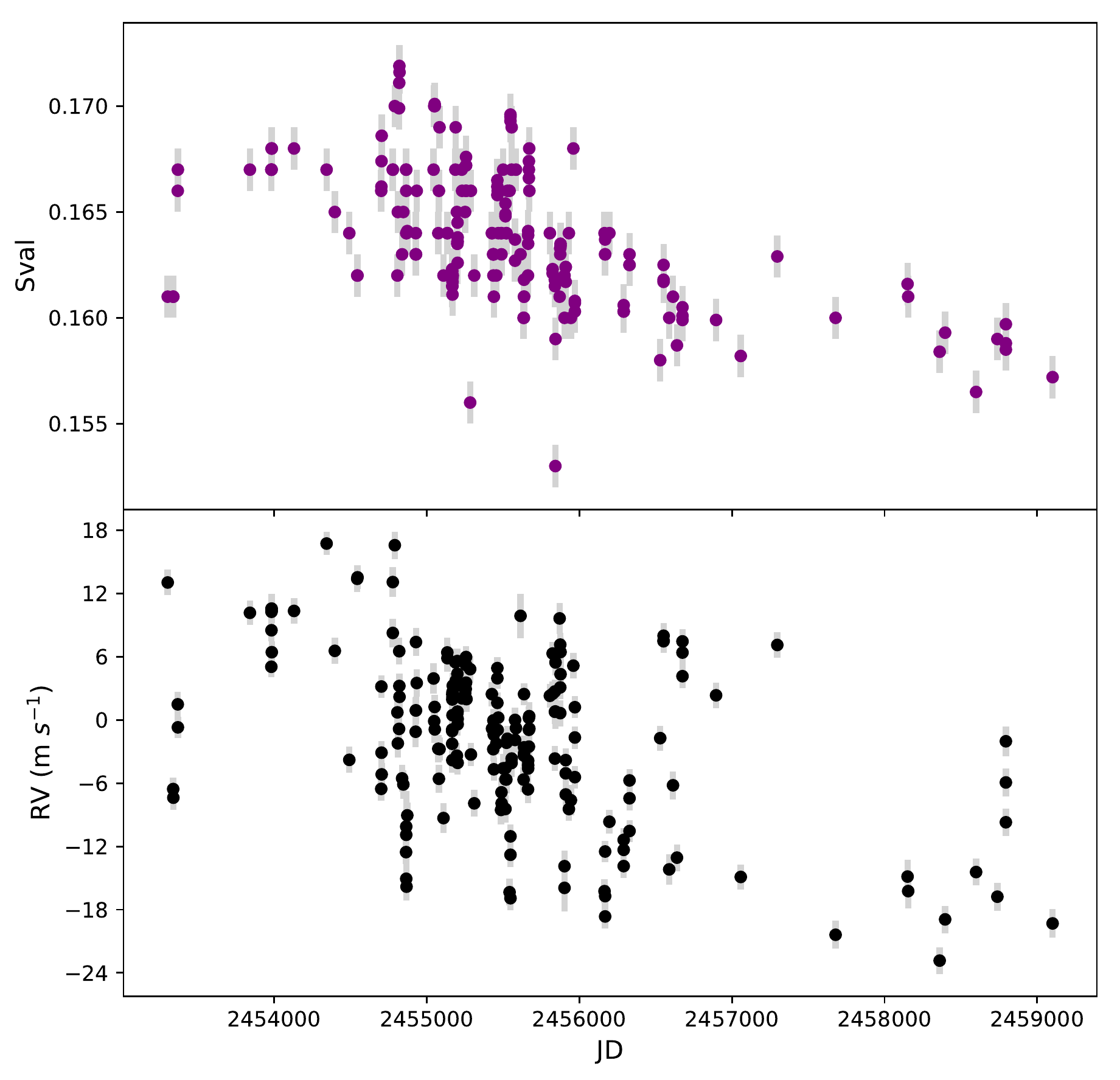}
\caption{HIRES post-upgrade RV and S-value activity timeseries for HD 34445, with the giant planet RV model subtracted. Note that these two datasets share a negative long-term trend, which we believe accounts for the claimed 5,700-day planet in the system.}
\label{fig:34445_timeseries}
\end{center}
\end{figure*}

\begin{figure*}[ht!]
\begin{center}
\includegraphics[width=\textwidth]{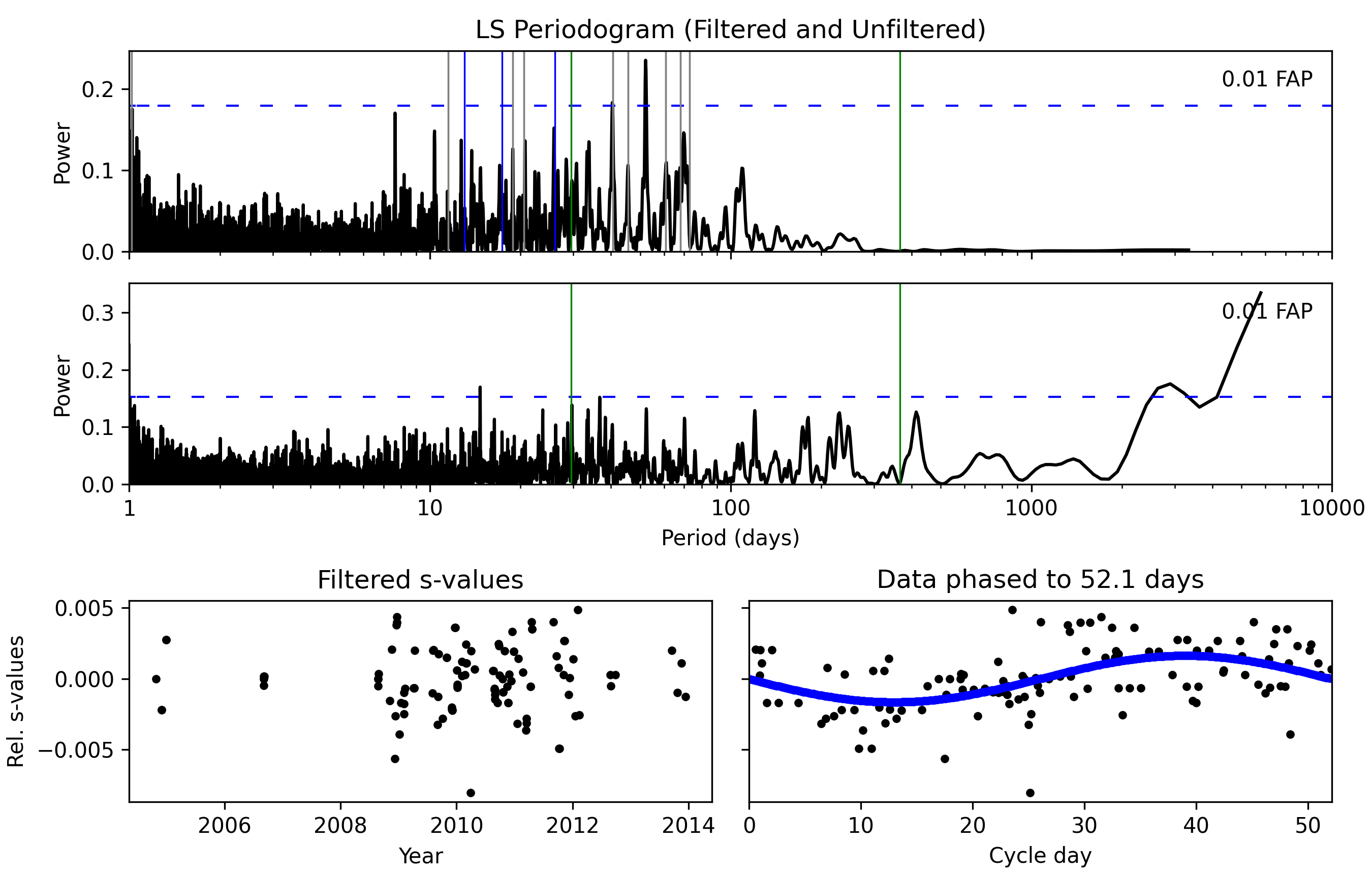}
\caption{Stellar rotation analysis of HIRES S-values for HD 34445. The top panel shows a Lomb--Scargle periodogram of the S-values after we applied a high-pass filter to them, to remove the impact of the long-period magnetic activity cycle. The middle panel shows a periodogram of the raw S-values. The top panel shows significant periodicity around 52.1 days. The bottom-left panel shows the filtered S-values, while the bottom-right panel shows the filtered S-values phased to 52.1 days; there appears to be a coherent signal at this period, implying stellar rotation with this period. This led us to label the 49 day claimed planet as a false positive, since there is insufficient evidence to distinguish it from stellar rotation.}
\label{fig:34445_rotation}
\end{center}
\end{figure*}

\begin{figure*}[ht!]
\begin{center}
\includegraphics[width=\textwidth]{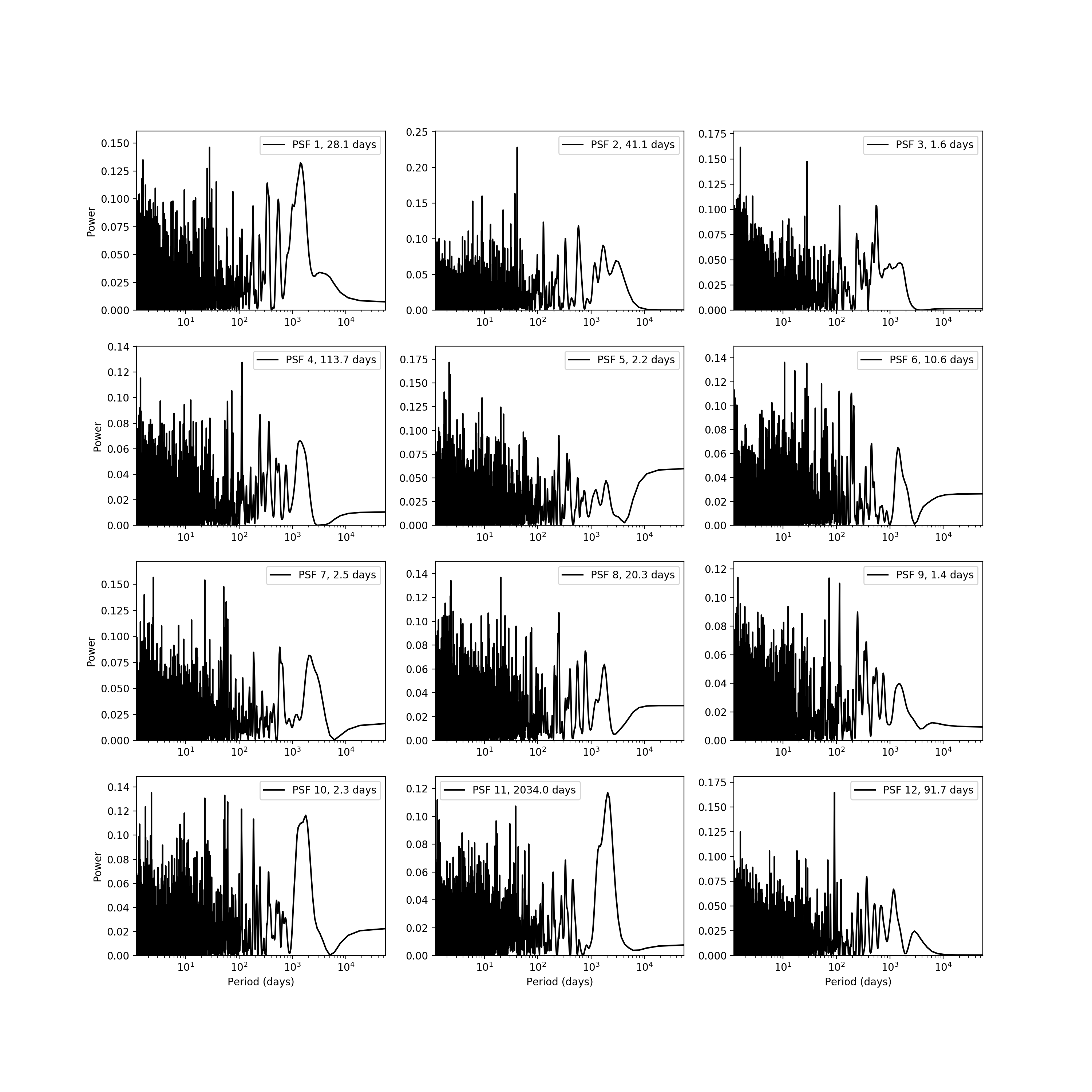}
\caption{PSF Lomb--Scargle periodograms for HD 34445. Each panel corresponds to a Doppler code PSF fitting parameter.}
\label{fig:34445_psf_periodogram}
\end{center}
\end{figure*}

\begin{figure*}[ht!]
\begin{center}
\includegraphics[width=\textwidth]{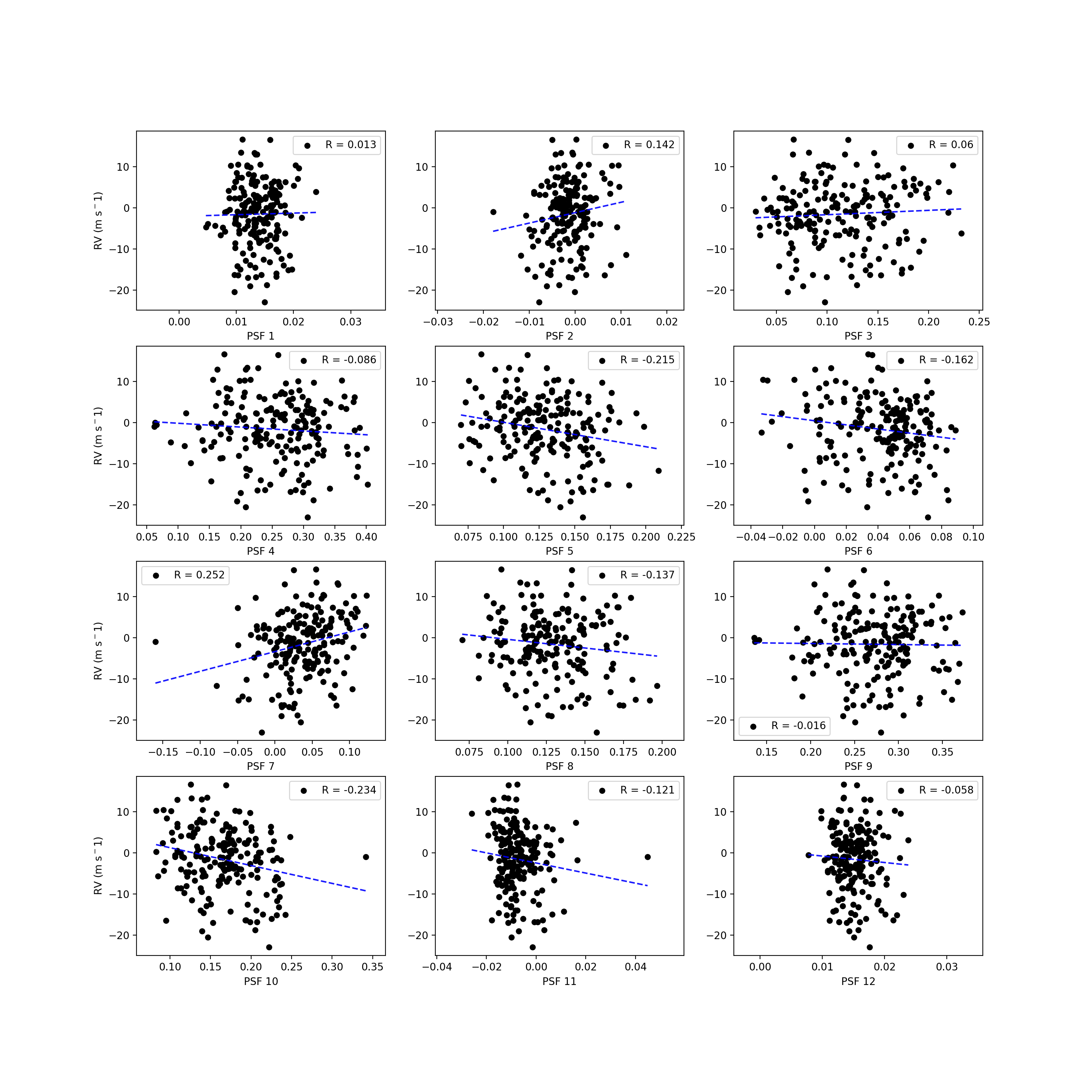}
\caption{PSF correlation plots for HD 34445, without the RV signature of the star's giant planet. Each panel corresponds to a Doppler code PSF fitting parameter, with PSF value on the x-axis and RV without the giant planet signature on the y-axis. Dashed blue lines are least-squares linear fits. R is the Pearson correlation value; multiple PSF parameters have $|R|$ $>$ 0.15.}
\label{fig:34445_psf_correlation}
\end{center}
\end{figure*}

\cite{Howard10} reported a giant planet orbiting this star at a period of 1049 days. \cite{Vogt17} reported five small planets, claiming evidence in LCES-derived HIRES radial velocities. \texttt{RVSearch} detected the giant planet and three of the five small planet claims, as seen in the summary plot shown in Figure \ref{fig:34445_summary}. The longest-period candidate among the five, not modeled as a Keplerian here, clearly correlates with HIRES S-values; we model this signal with a linear trend, for simplicity. Figure \ref{fig:34445_timeseries} juxtaposes the HIRES S-values and corresponding RVs, minus the Keplerian signal of the system's giant planet. As for the three other periodic signals that we detect, two are likely HIRES systematics and one is likely stellar rotation We detected significant periodicity at 52 days in the HIRES S-value measurements as seen in Figure \ref{fig:34445_rotation}, and determined that 52 days is the likely stellar rotation period of HD 34445. This places our weak detection of the 49 day claimed planet candidate under suspicion, and we have labeled it as a false positive in our catalog. There is also evidence of semiannual HIRES systematics, as seen in Figure \ref{fig:34445_psf_correlation}, which shows the correlation between HIRES RVs minus the giant planet signature and PSF parameters, and in Figure \ref{fig:34445_psf_periodogram}, which shows periodograms of each PSF parameter time series. Multiple PSF parameters correlate ($|R|$ $>$ 0.15) with the RV residuals, and multiple parameters show periodicity around one-third and one-fourth of a year. The two claimed planets at 118 and 215 days are close to one-third and one-half of a year, respectively, and show weak and equal-strength signatures in their \texttt{RVSearch} periodograms, as seen in Figure \ref{fig:34445_summary}. Therefore, we have labeled these signals as false positives in our catalog.

\end{document}